\documentclass[%
 amsmath,amssymb,
 aps,
rmp,
]{revtex4-2}

\usepackage{graphicx}
\usepackage{dcolumn}
\usepackage{bm}
\usepackage[english]{babel}
\usepackage{xcolor}
\colorlet{BLUE}{blue}

\usepackage{natbib}
\renewcommand*{\addcontentsline}[3]{\addtocontents{#1}{\protect\contentsline{#2}{#3}{}}}

\begin{document}

\title{Ultrafast Electron Diffraction: Visualizing Dynamic States of Matter}

\author{D. Filippetto}
\affiliation{Lawrence Berkeley National Laboratory, Berkeley, California, 94720, USA}%

\author{P. Musumeci}
\affiliation{Department of Physics and Astronomy, University of California at Los Angeles,
Los Angeles, California 90095, USA}%

\author{R. K. Li}
\affiliation{Department of Engineering Physics, Tsinghua University, Beijing 100084, China}%
\affiliation{Key Laboratory of Particle and Radiation Imaging, Tsinghua University, Ministry of Education, Beijing 100084, China}

\author{B. J. Siwick}
\author{M. R. Otto} 
\affiliation{Centre for the Physics of Materials, Department of Physics and Department of Chemistry,
McGill University, Montreal, Quebec H3A 2T8, Canada}%

\author{M. Centurion}
\author{J.P.F. Nunes}
\affiliation{University of Nebraska-Lincoln, 855 N 16th Street, Lincoln, Nebraska 68588, USA}%

\begin{abstract}
Since the discovery of electron-wave duality, electron scattering instrumentation has developed into a powerful array of techniques for revealing the atomic structure of matter. Beyond detecting local lattice variations in equilibrium structures with the highest possible spatial resolution, recent research efforts have been directed towards the long sought-after dream of visualizing the dynamic evolution of matter in real-time. The atomic behavior at ultrafast timescales carries critical information on phase transition and chemical reaction dynamics, the coupling of electronic and nuclear degrees of freedom in materials and molecules, the correlation between structure, function and previously hidden metastable or nonequilibrium states of matter. Ultrafast electron pulses play an essential role in this scientific endeavor, and their generation has been facilitated by rapid technical advances in both ultrafast laser and particle accelerator technologies. This review presents a summary of the remarkable developments in this field over the last few decades. The physics and technology of ultrafast electron beams is presented with an emphasis on the figures of merit most relevant for ultrafast electron diffraction (UED)
experiments.  We discuss recent developments in the generation, manipulation and characterization of ultrashort electron beams aimed at improving the combined spatio-temporal resolution of these measurements. The fundamentals of electron scattering from atomic matter and the theoretical frameworks for retrieving dynamic structural information from solid-state and gas-phase samples is described.  Essential experimental techniques and several landmark works that have applied these approaches are also highlighted to demonstrate the widening applicability of these methods. Ultrafast electron probes with ever improving capabilities, combined with other complementary photon-based or spectroscopic approaches, hold tremendous potential for revolutionizing our ability to observe and understand energy and matter at atomic scales.

\end{abstract}

\maketitle

\tableofcontents

\newpage

\section{Introduction}
\label{sectionI}

The discovery of the wave nature of the electron at beginning of the 20th century \cite{davisson_reflection_1928,thomson1928diffraction,Nobel} marked the start of a new era in the human quest for an atomic-level perspective on the architecture of the microscopic world.
Since then, the development of scientific tools exploiting the sub-\AA~imaging power of electron waves and their strong interaction with matter has seen rapid growth, starting with the invention of the transmission electron microscope (TEM) by Ruska in 1932 \cite{Ruska_1932}. Today, electron diffraction and microscopy are primary enablers of research and development in many scientific disciplines including chemistry, biology, physics and material sciences as well as in many industries. 

Over the years, continuous improvements in charged particle optics \cite{aberrationCorrection_beck_1979, Haider1998, Rose1990, Scherzer1947, Haider1995}, detectors, and new algorithms, have culminated in spatial resolution well below atomic spacing in matter and approaching the limit set by lattice vibrations \cite{Chen21_resolutionlimit}. In diffraction mode, electron optics can form beams able to illuminate areas well below 1~nm. 
These spectacular developments indicate that there is less to gain from
further improvements to spatial resolution alone than there once was, and
other frontiers in instrumentation development are beginning to emerge or see
renewed interests. These include improving elemental contrast, \emph{in-situ}
investigations in diverse sample environments (liquid and gas) and under
tunable conditions of temperature, pressure, as well as enhanced
time-resolution to interrogate systems far from equilibrium 
\cite{zhu2015future}. At the temporal resolution frontier, the overarching 
goal is to make the dynamic processes in materials across the sub-\AA~to 
micrometer length scales directly accessible, ``while they are occurring", 
under non-equilibrium conditions. This goal has become a reality by combining 
the atomic-scale information that can be obtained using electrons, with the 
femtosecond (10$^{-15}$ s) time resolution afforded by ultrafast laser 
technology. This review seeks to provide an account of the development of 
temporally-resolved electron diffraction to date, with a focus on the fundamentals of pulsed electron beams and their applications to visualizing dynamic, non-equilibrium states of matter from the analysis of diffraction patterns. 

Time-resolved electron scattering emerged first as a new scientific technique for structural dynamics in the early 1980s~\cite{Mourou1982}. The development of chirped pulse amplification and ultrafast optical laser systems~\cite{strickland_compression_1985} enabled the generation of short bursts of photo-electrons almost perfectly synchronized with suitable pump pulses to initiate or trigger dynamics in a specimen. Prior to the use of ultrafast laser-driven  photoemission, beams used in time-resolved electron microscopes were emitted via thermal or field emission. Time-resolution in these instruments was determined by the switching speed of mechanical or electronic shutters used to modulate the electron emission or shorten the exposure times of detector cameras and was limited to the 100 nanosecond to microsecond scale or above ~\cite{Ischenko1983, Bostanjoglo1987}. The absence of temporal structure in the beam and the lack of fast triggers for pulsed electron emission and specimen excitation, precluded access to the fastest time scales restricting conventional electron scattering instrumentation to the study of in-equilibrium systems by static images, diffraction patterns and spectra.
When technological developments provided direct access to the observation of the most fundamental processes in materials as they occur, they naturally ignited a revolution in research labs around the world ~\cite{Sciaini2011, king_JAP2005, Zewail_2010, miller2014, Musumeci:springer}. Sub-picosecond time scales unlocked access to fundamental dynamical processes in condensed matter and chemistry, such as nanoscale heat transfer, phonon transport and chemical bond formation, while the sub-atomic electron wavelength and the strong electron-matter interaction cross section enabled atomic-scale recording of dynamical processes such as irreversible phase transitions in solids \cite{siwick_atomic-level_2003}, the formation of molecular bonds~\cite{Ihee2001}, and very recently, hydrogen bond dynamics in liquids \cite{yang21_water, Lin21_ionizedwater}.

\begin{figure}[ht]
\centering
\includegraphics[width=0.8\columnwidth]{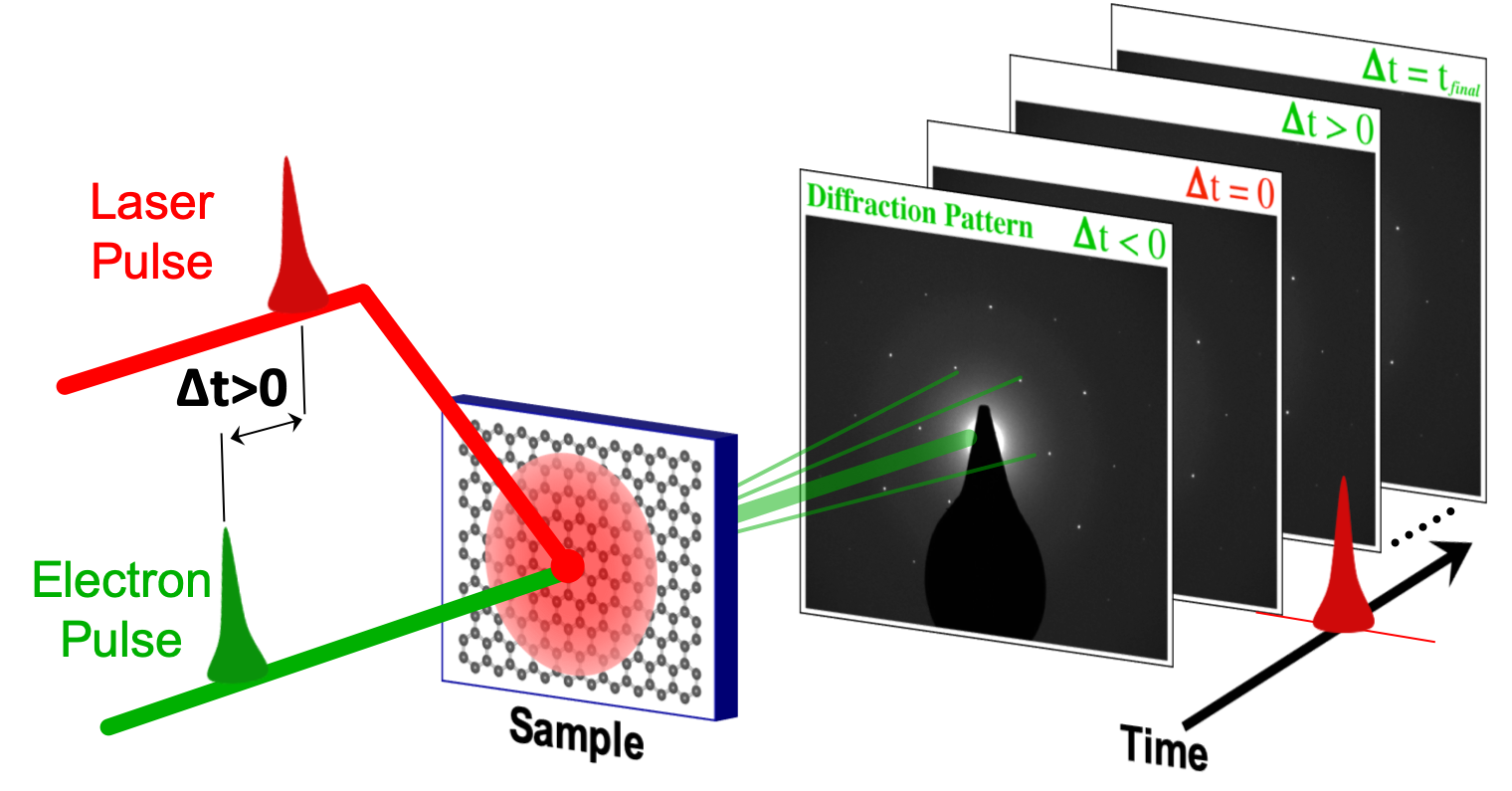}
\caption{Conceptual schematic for a pump-probe UED setup in transmission geometry.}
\label{fig:methodology}
\end{figure}

Ultrafast electron scattering is a rapidly growing cross-disciplinary field, drawing from decades of instrument developments in the physical and energy science areas, such as electron microscopy, particle accelerator and laser technology, condensed matter physics and ultrafast chemistry. Atomic-level information can be retrieved via different operating modes such as microscopy, diffraction and spectroscopy, isolating specific electron-matter interaction channels. Elastic and inelastic scattering processes encode sample information respectively on the angle and energy of the scattered electrons, while the specific electron optics setup determines the mapping of the electron parameters into the detector plane, commonly energy, angle (momentum transfer) or real space. Furthermore, the geometry of the interaction and the detector collecting angle can be optimized for the study of surface structures in bulk materials (reflection mode) or for the characterization of bulk structure in thin films, liquids and gases (transmission mode). This review will mainly focus on the technological and scientific advancements in transmission ultrafast electron diffraction (UED), which has seen a very rapid increase in interest over the last decade. Sustained by scientific discoveries of increasing impact, UED is now considered an established technique in the ultrafast sciences. 
However, it is worth noting that the vast majority of techniques discussed here can be directly applied to the other operating modes mentioned above. 
Throughout the manuscript, the topics are presented without any assumption on the probe electron beam energy, whose dependence is explicitly derived and discussed where needed. Such approach extends the relevance of the treatment proposed to UED beamlines with probe energies in the keV-to-MeV range. Low energy, eV-scale electron diffraction (LEED) are not included, since they are not commonly used in transmission mode, and therefore face a different set of challenges.

A conceptual schematic of the transmission UED technique in pump-probe geometry is summarized in Fig.~\ref{fig:methodology}. A short (compared to the relevant timescales) optical pulse impinges on the specimen at a time $t_0$, initiating the process of interest over a selected region. A paired electron pulse is spatially overlapped with the optical pulse at the sample and illuminates the probed area at a time $t_e$, with a delay of $\Delta t=t_e-t_0$. Diffraction patterns are acquired as $\Delta t$ varies from negative to positive values, and provide temporal snapshots of the atomic structural evolution from the initial equilibrium, through the transient, up to a final equilibrium state, which may be identical to the initial state or different. 

A short summary of the structure of this review article follows. After reviewing fundamental concepts in diffraction in Sec.~\ref{sect1.a}, we will define a common metric for discussion and comparison of electron sources that will be used throughout the article (Sec.~\ref{sec:brightness}), and briefly compare the different operating modes (Sec.~\ref{sect1.differentmodalities}) in terms of electron beam requirements. The scientific niche of UED setups will be discussed as introductory motivation to the following Sec.~\ref{sectionII}, which describes in details the state-of-the-art techniques for electron generation (Sec.~\ref{sectionII.b}), beam dynamics (Sec.~\ref{sectionII.c}), acceleration technologies (Sec..~\ref{section:acc_tech}), and spatio-temporal control of femtosecond electron beams including detection (Sec.~\ref{sectionII.d} and~\ref{sectionII.e}).
Sections~\ref{sectionIII} and ~\ref{sect:GUED} discuss respectively the case of solid state and gas-phase targets. After an overview of the main processes of interest we clarify sample requirements and describe the interaction geometry. We then review the main techniques and challenges in data analysis, providing insightful information on the requirement for source stability and reliability.
We then conclude with future prospects for UED techniques in Sec.~\ref{Outlook}.

\subsection{Electrons as probes of matter}
\label{sect1.a}
The usefulness of electron diffraction stems from the large amount of information about the sample atomic-scale structure that can be extracted from a typical diffraction pattern. In order to understand the basic principles of electron scattering, both particle and wave aspects of the nature of electrons need to be considered \cite{reimer2013,spencebook,carterTEMbook}. Diffraction effects in particular result from the scattering of electron waves of characteristic de Broglie wavelength $\lambda = h/p$, where $h$ is the Planck constant, $p = m c \beta \gamma$ is the electron momentum and $m$ and $c$ are the electron rest mass and speed of light, respectively. $\beta = \sqrt{1-1/\gamma^2}$ is the electron velocity normalized to $c$. In more quantitative terms, the de Broglie wavelength for 4 MeV (100 keV) electrons is $\lambda = 0.277 (3.701)$ pm, which highlights the potential of using electrons to achieve atomic-scale spatial resolution.

When such an electron wave is incident on a target, the scattered wave can be described by the complex amplitude $f(\theta,\phi)$, which indicates the probability of finding a scattered electron at angle $\theta$ and $\phi$ with respect to the incident direction. Tying together particle and wave approaches to electron scattering, this scattering amplitude depends on the detail of the interaction between the electron and the target and is related to the the differential scattering cross section as $\frac{d\sigma}{d\Omega}=|f(\theta,\phi)|^2$. In the first Born approximation (kinematic scattering), we can write the amplitude of the scattered wavefunction in the direction $\mathbf{k'}$, where $\mathbf{k - k' = s}(\theta,\phi)$ as the Fourier transform of the target scattering potential, $V(\mathbf{r})$:
\begin{equation}\label{eqn:e_form_factor}
f(\mathbf{s}) = -\frac{m}{2\pi\hbar^2}\int \textup{d}\mathbf{r}V(\mathbf{r})\exp\left(-i\mathbf{s}\cdot \mathbf{r}\right).
\end{equation}
where the momentum transfer magnitude $|\mathbf{s}|= \frac{4 \pi}{\lambda} \sin \frac{\theta}{2}$.

In the case where the target is an atom, the largest contribution to the elastic scattering amplitude will be the Rutherford scattering from the atomic nucleus with a smaller contribution from the surrounding electrons. Following Salvat et al. \cite{salvat1987analytical,salvat1993elastic}, it is customary to express the  (azimuthally symmetric) elastic scattering from an atom with atomic number $Z$ in terms of the momentum transfer $s$ as

\begin{equation}
\frac{d\sigma}{ds} = \frac{4 Z^2}{s^4 a_0^2}\frac{1-\beta^2\sin^2 \frac{\theta}{2}}{1-\beta^2} \left(1 - F(s)^2 \right)^2
\label{cross_section}
\end{equation}

where $a_0$ is the atomic Bohr radius, and $F(s) = \sum_i A_i \frac{\alpha_i^2}{s^2+\alpha_i^2}$ is a function which depends on the approximation details of the screened atomic potential. The sum over the index $i$ can include as many terms as desired for improved accuracy. As an example for silver we have $A_i = [0.25,0.62,0.13] $ and $\alpha_i = [15.59, 2.74, 1.14]$ \r{A}$^{-1}$.

\subsubsection{The role of electron energy in electron scattering} 
\label{sect1.a.1}

\begin{figure}[ht]
\includegraphics[width=1\columnwidth]{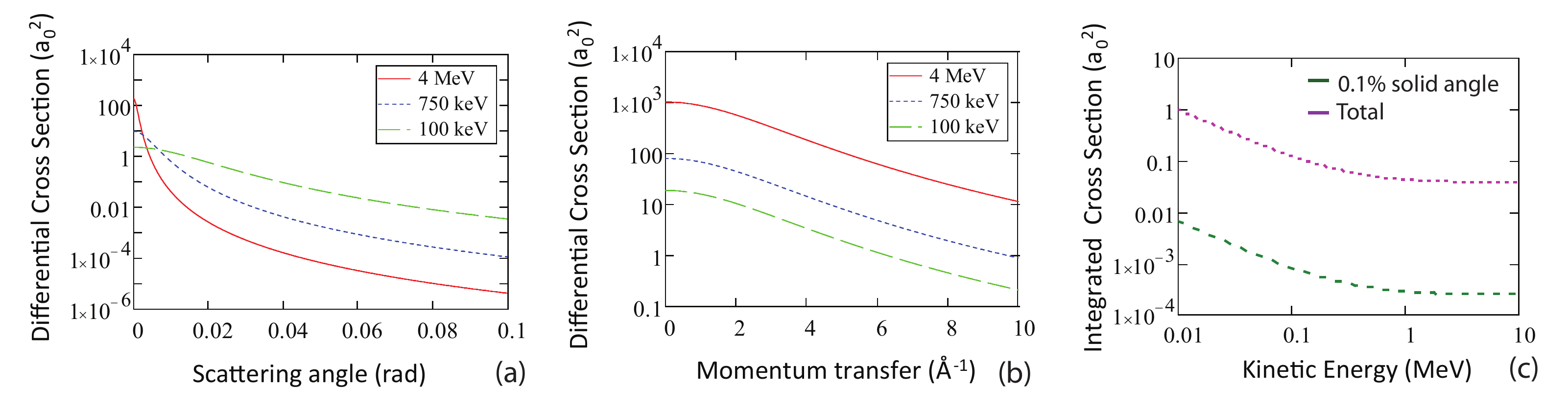}
\caption{Differential elastic scattering cross section vs. (a) scattering angle and (b) momentum transfer for 100 keV, 750 keV and 4 MeV electrons using Eq. \ref{cross_section}. 
Panel (c) shows the resulting integrated cross section over the entire solid angle (dashed) and over a small (0.1$\%$) interval around the momentum transfer $s=$~5 \AA$^{-1}$.}

\label{Fig:MeVUEDcrossection}
\end{figure}

It is instructive to plot (Fig.~\ref{Fig:MeVUEDcrossection}) the differential cross section vs. scattering angle (a) and momentum transfer (b) for various electron energies typically employed in UED beamlines \cite{BNL_UED}. The differential cross section vs. momentum transfer increases proportionally to $\gamma^2$ for relativistic electrons, essentially due to the scaling of the incident momentum of the particles. To calculate how many electrons are scattered within a given angular range, one needs to integrate the differential cross section over the detector collection angle. Some care should be taken here as the angles corresponding to a given $s$ depend on the incoming electron energy. So for example if we are interested in the information around $s=$~5 \AA$^{-1}$, we'd have to collect the scattered intensity in an interval around 29 mrad for 100 keV electrons and 2.2 mrad for 4 MeV electrons. The results of this integration are shown in Fig. \ref{Fig:MeVUEDcrossection}(c) which clarifies that the number of scattered electrons (integrated over the entire solid angle, or even just in a small angular interval around a region of interest) is nearly an order of magnitude smaller for 4 MeV than for 100 keV.

The total integrated cross section can be used to calculate the elastic mean free path, i.e. the statistical average distance of propagation inside the sample over which the electrons will undergo one scattering event as
$\frac{1}{n \sigma}$
where $\sigma$ is the integrated cross section and $n$ the density of scatterers in the material under study. Directly resulting from the scaling in Eq. \ref{cross_section}, illustrated in the plots of Fig. \ref{Fig:MeVUEDcrossection}, elastic mean free paths for higher energy electrons are significantly longer than for lower energy particles in the same material. For example, in an Al sample, the elastic mean free path is 38 nm at 100 keV and 250 nm at 4 MeV. For higher energy electrons, this allows the use of thicker samples, or alternatively yields lower number of scattering events for equal thickness of materials.

In cases where the mean free path is shorter than the thickness of the specimen, then it is likely that electrons would undergo more than one scattering event. In order to quantitatively extract information from the diffraction pattern, one must go beyond the simple kinematical approximation (one scattering event per electron) and utilize the more complex dynamical diffraction theory \cite{ZuoSpence,wang2013}. 

\subsubsection{Scattering from gaseous targets} 
\label{GEDtheory}
If the sample is made up by a large number of scattering targets (atoms), the total scattering amplitude will be the sum of the individual waves. The so called scattering form factor $F$ can then be written using the independent atom model as the sum of the atomic scattering factors $f_j$ from all the atoms in the with atomic coordinates $\mathbf{r}_j = (x_j,y_j,z_j)$ multiplied by a phase factor which takes into account the difference in phase between the scattered waves in terms of the momentum transfer vector $\mathbf{s}$ 
\begin{equation}
F(\theta) = \sum_j f_j(\theta) e^{i\mathbf{s} \cdot \mathbf{r}_j}
\label{Eq:structurefactor}
\end{equation}

In gas phase electron diffraction, high energy electrons (keV to MeV) elastically scattered from an ensemble of molecules produce an interference pattern on a detector, from which structural information on the molecule can be retrieved. The total scattering intensity can be obtained by the incoherent sum of the scattering from each molecule since the transverse coherence of the electron beam is typically smaller than the distance between molecules. 
For randomly oriented molecules, averaging over all possible orientation results in a scattered intensity only dependent on the polar angle (circular symmetry diffraction pattern) and that can be written as a function of the momentum transfer magnitude $s$ as $I_\mathrm{Total}(s) = I_\mathrm{A}(s) + I_\mathrm{MOL}(s)$. We can separate the contributions to the total scattering in two terms: the first is atomic scattering term $I_\mathrm{A}(s) = \sum_{m=1}^N  f_m^*(s) f_m(s)$ and contains no structural information and only depends on the atoms present in the molecule.; the second term, known as molecular scattering, can be written as
\begin{equation}
    I_\mathrm{MOL}(s) = \sum_{m=1}^N \sum_{n=1, m \neq n}^N f_m^*(s) f_n(s)  \frac{\sin(s r_{mn})} {sr_{mn}}
\end{equation}
where $N$ is the number of atoms in the molecule, and $\mathbf{r}_{mn}$ is the distance vector from atom $m$ to atom $n$ (assuming static molecular structure), and contains the interference between all atom pairs in the form of a sinusoidal modulation in the intensity of the diffraction pattern. 

For ease of analysis and to compensate for the fast decrease in scattering intensity with $s$, the modified scattering intensity is used: 
        \begin{equation}
            sM(s) = \frac{I_\mathrm{MOL}(s)} {I_{A}(s)} s
        \label{Eq:sM}
        \end{equation}

The most straightforward method to extract structural information from diffraction data is to Fourier (sine) transform the scattering intensity into a Pair Distribution Function ($PDF$) \cite{Hargittai1988}. The position of peaks in the PDF reflects interatomic distances in the molecule, with peak amplitudes proportional to the density  (in the case where there are multiple atom pairs with overlapping distances) and the product of the scattering amplitudes from each atom in the pair, while it is inversely proportional to the distance $r$. In practice, the diffraction pattern is only measured up to a maximum value $s_\mathrm{Max}$, resulting in a truncated $sM(s)$. To avoid introducing artifacts into the PDF from the sine transform of a truncated signal, a damping factor $k$ is added as:
        \begin{equation}
            PDF(r) = \int_{0}^{s_\mathrm{Max}} sM(s) \sin(sr) e^{-ks^2} ds
        \end{equation}        
where $r$ is the real space distance between atom pairs.

\begin{figure}[ht]
\centering
\includegraphics[width=1\columnwidth]{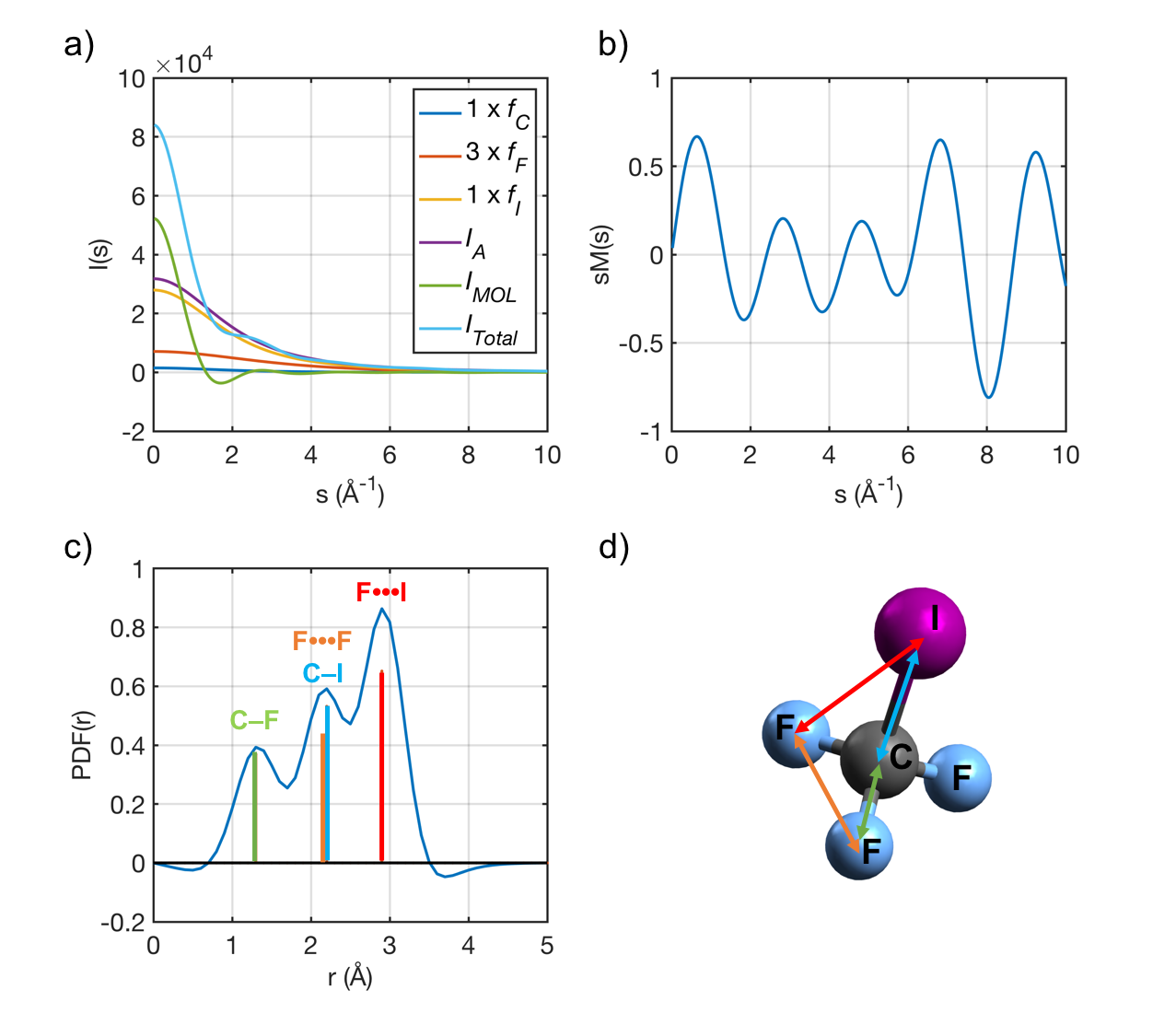}
\caption{Simulated gas-phase electron scattering for CF$_{3}$I showing the a) relative contributions of the each atom type to the atomic terms, and contributions of the atomic and molecular terms to the total scattering, b) simulated $sM(s)$ and c) PDF$(r)$ and d) a depiction of the inter atomic distances in the molecular color-coded to the peaks in the PDF$(r)$.}
\label{fig:GUED_scat_terms}
\end{figure}

The spatial resolution of the measurement is strictly defined by the width of the peaks in the $PDF$, and thus depends only on the value of $s_\mathrm{Max}$. Note that this value determines whether two nearby distances can be resolved in the $PDF$, but it does not determine the precision with which any individual distance can be determined. Finding a distance is equivalent to finding the center of the peak, which typically can be done to a value much smaller than the width of the peak, and depends strongly on the SNR of the measurement. Figure \ref{fig:GUED_scat_terms} shows the relative contributions of the molecular and atomic scattering terms to the total simulated scattering signal of CF$_{3}$I and corresponding $sM(s)$ and PDF$(r)$.

\subsubsection{Scattering from crystals} 
\label{SSEDtheory}

Consider the case of a beam of electrons with wavevector $\mathbf{k}$ incident on a perfect, infinite single crystal consisting of periodically arranged \textit{unit cells}, which defines the smallest repeating atomic arrangement within the material. The crystal can be described as a sum over all the $\mathbf{\alpha}$ atom positions within a unit cell, $\mathbf{r}_{\alpha}$, and an infinite sum over all the unit cell coordinates $\mathbf{R}_n$.  With these definitions the scattering potential of the entire crystal can be written as~\cite{Ashcroft1976a, wang2013, warren1990}
\begin{equation}\label{eqn:crystal_potential}
V(\mathbf{r})=\sum_n\sum_{\alpha}V_{\alpha}\left(\mathbf{r}-\mathbf{R}_n-\mathbf{r}_{\alpha} \right),
\end{equation}  
where $V_{\alpha}$ is the potential of atom $\alpha$ in unit-cell $n$. The periodicity of $V(\mathbf{r})$ ensures that the form of $V_{\alpha}(\mathbf{r}-\mathbf{R}_n-\mathbf{r}_{\alpha})$ is identical for a given pair of $n$ and $\alpha$ values.

Generalizing Eq. \ref{eqn:e_form_factor}, we can write the scattering amplitude at wavevector $\mathbf{k'}$ in terms of the momentum transfer\footnote{In literature focusing on solid-state samples, the momentum transfer is commonly denoted by $\mathbf{q}$.  The notation $\mathbf{s}$ is maintained here for consistency with other sections of this review.} $\mathbf{s}=\mathbf{k} - \mathbf{k'}$ in the single scattering (or kinematic) limit as the Fourier transform of the scattering potential $V(\mathbf{r})$:
\begin{eqnarray}\label{eqn:FT_crystal_potential}
f(\mathbf{s}) &=& 
\sum_{\{\mathbf{G}\}}\delta(\mathbf{s}-\mathbf{G})\sum_{\alpha}V_{\alpha}(\mathbf{s})\exp(-i\mathbf{s}\cdot\mathbf{r}_{\alpha}).
\end{eqnarray}
which can be understood as the as the product of the structure form factor $F$ which contains the details of the unit cell atomic composition, and the lattice or shape factor $G$ \cite{reimer2013} which depends on the shape and external structure of the crystal.

In writing Eq.~\eqref{eqn:FT_crystal_potential} we assume infinite crystal structure, and therefore the mathematical identity $G = \sum_n\exp(-i \mathbf{s}\cdot\mathbf{R}_n)=\sum_{\{\mathbf{G}\}}\delta(\mathbf{s}-\mathbf{G})$ has been applied. The reciprocal lattice vectors  $\mathbf{G}=h\mathbf{a}^*+k\mathbf{b}^*+\ell\mathbf{c}^*$ describe the periodicity of the crystal in reciprocal space and satisfy $\mathbf{G}\cdot\mathbf{R}_n=2\pi\times\textup{Integer}$ \cite{Ashcroft1976a}.   Eq.~\ref{eqn:FT_crystal_potential} demonstrates the well known \textit{Laue condition} for single crystal diffraction, which states that scattering amplitude is only non-zero when $\mathbf{s}=\mathbf{G}$;  the Bragg peaks of a diffraction pattern. 

If the crystal is not infinite, the delta function must be replaced by the finite sum over the unit cells. For example, considering a crystal with $N$ planes spaced by distance $d$, we have
\begin{equation}
G = \left(\frac{\sin (s^* N d)}{s^* d} \right)
\label{Eq:latticefactor}
\end{equation}
where $s^* = |\mathbf{s}-\mathbf{G}|$ is the deviation from the perfect Laue condition (excitation error).

The amplitude of the lattice factor $G$ is particularly important. If electrons are scattered by $N$ unit cells, at the Bragg peaks (i.e. $s^* = 0$ in Eq. \ref{Eq:latticefactor}), the lattice factor $G$ is responsible for a $N$ times increase in the scattered wave amplitude with respect to single atom case. The corresponding scattered intensity increases by a factor of $N^2$. This Bragg enhancement factor can be very significant (i.e. in excess of $10^5$ even for small microcrystalline samples). In this simplified picture, the angular width of the Bragg peaks just depends on the number of atomic planes in the sample (i.e. the shape factor of the target). In practice, as we will see below in the coherence length section, there are many other effects that must be taken into account in the width of the Bragg peaks including the angular distribution and energy spread in the probing electron wavepackets.  For the nanometer thick single crystal specimens used in UED, the measured width of a Bragg peak in the direction of the film thickness is typically determined by the finite size effects described above, while the measured width of a Bragg peak in the plane of the thin specimen is typically determined by instrumental broadening associated with the illuminating electron beam parameters.  

In Eq.~\eqref{eqn:FT_crystal_potential}, $V_{\alpha}(\mathbf{s})$ is simply proportional to the \emph{atomic form factor} $f_{\alpha}$ which is the normalized Fourier transform of the atomic potential for an isolated (spherically symmetric) atom $\alpha$. While the assumption of spherical symmetry often provides the starting point for crystallographic calculations, it is important to keep in mind that chemical bonding in the solid will modify the symmetry of the atomic scattering factors somewhat and can lead to observable effects in diffraction experiments. The \textit{crystal structure factor}, defined as $F_0(\mathbf{s}=\mathbf{G}) = \sum_{\alpha}V_{\alpha}(\mathbf{G})\exp(-i\mathbf{G}\cdot\mathbf{r}_{\alpha})$~\cite{fultz2012}, determines the scattering amplitude into the Bragg peak located at $\mathbf{s}=\mathbf{G}$, and depends sensitively on the relative position of atoms in the unit cell. 

The intensity of electron scattering as a function of $\mathbf{s}$, the quantity measured by an electron imaging detector, is~\cite{wang2013}:
\begin{equation}\label{eqn:scattering_intensity}
I(\mathbf{s}) \propto G^2\left(\mathbf{s}-\mathbf{G}\right) \sum_{\alpha}\sum_{\beta}V_{\alpha}(\mathbf{s}) V_{\beta}(\mathbf{s})\exp\left(-i\mathbf{s}\cdot(\mathbf{r}_{\alpha}-\mathbf{r}_{\beta})\right) 
\end{equation} 
The phase of the scattering amplitude is lost by intensity detection, resulting in the well known \textit{phase problem} of crystallography.
The result in Eq.\eqref{eqn:scattering_intensity} can be generalized in a straightforward manner to polycrystalline samples by appropriate integration of Eq.~\eqref{eqn:scattering_intensity} as described in detail by~\citet{Siwick2004}.

\begin{figure}[ht]
\includegraphics[width=0.7\columnwidth]{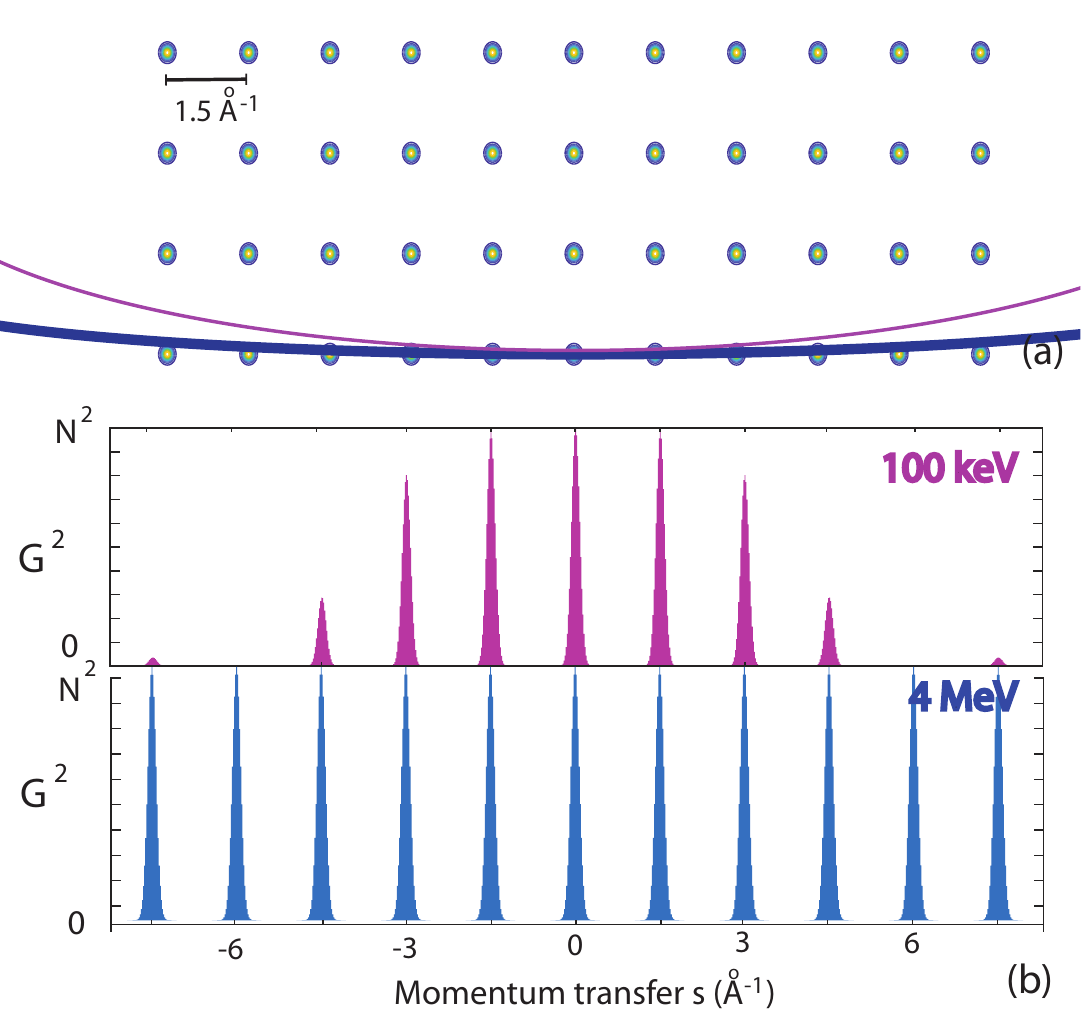}
\caption{Ewald sphere construction for diffraction from a crystal using 100 keV and 4 MeV electrons. The reciprocal lattice spacing is set by the crystal lattice constant. The volume of a reciprocal lattice 'point' is determined by the size of the crystal.}
\label{Fig:MeVEwaldsphere}
\end{figure}

The Ewald sphere construction is often used to graphically represent the Laue condition, describing which reciprocal lattice points (or diffraction peaks) will be seen in a diffraction pattern in a specific scattering geometry (i.e. crystal orientation with respect to the incident electron wavevector). We will use this construction here to illustrate how the electron deBroglie wavelength, $\mathbf{\lambda}$ (or beam energy), influences diffraction. However, the impact of other beam parameters, like the spread in electron beam energy and divergence angle, can also be understood using this construction.
The Ewald sphere is drawn on top of the crystal's reciprocal lattice with a radius of 1/$\mathbf{\lambda}$ and an orientation determined by the incident beam angle with respect to the crystallographic axes. This is shown in a simple geometry for a hypothetical simple cubic crystal at two beam energies in Fig. \ref{Fig:MeVEwaldsphere}. For elastic (Bragg) scattering both incoming and scattered beams lie on this sphere, thus the Laue condition for diffraction is only satisfied when the Ewald sphere cuts through a reciprocal lattice point.  Note that the curvature of the sphere is inversely proportional to the wavelength of the incident radiation. Since the deBroglie wavelength of electrons is 3.88 pm at 100 keV, but only 0.39 pm at 10 MeV, the Ewald sphere at 100 keV has 10 times higher curvature.  The flatter the Ewald sphere, the larger the number of reciprocal lattice points that can intersect with the sphere at large momentum transfer (or scattering angle).  This is an important advantage for MeV electron probes in terms of the scattering efficiency for higher order Bragg peaks, but even at 100 keV the Ewald sphere for electron scattering is already approximately 25 times flatter than it is for hard xray scattering (using 100 pm xrays).

However, there is a practical consideration resulting from the scaling of the de Broglie wavelength with electron energy and the resulting scattering angle which is much smaller for relativistic electron energies. For example, consider a set of crystalline planes separated by $d = 2$ \r{A}, the Bragg angle for 4 MeV (100 keV) electrons is 0.7 (9) mrad. This has strong implications on the experimental setup of the distance from the sample to the detector or diffraction camera length (which needs to be proportionally longer in the relativistic case in order to allow for the scattered electrons to physically separate from the unscattered ones, assuming no magnifying electron optics between the sample and detector), but importantly bears no effect on the attainable quality of the pattern as explained below.

\subsubsection{Coherence length and reciprocal space resolution in UED} 
\label{section_coherence_length}

In order to form a diffraction pattern, a large number (a beam) of probe electrons is used to illuminate the target. In Bragg scattering, if one wants to distinguish the scattered particles from the undiffracted ones, it is essential that the scattering angle $2\theta_B$ be much larger than the uncorrelated spread of the divergence angles in the beam at the sample. In the root-mean-square sense this can be expressed as $\sigma_{\theta}$ (i.e. $\sigma_\theta \ll 2\theta_B$). Note that any angular divergence correlated with position (for example due to a converging or diverging beam) can be removed by the transport optics and does not play a role in the diffraction contrast. 

For polycrystalline or gas/liquid phase samples, where the diffraction pattern is a series of concentric rings due to the random orientation of the grains, it is customary to introduce as figure of merit for resolution $\mathbb{R} = R/\Delta R$ where $R$ is the radius of the diffraction ring on the detector screen and $\Delta R$ is the smallest distance between two neighboring rings which can just be discriminated at the detector. Note that the position on the detector screen is simply proportional to the scattering angle so that $\mathbb{R}$ can also be interpreted as the inverse of the relative reciprocal space resolution, i.e  $\mathbb{R} = R/ \Delta R = s/\Delta s$. A typical TEM operating in diffraction mode achieves $\mathbb{R}>10^2$ or more for static images. For UED, a resolving power of $\mathbb{R} > 10$ guarantees a good quality diffraction pattern and provides enough resolution to adequately resolve typical ultrafast structural rearrangements.
The experimental value of $\mathbb{R}$ is affected by multiple factors, such as the electron beam angular and energy spread, and the spatial resolution of the detector, as it will be discussed in detail in the following sections. In most diffraction setups the uncorrelated beam divergence is the dominant limiting factor in the resolving power of the diffraction camera~\cite{Grivet:2013vaba}, so one can write $\mathbb{R} = \lambda/2d \sigma_\theta \approx \theta_B / \sigma_\theta$. It is useful to note that the value of $\mathbb{R}$ is independent on the beam energy, as both components of the ratio above are proportional to $\propto 1/\beta\gamma$.
Note that the absolute reciprocal space resolution is simply $\Delta s$. This quantity determines the longest range order which can be observed in the diffraction pattern. In practice, this corresponds to effectively how small the electron beam can be made on the detector screen. 

The importance of the beam divergence at the sample in UED is encoded in the concept of coherence length $L_c$ which is an equivalent figure of merit for diffraction contrast. In standard optics the coherence length indicates the extent of the coherent portion of the illumination (i.e. the spatial extent over which the phase of the illuminating beam wavefunction is correlated). For example, for an incoherent source, with no optics between the source and the sample, the Van Cittert-Zernike theorem defines the coherence length as the wavelength divided by the angle subtended by the source \cite{bornwolf}. In a UED beamline the definition must take into account that the beam from the electron source is magnified and refocused before illuminating the sample. One can show in this case that the the visibility of interference fringes from two scattering centers (or planes) separated by a distance $d$, depends on the ratio between $d$ and the transverse coherence length as $L_c=\lambda/2\pi\sigma_\theta$ \cite{kirchner:coherence,tsujino:coherencemeasurement} where $\sigma_\theta$ is the uncorrelated beam divergence at the sample. This is important since, as we have discussed above, the spatially periodic arrangement of the atoms in a crystal allows for a large enhancement of the diffraction signal, but if the beam phase front is not coherent over multiple unit cells of the structure under study, then no constructive interference can be developed and the visibility of the diffraction peaks is strongly reduced. In the limit that the coherence length is smaller than a unit cell, the Bragg peaks disappear. Note that this strong dependence suggests the use of diffraction pattern visibility as a sensitive quantity to measure of the beam divergence \cite{yang:diagnostics}. The visibility of the Bragg interference peaks also depends on the longitudinal coherence properties of the beam, but in typical UED setups the longitudinal coherence length i.e. $L_l = \lambda /( 2\pi \frac{\delta (\beta \gamma)}{\beta \gamma})$, even for energy spreads as high as 1 $\%$, is often much longer than the differences in optical path length for the diffracted beams and so hardly contributes to the sharpness of the diffraction pattern.

\begin{figure}[ht]
\includegraphics[width=0.95\columnwidth]{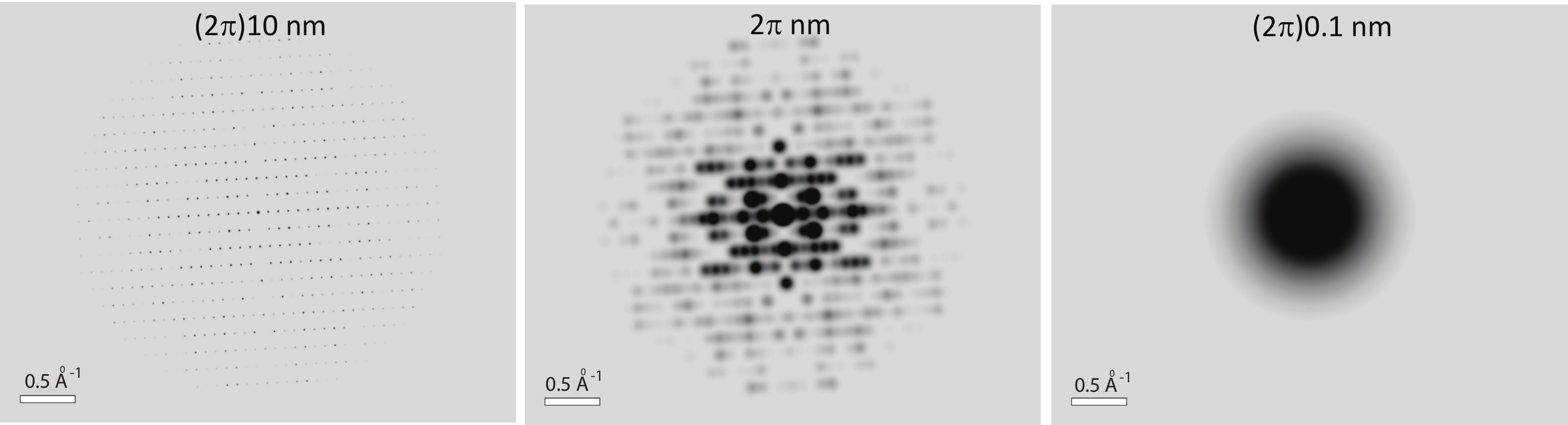}
\caption{Simulated diffraction patterns of a Salicylic acid (aspirin) crystal for electron probe beams having coherence lengths of (2$\pi$) 10 nm, 1 nm and 0.1 nm respectively.}
\label{aspirin}
\end{figure}

To illustrate the impact of beam coherence on the quality of the diffraction pattern, we show in Fig.~\ref{aspirin} simulated diffraction patterns from salicylic acid (aspirin) molecule for different coherence length values, ranging from 62.8 nm to 0.628 nm. The unit cell vector lengths for this crystal lattice are [11.3, 6.5, 11.3] \AA~\cite{wheatley:aspirin}. It is clear that much more detailed information on the crystal structure can be extracted from the pattern to the left.

To compare different electron beamlines, it is also useful to normalize the coherence length to the electron beam size at the sample $\sigma_x$ and define a relative coherence length
\begin{equation}
l_c = \frac{L_c }{\sigma_x}
\label{relative_coherence_length}
\end{equation}
Indeed beam divergence can be  controlled by the electron optics before the sample, and the coherence length can be adjusted, while the relative coherence length is an intrinsic beam property and effectively can be thought as the fraction of the beam which participates in coherent scattering.

A final point related to the study of sensitive materials is related to the damage effects associated with the bombardment of the sample by high energy electrons. The main mechanism involved is ionization damage (radiolysis), in which valence or inner-shell electrons within the specimen are excited by inelastic scattering events either directly breaking a
chemical bond, or indirectly by secondary electron emission \cite{egerton:outrun}. In order to evaluate the relative importance of these effects one needs to compare the elastic to inelastic mean free path as well as the energy deposited per scattering event. After taking all of this into account, it turns out that the overall damage is not particularly sensitive to the electron energy. 
In addition, it should be mentioned here the possibility for irreversible specimen damage associated with the knock-on effect. This is a rare occurrence where collision between an incident electron and an atomic nucleus create an atomic vacancy \cite{egerton:knockon}. The onset of this effect depends on the atomic species, but generally is above 80 keV. Due to the steep energy dependence, it had been one of the causes of the progressive disappearance of high voltage (MeV) electron microscopy (accelerated by the resolution improvements at lower voltage resulting from aberration correction implementation). In high energy UED, the Bragg enhancement effect (spatial averaging over the sample) allows to utilize a much lower dose to acquire a diffraction pattern and significantly reduces this problem. As an example, while to acquire a high-contrast nm-spatial resolution image a dose of 100 e-/nm$^2$ would be required, the typical doses for high energy UED are 10$^6$ e-/ 10 $\mu$m$^2$ which is 10$^4$ times smaller. Furthermore, novel setups developed in the last few years hold the promise of full diffraction signal acquisition faster than any structural change due to damage (i.e. in few tens of fs), with an approach similar to the diffract-and-destroy technique employed in 4th generation light sources~\cite{Spence:2008ibba}.

\subsubsection{Electron vs. X-ray scattering}

It is useful at this point in order to better appreciate the opportunities enabled by the development of ultrafast electron scattering to draw a comparison with x-ray scattering techniques. In particular, there is often a debate in the comparison of the effectiveness of probing with electrons or x-rays, even though the information extracted from these different technologies is mostly complementary.

Aside from significant difference in the size and cost of electron and x-ray machines \cite{Carbone:2012ghba}, there are two main differences in the interaction with matter. The first one is that elastic scattering of X-rays from matter is relatively weak due to the very small cross section for photon interaction with charged particles (Thompson cross-section) \cite{warren1990}. To make a quantitative comparison, considering the same momentum transfer $s$ = 10 \AA$^{-1}$, the Rutherford cross section is more than 5 orders of magnitude larger than the x-ray cross section for elastic scattering. This implies that 5 orders of magnitude less electrons generate an equal diffraction signal when illuminating a target with the same number of scattering centers. It is no surprise that electrons are then the preferred choice anytime the number of scatterers in the target is small (gas phase, membrane protein crystals, 2D and quasi-2D materials, etc.). 

Owing to their higher cross section, electrons have significantly shorter penetration depth than hard x-rays, with important consequences on the sample thickness of choice and on the detector technology. The value of the probe beam penetration depth is an important factor in designing pump-probe experiments. An ideal excitation (absorbed fluence/layer) would have a uniform profile throughout the sample thickness. On the other hand, perfect uniformity is only reached with negligible absorption, i.e. negligible excitation. Therefore a sample thickness roughly equal to one absorption length at the excitation wavelength can be considered a good tradeoff between uniformity and pumping efficiency. Typical electron elastic mean free path values limit sample thickness for UED in the tens-to-hundreds of nanometers (depending on electron energy and atomic composition).Such values are a good match for optical radiation in a metal, while insulators and semiconductors can have absorption depths up to cm-scale. 
For x-rays (non-resonant, hard and soft) the penetration depth mostly depends on the form factor, i.e. how heavy the elements are, but it is typically on the scale of cm or longer. For soft x-rays, there is an additional situation when one goes into resonant absorption. There, the elemental absorption becomes extremely strong and the penetration depth short and in some cases comparable with visible light \cite{lindenberg2000time}. A different situation occurs when pumping in the THz regime of great interest for material science where the pump penetration depth is significantly longer~\cite{sie2019ultrafast}.

Furthermore, the difference in  wavelength of the probing particles leads to key differences in the experimental data.  
An X-ray photon energy of 1-10~keV corresponds to a wavelength in the range of 1-10~\AA, while electrons with energies typically used in UED exhibit  wavelengths in the picometer-range, with a dramatic difference in the curvature of the Ewald sphere between the two cases. As a consequence, X-rays provide excellent momentum resolution in reciprocal space within a narrow range, i.e.typically only few spots per diffraction pattern. Conversely, each electron diffraction pattern typically includes a large number of spots/rings/diffraction features from which more information can be retrieved \cite{Yang2018_CF3I}. In addition, the technological development of high quality X-ray optics significantly lags its electron counterpart, and related to this, the focusability of X-ray and electron beams is very different. While the latter can be easily focused down to spot sizes well below 100~nm, typical spot sizes at state-of-the-art XFELs are still in the micrometer range. 

Another important difference relates to the amount of energy deposited in the sample for a single inelastic scattering event. X-rays are fully absorbed, depositing their entire energy into the sample, while electrons typically only release a small fraction of their energy in a collision. In fact it has been pointed out by Henderson \cite{henderson2004realizing} that per elastic scattering event electrons deposit as little as 1/1000 of the energy of x-rays in the sample. Especially for sensitive biology-relevant samples this might be an important advantage. The same paper also points out that the inelastic scattering cross section of soft x-rays has the same order of magnitude than the elastic cross section for high energy electrons. This suggests the fascinating possibility of drawing complementary information using potentially the same samples pairing up UED and inelastic scattering techniques from soft x-ray beamlines.

Finally, with the advent of X-ray lasers \cite{emma2010first}, fully transversely coherent ultrashort x-ray pulses can be available enabling coherent diffraction imaging algorithms to replace the role of the optics in retrieving real-space images of the sample \cite{miao1999extending}. In short-pulse electron scattering instrumentation, as it will be discussed below, this limit is still very far from reach and only partially coherent electron beams have been used to date. 

\subsection{Electron beam brightness} 
\label{sec:brightness}

In this section we introduce a metric for measuring the ability of a specific setup to deliver high density electron beams, and for comparing different instruments. The definitions introduced below will be used throughout the article to elaborate on the capability of an electron beam to perform specific experiments or provide the required spatial and temporal resolution.

\subsubsection{The electron beam concept}

Adding temporal resolution to electron scattering experiments requires the formation of an electron bunch, i.e. a three-dimensional charge distribution well defined and limited in space and time. Such electron \textit{beam} can be defined by the sum of isolated electrons correlated in time by periodic emission (stroboscopic approach)~\cite{baum2013physics}, or by a set of electrons tightly packed in a small volume (single-shot setups), traveling together along a preferred direction. In both cases, the level of confidence by which one can describe the temporal contours of the beam will set the basis for the definition of temporal resolution $\tau_{res}$ in a ultrafast experiment. 
In conventional continuous sources electrons are emitted at random times and, therefore, no temporal information can be extracted without further manipulation of the electron stream. 
A quality metric for such sources is provided by the five-dimensional beam brightness $\beta_{micro}=\frac{4i_e}{(\pi d_0\alpha_0)^2}$ \cite{TEM:book}, a measure of the average current $i_e$ per unit of source size $d_0$ (full beam diameter at crossover) and solid angle of emission $\alpha_0$ (semi-angle of emission at crossover). In absence of downstream beam acceleration, $\beta_{micro}$ is a constant of the motion along the electron beamline/column, that is if one desires a smaller spot size, a larger beam divergence is unavoidable. 

If the beam spatial and angular distributions are not uniform, a more general definition of beam diameter and angular spread is needed. Using the statistical framework, we introduce the generalized standard deviations of the beam along a specific direction, also known as root mean square moments of the distribution about its mean, rms hereafter~\cite{rhee_RMSinvariance_1986}. 

For the case of pulsed electron sources, a distinction between average and peak current needs to be made, the latter describing a local property of the individual bunch of electrons in a longer bunch train, and defined as the instantaneous rate of change of the beam charge. The resulting peak and average brightness values will bear different information, the former describing the ability of a particular setup of performing single-shot measurements, and the latter providing information on experiment recording times. Unless specified otherwise, the quantities defined in the following of this section will relate to isolated bunched beams. 

\subsubsection{Beam phase space and brightness definitions}
\label{brightness}

The key distinction between pulsed and continuous wave beams is the role played by the longitudinal parameters in the experiments. Similarly to transverse focusing in which beam size and divergence can be trade-off for each-other, longitudinal compression allows to manipulate pulse length and energy spread to achieve optimal resolution. A modified metric for pulsed source quality which includes both the transverse and longitudinal degrees of freedom, is obtained by introducing the concepts of six-dimensional phase space and six-dimensional brightness. 

From a classical mechanics standpoint a set of $N$ particles represents a system with a total of 6N degrees of freedom, including each particle coordinates in space $r_i$ and their relative conjugate momenta $p_i$. In most cases of interest the temporal evolution of such system can be described by an Hamiltonian which, in turn describes the evolution of a unique trajectory in the 6N-dimensional space defined by the full system degrees of freedom. The number of dimensions can be reduced back down to six if particle-particle interactions can be neglected or described by a mean field approximation, resulting in a description of the electron beam as a clustered set of points in the hyper-volume $\mathcal{V}_6$, called 6D phase space for each instant in time.
A key concept in this description of electron beams is represented by the phase space charge density $\rho_6(\mathbf{ r},\mathbf{p},t)$, also called microscopic six-dimensional brightness~\cite{Brightness_refined_1992}, defining the charge distribution in the phase space $dQ = \rho_6\delta\mathcal{V}_6$.

Although the shape of the distribution changes with time, the Liouville theorem states the invariance of its total volume during motion under the the assumption of Hamiltonian evolution. The six-dimensional beam brightness is therefore a constant of motion.

In the special but not uncommon case of decoupled motion between the different planes, the 6D volume can be written as $\mathcal{V}_6=\mathcal{A}_x\mathcal{A}_y\mathcal{A}_z$, where $\mathcal{A}_i$ is the phase space area in the $(i,p_i)$ plane ($i=x,y,z$). If we use second order moments of the distribution to describe the area enclosed by the beam, then $\mathcal{A}_i$ takes on the meaning of normalized rms emittance $\epsilon_{n,i}$. 

\begin{figure}[ht]
\centering
\includegraphics[width=1\columnwidth]{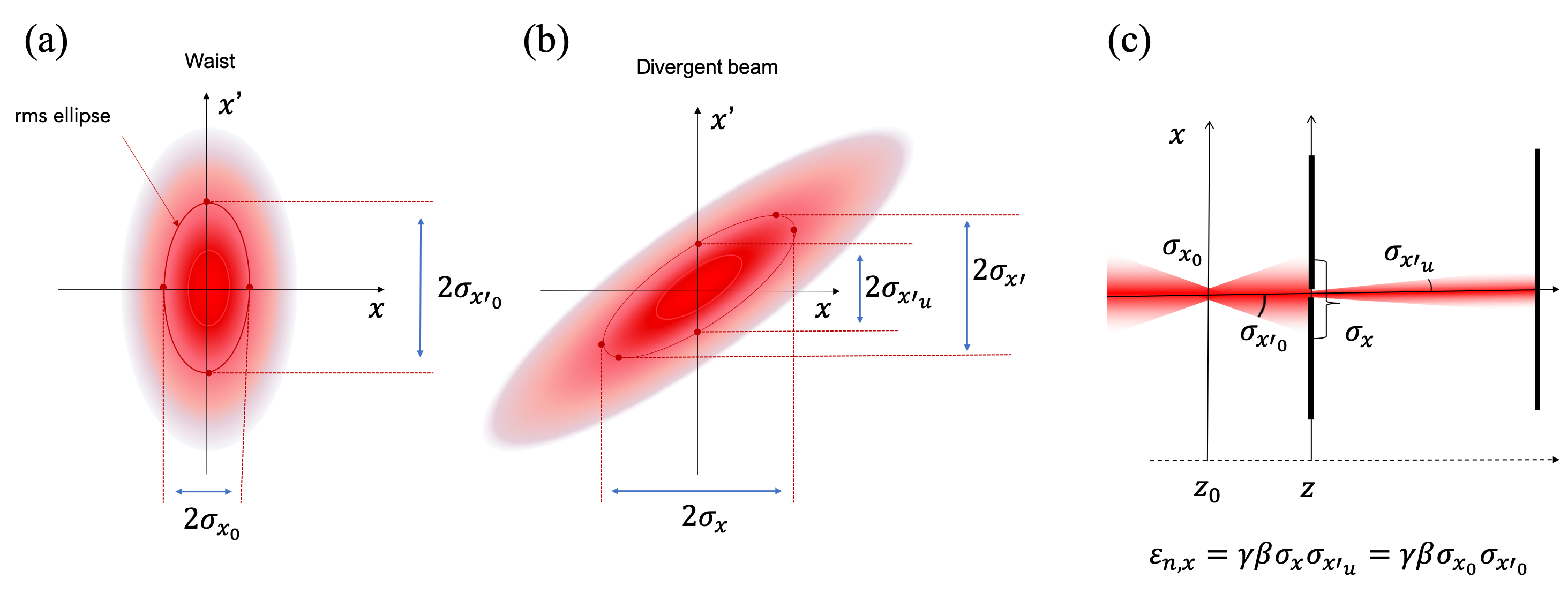}
\caption{Schematic visualization of rms beam properties and emittance. The elliptical contours represents the beam density in phase space (a) and (b). In (c) the concept of uncorrelated beam divergence (and its relation with the beam emittance) is clarified.}
\label{fig:PSconcept}
\end{figure}

It is often convenient to express the beam properties in terms of the angle of the particle trajectory with respect to the propagation direction $z$, $x'=\frac{p_x}{p_z}$. Considering a beam waist at a position $z_0$ as shown in Fig.~\ref{fig:PSconcept}(a), the normalized transverse rms emittance in the $(x,x')$ plane can then be written as $\epsilon_{n,x}=\gamma\beta\sigma_{x_0}\sigma_{x'_0}$, where $\beta$ and $\gamma$ are the relativistic Lorentz factors. In the more general case depicted in Fig.~\ref{fig:PSconcept}(b), the emittance calculation at a plane $z$ will need to account for correlations $\sigma_{xx'}$ in the plane, and the equation becomes: $\epsilon_{n,x}=\gamma\beta\sqrt{\sigma^2_{x}\sigma^2_{x'}-(\sigma_{xx'})^2}$. Introducing the uncorrelated transverse rms spread in divergence $\sigma_{x'_u}$ simplifies the general equation back to the product of two terms, $\epsilon_{n,x}=\gamma\beta\sigma_{x}\sigma_{x'_u}$. Figure ~\ref{fig:PSconcept}(c) clarifies the physical meaning of uncorrelated divergence at a position $z$ along the beam path, equivalent to  $\sigma_{\theta}$ introduced in Sec.\ref{section_coherence_length}. The uncorrelated divergence is a key parameter in UED experiments, determining the beam transverse coherent length and the reciprocal space resolution.

In case of uncoupled dynamics, the rms six-dimensional brightness can be written as:   
\begin{equation}
B_{6D}= \frac{Ne}{\epsilon_{n,x}\epsilon_{n,y}\epsilon_{n,z}}=\frac{I_{rms}}{\epsilon_{n,x}\epsilon_{n,y}\frac{\sigma_E}{mc}} 
\label{B6D}
\end{equation}
where we assume no time-energy correlation in the bunch and $\epsilon_{n,z}=\sigma_z\frac{\sigma_{p_z}}{mc}\approx c\sigma_t\frac{\sigma_E}{mc^2}$, and $I_{rms}=Ne/\sigma_t=\eta I_{peak}$, with $I_{peak}$ being the maximum current within the pulse, $N$ the number of electrons in the bunch, and $\eta$ a numerical value depending on the shape of the temporal distribution ($\eta=\sqrt{2\pi}$ for a gaussian temporal profile). 

Depending on the specific application, it is common to introduce different brightness definitions which better capture the key beam properties.
In typical ultrafast electron diffraction experiments electron beam's transverse emittance rather than the energy spread dominates the minimum beam size at the sample and the resolution in reciprocal space. 
In this case we can then consider the five-dimensional brightness to be more representative of the effectiveness of the electron beam to carry out an experiment, $B_{5D}= \frac{I_{rms}}{\epsilon_{n,x}\epsilon_{n,y}}$. 
This parameter is directly proportional to the $\beta_{micro}$ defined above and used in microscopy. The proportionality factor depends on the details of the charge distribution (for example uniform, gaussian, parabolic). There is also an additional factor $(\gamma\beta)^2$ which is used to make $B_{5D}$ invariant under particle acceleration. On the other hand, this value can be increased by longitudinal beam compression, which increases the beam peak current at expenses of energy spread. 

Lowering further the number of dimensions, one can define a brightness in the transverse planes, called four-dimensional brightness and  defined as:
\begin{equation}
B_{4D}=\frac{Ne}{\epsilon_{n,x}\epsilon_{n,y}}
\label{B4d}
\end{equation}

This metric results particularly useful when balancing trade offs between temporal and spatial resolution in time-resolved electron scattering. Larger values of $B_{4D}$ result in better diffraction pattern contrast and higher spatial resolution. One simple way to increase $B_{4D}$ is by starting with a longer pulse length, which would increase the charge at expenses of temporal resolution. 
Assuming no coupling between longitudinal and transverse planes, the four-dimensional brightness is set at emission and remains constant during transport and acceleration.

\subsubsection{Quantum limit of beam brightness}

The fermionic nature of the electrons limits the number of electrons that can occupy the same phase space area through the Pauli exclusion principle. This sets a value for the  maximum phase space electron density which can be derived starting from the uncertainty principle, stating that $\sigma_x\frac{\sigma_{p_x}}{mc}\geq\frac{\lambda_c}{4\pi}$, providing the volume of a coherent state in phase space~\cite{callaham_quantum-mechanical_1988, zolotorev:degas}. Here $\lambda_c$ is the Compton wavelength of the electron. The final quantum limited rms brightness can be written:
\begin{equation}
B^q_{6D}=2e\left(\frac{2\pi}{\lambda_c}\right)^3 
\label{B5dmicroscopy}
\end{equation}

The ratio between the beam six-dimensional brightness and the quantum limited brightness defines the beam degeneracy parameter $\delta=\frac{B_{6D}}{B^q_{6D}}$, a measure of the source quality with respect to the ultimate physical limit. In the case of a unpolarized source, $\delta_{max}=1$. Typical values of $\delta$ for state-of-the-art electron sources range from $10^{-2}$ of single-atom emitters to $10^{-6}$ of large-area photo-emitters.

When normalized by the quantum-limited transverse brightness $B^q_{4D}=2e(\frac{2\pi}{\lambda_c})^2$, the four-dimensional brightness provides a direct measure of the source lateral coherence. Using the definition of beam normalized emittance, the relative coherence length (Eq.~\ref{relative_coherence_length}) can be rewritten as $l_c=\lambda_c/(2\pi\epsilon_n)$, and the normalized transverse brightness for a round beam (same emittance in $x$ and $y$ planes) then is:
\begin{equation}
\frac{B_{4d}}{B^q_{4d}}=\frac{N}{2l_c^2}=\frac{N_c}{2}
\label{B4d_norm}
\end{equation}
where $N_c=N/l_c^2$ is the number of electrons per coherent area in the beam.

\subsection{Different modalities of ultrafast electron scattering instrumentation: diffraction, imaging and spectroscopy} 
 \label{sect1.differentmodalities}
 
As an electron beam interacts with matter, a wealth of information related to the lattice and electronic structures, as well as their dynamics, gets encoded in the beam phase space, corresponding to changes of the momentum, energy, and intensity of the beam. Over the past century, various specialized methods and instruments utilizing electron probes have been developed, focusing on one or few types of changes in phase space, giving rise to various modalities of electron scattering instrumentation such as diffraction, imaging and spectroscopy~\cite{TEM:book, reimer2013, spencebook}. 

\begin{figure}[ht]
\centering
\includegraphics[width=1\columnwidth]{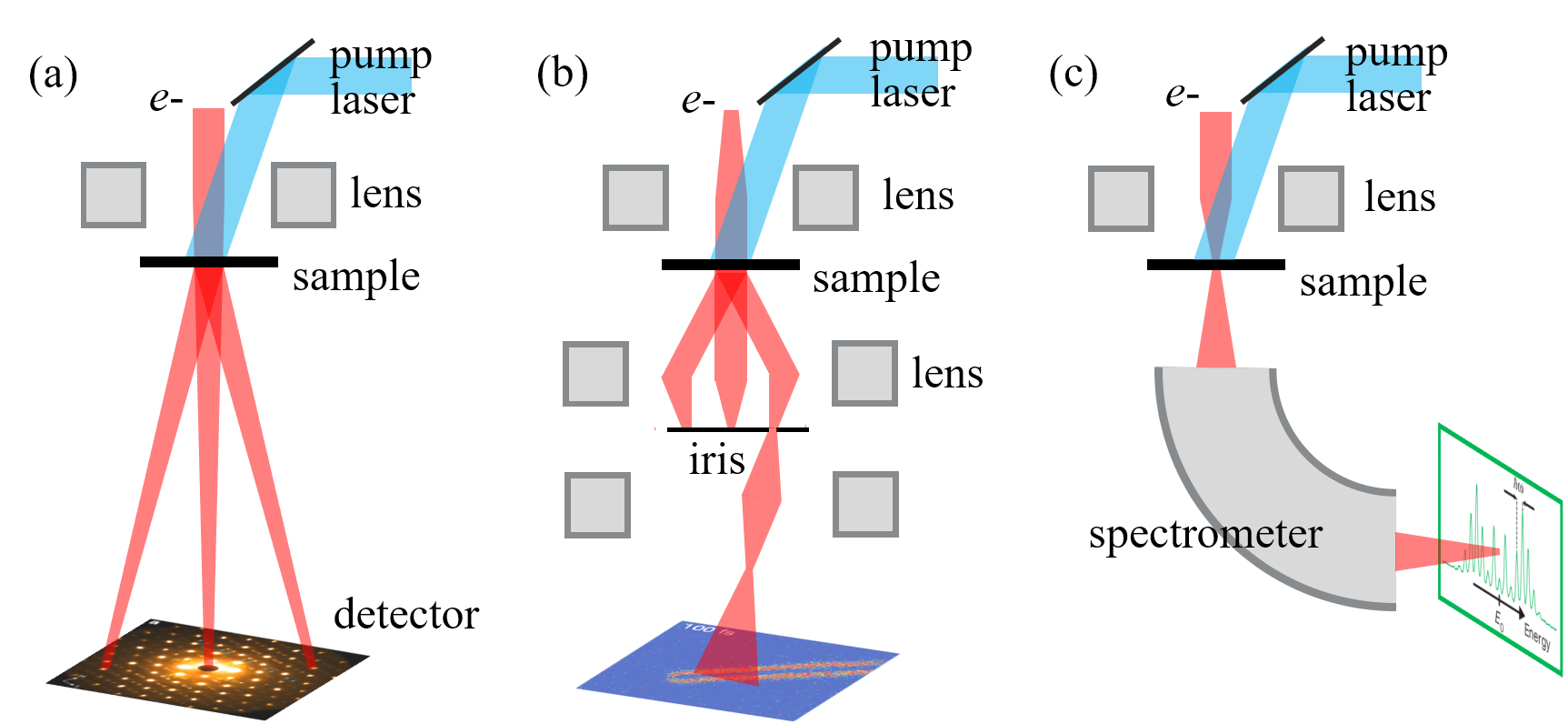}
\caption{Different modalities for ultrafast electron scattering instrumentation, including (a) diffraction, (b) imaging, and (c) spectroscopy. Exemplary images are adapted from references \citet{Kogar2020}, \citet{barwick:pinem}, and \citet{ropers15}, respectively.}
\label{fig:threemodalities}
\end{figure}

Referring to Fig \ref{fig:threemodalities}, the imaging modality allows the formation of real space images of an illuminated sample area using electron optics and apertures (TEM mode). Contrast in the image is introduced by filtering out the scattered electrons with momentum and/or energy changes using apertures to only transmit a certain region of the beam phase space (dark and bright field imaging). Imaging is most powerful to observe and track non-periodic features of interest. 

The quality of the image depends on many parameters, including those related to illumination (electron dose, relevant to the so called Rose criterion \cite{rose:1948}) and contrast (intrinsic beam divergence and energy spread). These quantities also contribute to the image blurring through the spherical and chromatic aberrations of the electron lenses. Considering the definition of the 6D beam brightness (Eq. \ref{B6D}),  $Ne$ is proportional to the total scattered and recorded signal and $\sigma_{x,y}$ are the transverse rms spot sizes of the probe. The spatial resolution and contrast are encoded in the rms beam divergence $\sigma_{x',y'}$ and the rms energy spread $\sigma_{E}$. The rms bunch length $\sigma_t$ sets the limit for the temporal resolution, so that very high beam brightness is required for this application. 

Accessible time scales in ultrafast electron imaging range from ns for single shot full field images \cite{LLNL:DTEM,Bostanjoglo_2002, Picher18} to fs in stroboscopic mode \cite{Zewail_2010, piazza_design_2013, feist_ultrafast_2017, flannigan16phonon, Houdellier18, ljq15}. Aiming at reaching enhanced capabilities, ultrafast imaging using electron beams with higher energy (MeV level) and potentially higher brightness is an area under intense development \cite{limusumeci:prapplied, xiang14uem, yang15uem, wan18uem, cesar16uem, lu_uem_2018, li2017femtosecond}, which drives innovative approaches to electron sources, beam optics, and operation schemes. A separate operating mode for imaging can be achieved by scanning a focused electron probe across the sample and recording the scattering signal for each position (STEM, 4D STEM, ptychography, and ultrafast nanodiffraction \cite{ji_nanoued_2019}). The advantage in this case is the opportunity to identify correlations between the material domains (easily identifiable in imaging mode) and the ultrafast changes in the unit cell (accessible in diffraction mode). 

Adding an energy filter at the end of the electron column enables observation of time-dependent changes in the electron energy loss spectrum (EELS) \cite{barwick:pinem, carboneeels09, ropers15}. The EELS signal is directly correlated to chemical and electronic properties of the specimen. EELS requires nearly monochromatic illumination, i.e. the beam energy spread $\sigma_{E}$ must be smaller than the energy feature to be resolved (from single eV to meV level, depending on the process).

Combining spectroscopy with diffraction one would provide access to momentum-resolved EELS carrying a great deal of information on the electronic structure of the sample. An important additional benefit of using ultrafast sources for EELS is that the time structure of the beam allows the possibility for more accurate energy measurements \cite{verhoeven2018high} by taking advantage of beam control techniques in the longitudinal phase space (for example using RF cavities as time-domain lenses) . Time-of-flight electron spectroscopy \cite{verhoeven2016time} is also enabled by having short electron bunches at the sample.

In comparison with these other modalities, diffraction requires no further electron optical elements between the sample and the detector. The signal is generated by the interference of elastically scattered electrons, i.e. those electrons with modified transverse momentum by negligible energy changes . This signal encodes the structural information averaged over the entire probed area. Benefiting from its simplicity, diffraction usually also allows for most quantitative correlation between the measured pattern and the structure of matter. 

For these reasons, UED has been the first modality of time-resolved electron scattering to receive attention, but advances in ultrafast electron beams from photoemission sources establish exciting new capabilities for imaging and spectroscopy as well. In this review we will focus on the recent developments in UED, with the understanding that the other modalities will likely take advantage of much of the technical progresses described below. In addition, mixed-modalities instruments, for example setups where ultrafast electron microscopy and UED can take place in the same modified TEM column \cite{feist_nanoscale_2018, Carbone:2012ghba, Sun15_TaSe2_CDW_UEM} are becoming more widely available for scientific discoveries.

\subsection{Scientific drivers for ultrafast electron scattering}
\label{sect.scientificdrivers}

\subsubsection{Solid State: ordering, excitation and emergent phenomena in materials}
\label{sec:ss_material_phenomena}

    Many of the central questions of materials physics relate to the complex interplay between charge, spin, orbital and lattice-structural degrees of freedom that give rise to the emergent macroscopic properties and ordered phases of materials~\cite{Basov2017, delatorre:rmp}. Since electron diffraction provides a `map' of the electrostatic potential of a crystal in reciprocal space~\cite{fultz2012}, as discussed above in Sec.~\ref{sect1.a}, the intensity of diffraction peaks are profoundly sensitive to the details of the lattice, charge and orbital order present in a material.  Only spin-specific ordering is relatively hidden from view with high-energy electron beams (even spin polarized ones) due to the relatively small differential scattering cross section between aligned and anti-aligned spins at high energies.  Magnetic structure peaks are not present in a UED pattern as they are in neutron scattering, however, rich information on magnetism in materials can be obtained with electron beams via imaging. Magnetic domain structure \citet{Park2010} and magnetic texture \citet{eggebrecht2017light, huang2020melting} dynamics are accessible to UEM when operated in Lorentz microscopy mode.  

    In addition to the static ordering of charge, spin, orbital and lattice degrees of freedom in materials, an understanding of the elementary excitations that are present --both collective and single particle-- and how these excitations couple/interact with one another is required for a fundamental understanding of the diverse phenomena and properties found in condensed matter. The interactions between collective excitations of the lattice system (phonons) and charge carriers, specifically, are of particular relevance and easily studied by UED.  These interactions are known to lead to superconductivity, charge-density waves, multi-ferroicity, and soft-mode phase transitions. Carrier-phonon interactions are also central to our understanding of electrical transport, heat transport, and energy conversion processes in photovoltaics and thermoelectrics. Phonons can themselves be intimately mixed in to the very nature of more complex elementary excitations, as they are in polarons or polaritons. Further, the coupling of spin and lattice systems can also be studied from the lattice perspective with UED.
    
\begin{figure}[t]
\includegraphics[width=1\columnwidth]{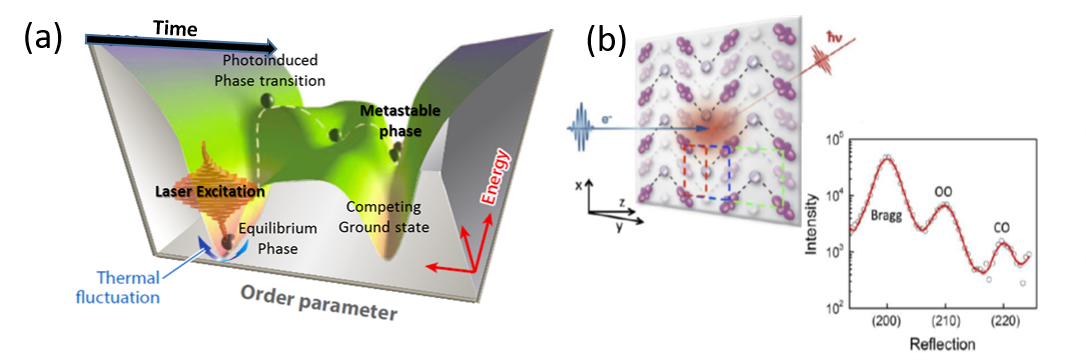}
\caption{Properties on demand: controlling the structure and properties of Quantum Materials with light.  a) Laser excitation can lead to a photoinduced phase transition on the material's free energy landscape, steering the system to a competing ground, metastable or transient state with dramatically different ordering and properties.  Some photoinduced phases can be completely inaccessible at thermal equilibrium.  Adapted from \citet{zhang2014dynamics} b) Schematic of a UED experiment on Manganite, which exhibits crystalline/lattice (Bragg), orbital (OO) and charge order (CO). Since the diffraction patterns of Manganite show separated peaks associated with each order, UED can follow their time-depedence and provide deep insignts into photoinduced phase transitions like that shown schematically in a).   Adapted from \citet{Li2016}. }
\label{fig:ued_intro_fig_1a}
\end{figure}

By tuning the excitation wavelength in the mid to far IR and THz (see for example~\cite{sie2019ultrafast}), UED tools can be used to follow the linear and non-linear behaviour of selectively driven phonon modes \cite{forst2011nonlinear, von2018probing}, and their coupling with other degrees of freedom. 
The development of bright ultrafast electron beams has opened up an enormous space for experimentation on the structure, dynamics and nonequilibrium properties of materials.  In some of its earliest manifestations, UED was used to probe strongly driven melting (order – disorder) transitions in materials, thanks to the ability of obtaining high quality diffraction patterns in a single shot.   More recently, strongly-correlated or quantum materials have been the target of study (see for example~\cite{Kogar2020},~\cite{Duan21_TiSe2_nature} and ~\cite{siddiqui_ultrafast_2021}). The non-equilibrium properties of quantum materials are particularly interesting because the interactions between lattice, charge, orbital or spin degrees of freedom are typically on par with electronic kinetic energy.  The presence of a `soup' of competing and collaborating interactions on similar energy scales tends to result in a complex free-energy landscape that can show many nearly degenerate ground states that each exhibit different ordering and properties.  Mode-selective excitations that modify the interplay between these DOF have been shown to result in dramatic transformations (Fig.~\ref{fig:ued_intro_fig_1a}~a)).  The associated changes in lattice, orbital and charge order can be followed directly with UED (Fig.~\ref{fig:ued_intro_fig_1a}~b)). The manipulation and control of material properties far from equilibrium with light offers almost completely untapped and unexplored possibilities for discovering novel states and phases of materials with exotic and transformative behaviours (see for example~\cite{reid_beyond_2018}, ~\cite{Sood21_VO2}, and~\cite{Mo22_compression}). This new 'properties on demand' frontier~\cite{Basov2017} is a \emph{Grand Challenge} for the fundamental sciences \cite{fleming2008grand} and complements the conventional means of materials discovery, which has been to explore the structural and compositional phase space that is accessible at thermodynamic equilibrium in the search for desirable properties example~\cite{mitrano2016possible}). Ultrafast pulsed electron beams provide the sophisticated tools of structural characterization on femtosecond timescale that are a basic requirement of such work. 

\subsubsection{Gas Phase: Uncovering the structure-function relationship behind photochemical reactivity}
Knowledge of how molecules responds to the incidence of light is essential to our understanding of nature and its fundamental processes, e.g. photosynthesis~\cite{cheng2009_photosynthesis}, vision \cite{Polli2010}, DNA photo damage~\cite{Schreier2007_DNAdamage}, as well as the technological development of light harvesting and storage devices~\cite{MansO2018_MOST}. The absorption of ultraviolet (UV) light by a molecule leads to its promotion to an electronically excited state.  The absorbed photon energy may be redistributed through the breaking of chemical bonds leading to photolysis, or through the coupling between Franck-Condon active and inactive modes leading to new vibrations. Alternately, structural rearrangement may result in a new molecular geometry in which the excited electronic state becomes degenerate with another electronic state. These geometries represent conical intersections, which provide an efficient pathway for radiationless decay between electronic states.\cite{Domcke_CIs} Electron scattering is perfectly suited to capture structural changes, as electrons interact with the Coulomb potential of the target system,\cite{Maxwell_GUED} and thus are sensitive to both changes in the position of the nuclei and the redistribution of electron density. UED experiments in the gas-phase have resolved coherent nuclear motions of vibrational wavepackets along both ground and excited states \cite{Yang:2016I2} and captured the photolysis~\cite{Wilkin2019,Liu2020} and ring-opening dynamics on the atomic scale \cite{Wolf2019} $i.e.$ with angstrom spatial resolution and temporal resolution approaching 100 fs. The main scientific driver for UED is to capture the structural dynamics that takes place as the photoexcited molecule returns to the ground state by following the coherent motion of nuclear wavepackets and redistribution of energy. The focus of the work so far has been on a) Investigating coupled-nuclear electronic motion in the excited state, b) Capturing relaxation dynamics: resolving reaction paths during the relaxation of molecules to the electronic ground state, and determining the structure and vibrational motions of intermediates and end products, c) Direct retrieval of three-dimensional structure from diffraction measurements.

 The observation of coupled electronic and nuclear rearrangements, arising from conical intersections, are key to understanding the conversion of light into mechanical and chemical energy. Many important photochemical processes, such as photosynthesis, retinal isomerization in vision, ultraviolet-induced DNA damage \cite{Crespo-Hernandez2004_DNA}, and formation of vitamin D \cite{Holick1987_previtaminD3} are governed by non-adiabatic processes taking place at conical intersections. The first spatially resolved observation of a wavepacket traversing a conical intersection was a recent landmark UED study of the photodissociation dynamics of trifluoroiodomethane, by Yang $et al.$, \cite{Yang2018_CF3I}, however, much remains to be learned, particularly in more complex molecules. While most UED experiments have focused on capturing nuclear motion, a recent studied has shown that electronic changes can also be retrieved from electron diffraction signals \cite{Yang2020_Pyridine}, which enables UED measurements to capture both electronic and nuclear changes, and measure time delays between electronic and nuclear motions.  

 The non-radiative relaxation of a system relies on the redistribution of internal energy into nuclear degrees of freedom as the molecule returns to the ground state. By spatially resolving the nuclear wavepacket motion from its inception in the excited state to its vibrational dephasing in the ground state, UED experiments can glean information into the mechanisms mediating the dissipation of internal energy.  A recent UED experiment probing the photoinduced ring-opening dynamics of 1,3-cyclohexadiene, CHD, a model for the photosynthesis of previtamin D3, using UED, revealed a coherent oscillatory rotation of the terminal ethylene groups in the ground state photoproduct 1,3,5-hexatriene on the ground state \cite{Wolf2019}.  UED has also successfully investigated structural dynamics triggered by dissociation in 1,2-diiodotetrafluoroethane, C$_2$F$_4$I$_2$ \cite{Wilkin2019} and 1,2-diiodoethylene, CH$_2$I$_2$ \cite{Liu2020}. Knowledge of the structure of a transient state in a reaction is key to the rationalization of chemical reactivity. The photodissociation reaction of C$_2$F$_4$I$_2$ produces the intermediate state C$_2$F$_4$I before dissociation of the second iodine atom to produce C$_2$F$_4$. The structure of the intermediate was determined first with picosecond resolution \cite{Ihee2002}, and later with femtosecond resolution \cite{Wilkin2019}.
   
 In gas-phase UED, the random orientation of molecules in the target volume results in the loss of structural information, which prevents the retrieval of three-dimensional structural information directly from the diffraction pattern alone. Controlling the angular distribution of the target molecules, more specifically alignment along a single axis, increases the information content of the diffraction patterns \cite{Centurion2016,YangCenturion2015} and has been shown to be sufficient to retrieve 3D structures from a combination of multiple diffraction patterns from molecules aligned by a femtosecond laser pulse \cite{Hensley2012,Yang2014_align,Yang2015_CS2}. In principle, by alignment of the molecules before excitation, it should be possible to retrieve the full time-dependent three dimensional structure of the evolving molecules, at least for simple structures \cite{NunesCenturionAAMOP}. This capability could greatly enhance the information content of UED experiments. 

Advances in the UED sources have, and will undoubtedly continue to be reflected in great strides in our understanding of photochemistry and photobiology. The technique has demonstrated its enormous impact in providing complementary information to laser-based spectroscopic methods that probe the electronic structure, and in combination with other methods can help to build a complete picture of the electronic and nuclear dynamics. Technological and methodology developments in gas-phase UED will soon allow for the study of large and more complex model systems and the study of classes of reaction across multiple systems. These will enable the rationalization of general rules for reactivity with the goal that molecules can be designed from first principles to fulfill a particular function. 

\section{Ultrafast probes for electron diffraction}
\label{sectionII}

\subsection{Overview of a general UED setup and operating modes}
\label{sectionII.a}

The consolidation of ultrafast electrons as probes of matter providing high spatial and temporal resolution is the result of concerted advancements in multiple scientific and technological areas. To start, the widespread adoption of photoemission for particle accelerator sources has revolutionized the field of high brightness electron beams which had already seen a leap forward with the invention of field-emission electron guns in the late 60s~\cite{FEgun_1968} with respect to thermal emission sources used earlier. For field-emission based guns, higher beam quality is achieved by minimizing the effective source size rather than by increasing the total current. In case of photoemission, the laser pulse triggers prompt emission of densely packed electron pulses. In this case, the temporal duration of emission is limited by the laser pulse length, thus reducing the effective duty cycle (ratio between emission time on and time off) by orders of magnitude when compared to continuous field or thermal emission sources. To compensate the ensuing reduction in average current, UED instruments commonly generate pulses with many electrons per bunch via emission from macroscopic flat photocathode surfaces, with typical sizes ranging from micrometers to millimeters. Here the angular spread of the emitted electrons is a key factor that sets the limit on the achievable beam brightness \cite{dowell_cathode_2010} and the large area enables the extraction of Ampere-scale instantaneous currents \cite{MaxCurrent}.

After extraction, preserving high beam quality to the sample becomes of upmost importance. The interactions of the electron beam with the environment and within itself via Coulomb forces can indeed broaden the pulse temporal distribution effectively resulting in degradation of the instrument temporal resolution. Progress in understanding the latter (Coulomb-broadening or space charge effects) in these instruments \cite{Siwick:2002hiba,Reed:2006jbbaca} has led to identifying the most efficient ways of avoiding or managing temporal stretching, i.e. limiting the propagation distance to the sample and/or rapidly accelerating the electrons to higher energies. The accelerating electric field and the final kinetic energy have then turned into key parameters of electron guns for UED applications. Cross-fertilization with the neighboring field of high brightness electron sources for high energy particle accelerators promoted the introduction of a variety of beam manipulation methods and technologies, expanding the parameter space and tailoring the beam phase space around the particular application. Examples include the use of radio-frequency (RF) accelerating cavities where electric fields approaching 100~MV/m can be used to quickly boost the energy of the electrons to the MeV range \cite{Wang:1996up, JKoreanPhys06}. RF fields can also be employed to reverse the space charge induced temporal expansion to retrieve very short bunches at the sample plane \cite{Oudheusden:2007foba,Chatelain2012, Gao_2012,otto_solving_2017,Gliserin2012}. RF-based deflecting cavities have been used as ultrafast streak-cameras \cite{Musumeci:2009kyba,Oudheusden:2010cfbaca}, high-speed beam blankers \cite{verhoeven2018high}, or in high resolution time-of-flight spectrometers \cite{verhoeven2016time}. A more recent example is the adoption of achromatic beam transport lines originally developed for synchrotron x-ray sources, to passively reverse the space charge induced expansion and at the same time reduce the time-of-arrival jitter of electron bunch at the sample \cite{kim:isochronous,Qi2020}.

In this fertile research environment different technological approaches sprung, with the shared ultimate goal of achieving ever improving spatio-temporal resolution. In many cases, custom instruments have taken the form of compact accelerator beamlines with flexible designs, equipped with a mix of electromagnetic, electrostatic and magnetostatic optical elements and insertable diagnostics stations \cite{musumeci2010high, Li:2009kjba, Murooka11, Manz:2015cwba, SLAC_first, ASTA_RSI_2015, BNL_UED, Chatelain2012, Filippetto2016, mancini_design_2012, Waldecker_2015, caoued2003, Fu_2014}. A parallel technological approach utilizes modified electron microscope columns to effectively take advantage of the unsurpassed lateral beam quality and electron optics of these setups \cite{feist_ultrafast_2017, RF_TEM_2018, flannigan14, Kuwahara_2012, Kwon_uem_2017, Houdellier18, Zewail_2010, cao_uem_2015, vanderVeen_20}. Such systems usually work in the single-electron emission mode to achieve sub-picosecond resolution and necessitate coupling with high repetition rate optical excitation of the sample to maintain an acceptable signal-to-noise ratio. In TEM-column instruments, it is relatively easy to achieve nanometer-scale spot sizes at the sample plane, and the large flux density (electrons/s/$m^2$) allows for the collection of nanoscale information from heterogeneous specimens. This approach has demonstrated successful, especially in the area of time-resolved electron nano-diffraction and microscopy~\cite{danz2021ultrafast,Flannigan_2016}.

\begin{figure}[ht]
\centering
\includegraphics[width=1\columnwidth]{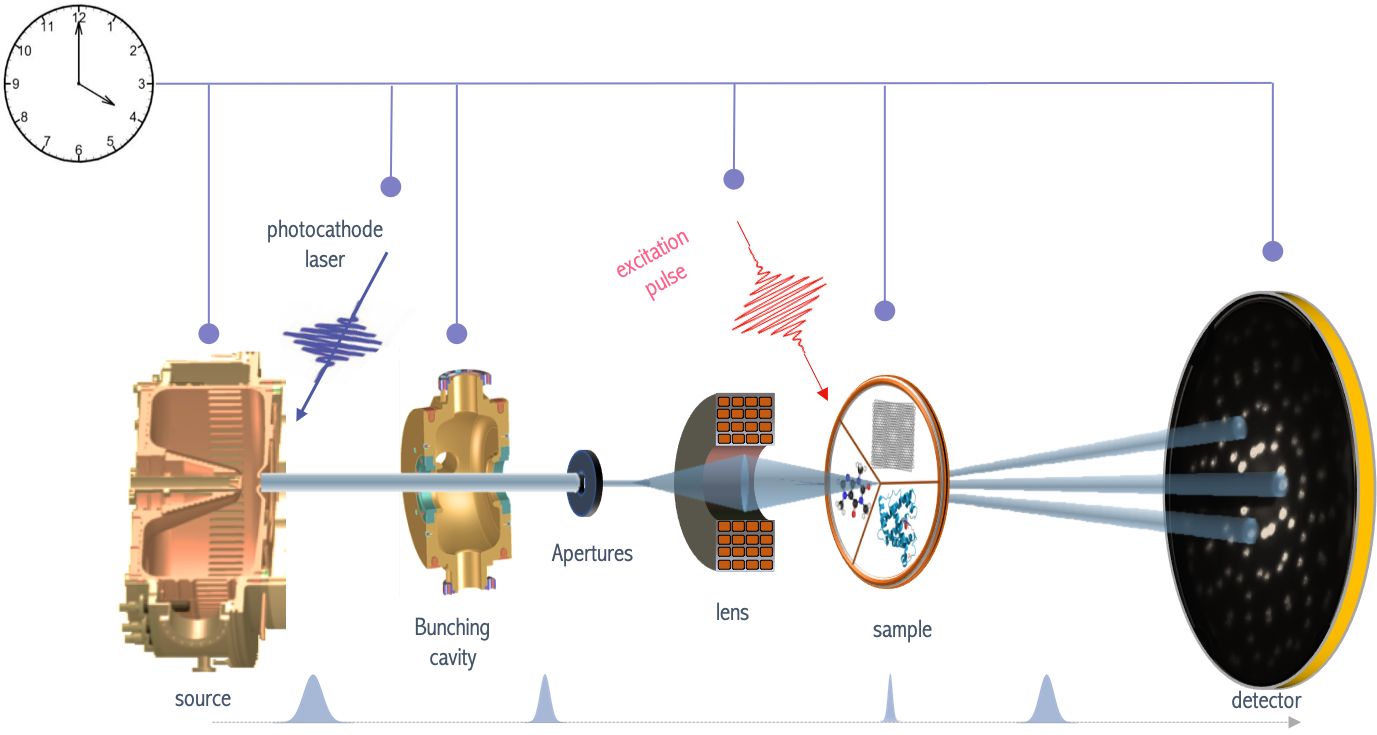}
\caption{Schematic of a general UED setup.}
\label{fig:setup}
\end{figure}

Figure~\ref{fig:setup} provides a general schematic of a UED beamline with all its components. The electron source consists of a photocathode and subsequent accelerating gap. Its geometry also provides an optical path for an ultrafast laser pulse to reach the photocathode, either by back or front illumination. Acceleration can be provided by static or time-varying electric fields~\ref{section:acc_tech}).Electron optics and collimation are used to tune sample illumination and reciprocal space resolution, and time-varying fields can be used for temporal beam compression (bunching). After the passage of the electron probe beam through the sample, the diffracted signal is detected downstream the sample plane.

In its most general configuration, a UED setup includes a timing and synchronization system, as schematically shown in Fig.\ref{fig:setup}. The generation of an electron pulse is temporally coordinated with downstream beamline subsystems via a timing distribution system consisting of opportunely generated and delayed trigger pulses. Such signals, electronically or optically distributed, initiate or terminate synchronous actions along the line, such as image acquisition or pulsed sample delivery systems. 

\subsubsection{Temporal resolution}

The overall temporal resolution is probably the single most important parameter in a UED setup, and it is described as a combination of multiple uncorrelated terms, including the excitation pulse length $\tau_{pump}$, the electron beam pulse duration $\tau_{probe}$, the velocity mismatch $\tau_{VM}$ (if applicable, see Sec.\ref{sect:GUED}), and the fluctuations $\tau_{\Delta pp}$ in temporal delay ($\Delta t$)   between the laser pump and the electron probe (see Fig.~\ref{fig:methodology}). A generally accepted metric for calculating and reporting the instrument temporal resolution of an instrument using Eq.~\ref{eq:temporalresolution} is that of Full-Width-Half-Maximum (FWHM hereafter).

Accelerating and bunching field amplitude and relative phase fluctuations cause shot-to-shot fluctuations of $\tau_{\Delta pp}$ (see Sec.~\ref{section:longitudinalDynamics} and \ref{section:timecompression}), and require precision phase synchronization  between the different sources is required (Sec.~\ref{section:clocking}). In the assumption that the same laser system is used to both generate photo-electrons and to excite the sample, we then have $\tau_{\Delta pp}=\tau_{\Delta_{eTOF}}$, where $\tau_{\Delta_{eTOF}}$ is the electroh time-of-flight (TOF) from cathode to sample. If instead more than one laser system is used in the experiment, a similar synchronization system is required between the different optical oscillators, and the jitters in arrival time of the laser to the cathode and to the sample plane would need to be taken  into account separately.

\begin{equation}
\tau_{res}=\sqrt{\tau^2_{pump}+\tau^2_{probe}+\tau^2_{VM}+\tau^2_{\Delta pp}}
\label{eq:temporalresolution}
\end{equation}

In the following we will provide an in-depth review of the state-of-the-art of each of the subsystems introduced above. 

\subsubsection{Electron packets: from single-electron to single-shot}
\label{SS}

The number of electrons interacting with the specimen required to obtain structural information varies by orders of magnitude, depending on the modality and on the specimen details. As an example, electron microscopy provides real-space local information, and therefore it requires high dose at the sample (10-100 electrons/(spatial resolution)$^2$). The requirement for number of electrons $N^{I}_e$ illuminating the sample  is usually in the range of $10^8$ to $10^9$. 
In electron diffraction on the other hand, the signal at the detector carries reciprocal space information integrated over the entire illuminated sample area. For solid-state specimens the signal is concentrated in few areas of the detector, usually spots or rings, as a consequence of the highly ordered atomic structure of the sample. Typically less than $N^{ED}_e\approx 10^6$ electrons are sufficient to obtain high quality (multiple Bragg spots) diffraction patterns from a thin (one elastic mean free path) solid-state sample~\cite{siwick_atomic-level_2003}.  The sample material (high Z atoms scatter more efficiently) and thickness (dynamical scattering effects can lower the signal on the Bragg peaks), play a role in the definition of $N^{ED}_e$, such as the density of the material itself. For electron diffraction on gas-phase targets, the value of  $N^{ED}_e$ is usually many orders of magnitude larger, depending on the gas density and types of atoms in the molecules.Furthermore, in UED experiments the transient signal are usually retrieved from the difference image between diffraction pattern before and after excitation. Hence the value of $N^{ED}_e$ will also depend on the magnitude of the signal to be detected. If the goal is to resolve 1$\%$-level changes in peak intensity, then Poisson statistics dictates at least 10000 electrons in the Bragg peaks analysed.

\begin{figure}[ht]
\includegraphics[width=0.7\columnwidth]{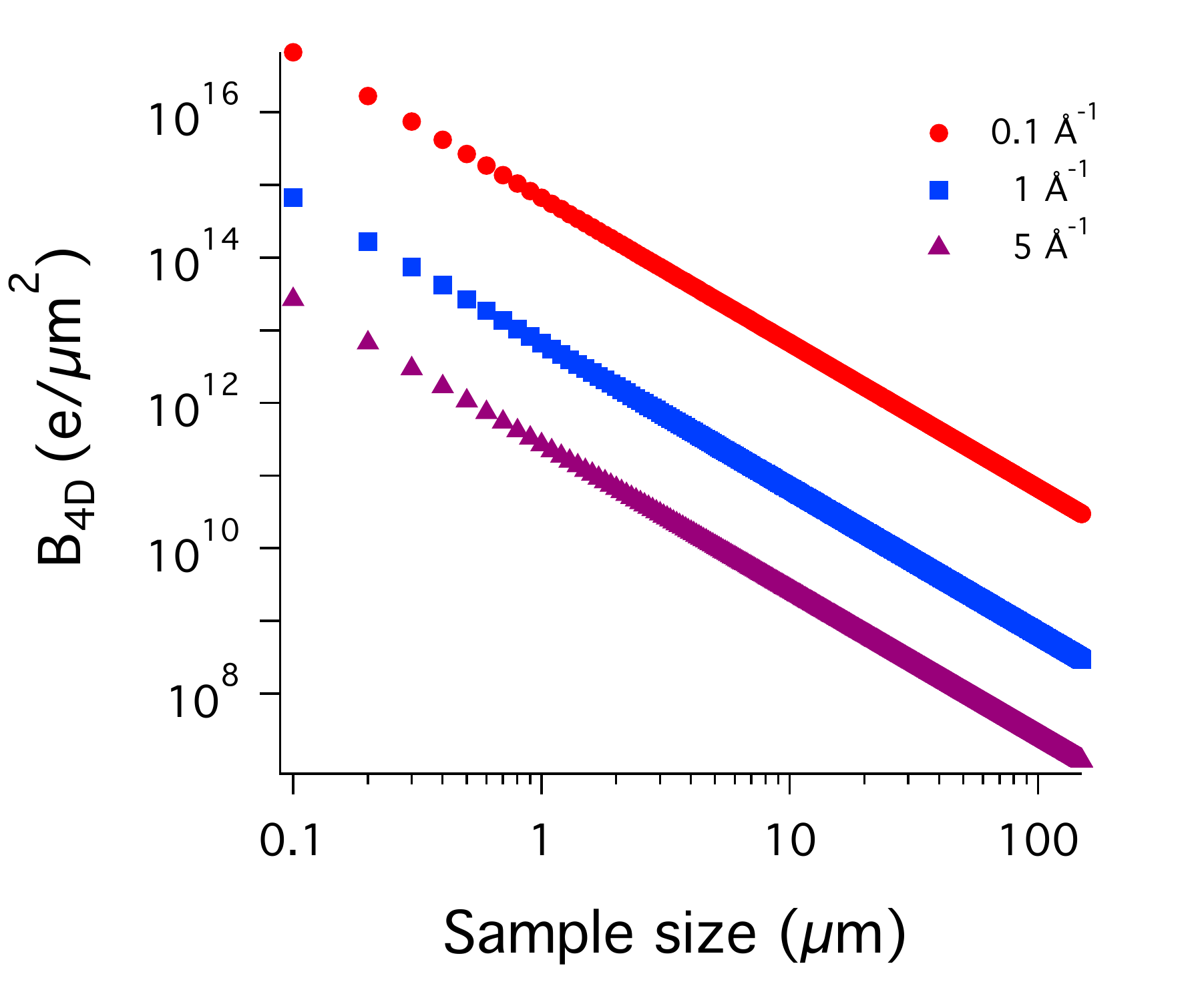}
\caption{Requirements of 4D brightness as function of sample size for different values of momentum transfer, from 0.1 to 5 \AA$^{-1}$. The calculations assume that the electron beam fully illuminates the sample area, $N^{ED}_e = 10^6$ and $\mathbb{R}=10$. Different curves relate to different momentum transfer values.}
\label{fig:B4dVsSS}
\end{figure}

When evaluating the feasibility of an experiment it is instructive translating electron diffraction requirements into constrains for the beam four-dimensional brightness. Electrons must be tightly confined spatially within the specimen boundaries, while maintaining a small angular spread for achieving good resolution in reciprocal space (and a large enough spatial coherence length). Using the definition of $\mathbb{R}$ from \ref{section_coherence_length}, the minimum required value for the 4d brightness (Eq.~\ref{B4d}) is equal to:

\begin{equation}
B^{min}_{4d} = e N^{ED}_e\left( \frac{2\pi\mathbb{R}_{min}}{ s d_{s}\lambda_c}\right) ^2
\label{B4Dmin}
\end{equation}
where $\mathbb{R}_{min}$ is the experimental target for resolving power at momentum transfer $s$, $d_s$ is the illuminated specimen lateral size (assuming circular symmetry for simplicity), and $\lambda_c$ is the Compton wavelength. Figure~\ref{fig:B4dVsSS} reports calculated values of four-dimensional brightness assuming $N^{ED}_e = 10^6$ needed to obtain diffraction patterns with adequate SNR, using $\mathbb{R}=10$ for different diffraction momentum transfer values. The illuminated sample size strongly affects the requirements on the electron beam, and can ultimately drive instrument design choices.

Experiment acquisition modalities can be separated in two broad categories: single-shot and multi-shot (stroboscopic) modes.The choice of the modality is often dictated by the details of the phenomenon under study. In a reversible process, the excited specimen can be cycled between identical initial and final states by a very large number of times, undergoing exactly the same dynamical process and allowing data integration over many shots. Other samples show enhanced sensitivity to the excitation and damage or a modified initial state develop after a finite number of pulses, limiting the total number of excitation events (partial reversibility). Finally, if the excitation pulse drives the system to an irreversible final equilibrium state different from the initial one, only the paired probe pulse will be able to capture the transition before the sample is permanently altered.

In line with the different types of processes, UED operation modalities span from single-electron to high-charge per bunch, and from one/few shots per second to millions, with fundamental impact on the instrument technology used, starting from the choice of the laser system and repetition rate, the electron source size and geometry, the transverse and longitudinal compression schemes, and the overall footprint of the setup. 

A key difference between the single and multi-electron beam modalities is the role of the beam self-fields (see Sec.~\ref{section:spacecharge}) in the beam dynamics. The so-called space-charge fields effect the bunch duration, the beam energy spread, and the total beam emittance of a multi-electron bunched beam, while single-electron pulses are only constrained by transverse and longitudinal emittance at emission~\cite{baum_2010_singleelectron}. Upon RF compression, for example, single-electron wavepackets can theoretically be squeezed down to well below 1 fs \cite{baum2013physics}. Since the longitudinal emittance is conserved, temporal compression does come at expenses of energy spread, but typical UED experiments can tolerate this. Another advantage of single-electron "beams" operations, is that the emission source can be arbitrarily small (and correspondingly higher beam brightness), due to the absence of external field screening from other electrons. As it will be more clear in Sec.~\ref{sourcesize}, such beams can  be focused down to nanometer-scale sizes at the specimen maintaining good transverse coherence length. 

Note that the concepts of beam size and angular spread in single-electron mode take the meaning of moments of distribution of the statistical ensemble represented by many single-electron beams, generated and transported through the beamline at different times. Although for an isolate electron one could define and measure angle and position to a better degree, a visible diffraction pattern is only formed upon accumulation of many electrons, and the overall resolution will still depend on the moments of the ensemble distribution. This issue could potentially be minimized via the combined used of fast single-electron detectors and time-stamping, although high precision non-invasive time-stamping methods for single-electron beams are still out of reach.
Finally it is also worth pointing out that, as a direct consequence of the statistical nature of photo-emission, the beam current in this configuration is in practice limited to much less than 1 electron per shot. Indeed, in order to maintain the spatio-temporal characteristics of the beam shot by shot, the generation of beams with more than one electron should be avoided. The photo-emission probability is described by Poisson statistics and, in order to ensure that the overwhelming majority of pulses contain only one electron, the average value of the distribution needs to be below 0.5 \cite{baum2013physics}. 

\subsection{Generation of electron pulses}
\label{sectionII.b}

Although a continuous electron stream can be temporally chopped or bunched by (a series of) RF cavities (see for example Sec.~\ref{sectionII.d}), most UED electron sources use short pulse lasers for generation of electron bunches by photoemission. When a laser beam impinges on a photocathode surface, single or multiphoton absorption can cause electrons in the material to gain enough energy to overcome the potential barrier at the interface and escape into the vacuum. The spatio-temporal format of the exciting laser pulse is nearly preserved in the photoemission process offering the opportunity to shape the initial electron beam distribution by controlling the properties of the illuminating laser. 

Photocathodes are evaluated by a few key parameters: the quantum efficiency $QE$, the mean transverse energy of emitted electrons $MTE$ \cite{karkare_effects_2015}, response time, and effective emission lateral size. The geometry of the emitting surface is also of importance. A small radius of curvature can be used locally enhance the external fields amplitude (DC, RF or optical). Larger radius of curvatures would not produce significant enhancement, but introduce transverse focusing or defocusing fields in the cathode vicinity, which would modify the downstream beam dynamics(Sec.~\ref{section:cathodecurvature}). 

\subsubsection{Quantum efficiency}
\label{section:QE}
The cathode quantum efficiency $QE$ is defined as number of emitted electrons per number of photons incident on the material, i.e. $QE=\frac{\hbar\omega}{e}\frac{Q}{E_{ph}}$, where $Q$ is the electron beam charge and $E_{ph}$ is the laser pulse energy. 
A theoretical expression for $QE$ in metals can be found by following the three step model model \cite{SPICER_1964}, and the QE can be directly related to the difference between laser photon energy $\hbar\omega$ and material work function $\Phi_W$ (i.e. to the electrons excess energy $E_{ex} = \hbar\omega-\Phi_W$). For photo-emission to happen, the electron first absorbs one (or more) photon, then travels to the surface avoiding scattering with other electrons, and lastly reach the vacuum interface with enough energy in the normal direction to overcome the potential barrier. Typical metals used as photocathode materials (Cu, Ag) have work function in the $4.5-5 eV$ range, with $QE$ values upon UV pulse illumination ranging between $10^{-5}-10^{-4}$. As a numerical example, using a Cu cathode with $10^{-5}$ QE, a laser pulse with 80 nJ  energy at 266 nm (third harmonic Ti:Sa laser) would suffice to generate $10^6$ electrons.

In presence of an externally applied electric field $E_0$ on the cathode surface, the total potential barrier is modified by the Schottky potential $\Phi_{Schottky}$~\cite{Schottky}. The resulting effective potential therefore becomes $\Phi_{eff}=\Phi_W-\Phi_{Schottky}$, where $\Phi_{Schottky} = \frac{e}{2}\sqrt{\frac{eE_0}{\pi \epsilon_0}}$.
In the approximation of constant electron density of state close to the Fermi level (where electrons are emitted from), and approximating the material temperature to zero, it can be shown that for small excess energies $QE\propto(\hbar\omega -\Phi_{eff})^2$~\cite{dowell_quantum_2009}. 
Besides lowering the work function, the applied field at the cathode plays an important role in determining the maximum charge and current density that can be extracted, as we will discuss later.

\subsubsection{Photocathode thermal emittance}
\label{section:MTE}

\begin{figure}[htb]
\centering
\includegraphics[width=0.6\columnwidth]{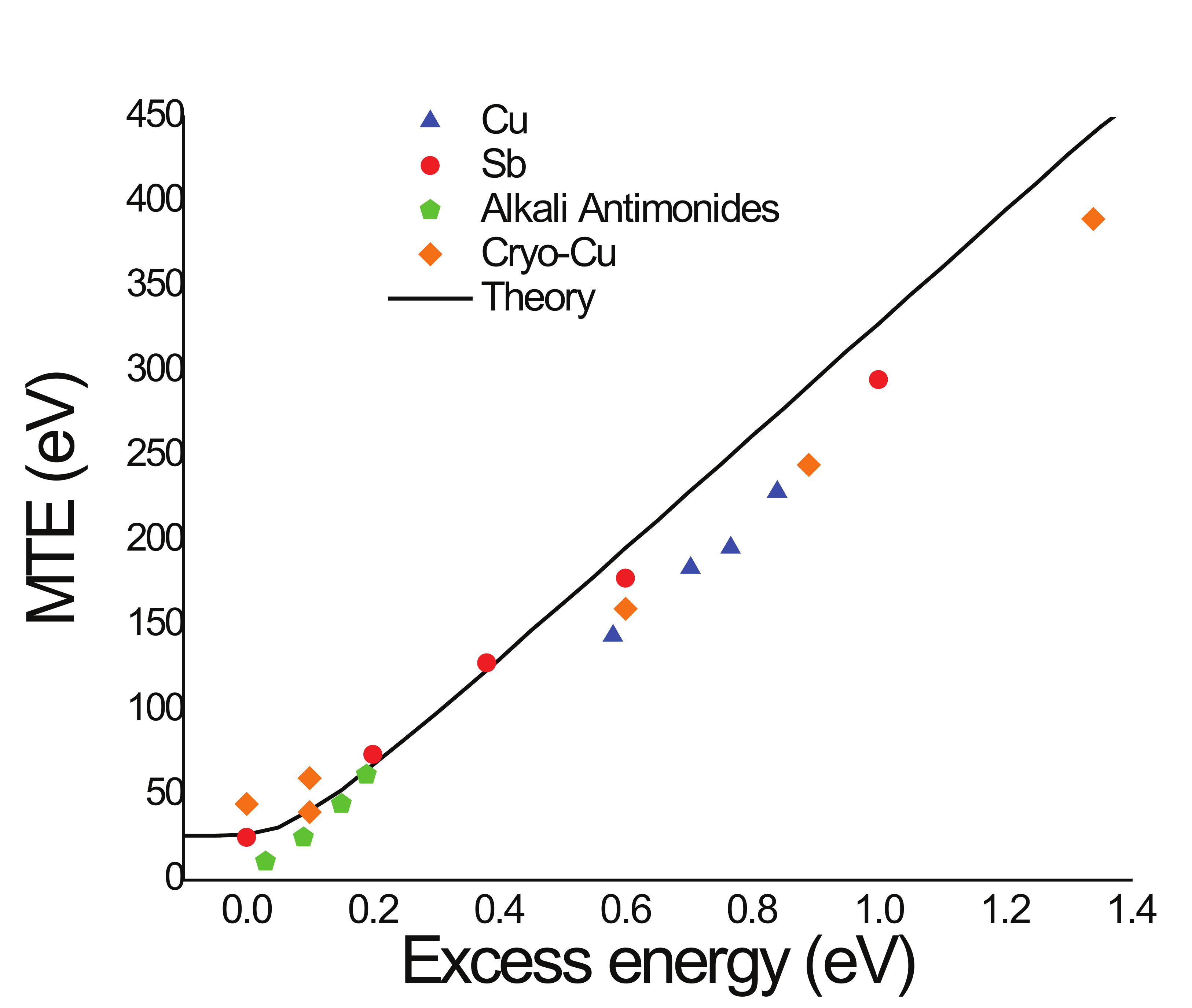}
\caption{MTE versus excess energy for different cathodes, and compared with theory. The minimum of MTE measured corresponds to ambient temperature (26 meV) adapted from \cite{musumeci_advances_2018}.}
\label{MTE}
\end{figure}

The mean transverse energy (MTE) of the emitted electrons determines the beam emittance and, therefore, plays a relevant role in determining the beam brightness. The beam normalized rms emittance at emission can be written as $\epsilon_n=\sigma_{laser}\sqrt{\frac{MTE}{m c^2}}$ \cite{karkare_effects_2015}. Using the same approximations for the density of states and the Fermi-Dirac distribution used above to calculate the $QE$, we can 
integrate the standard deviation of the particle transverse momentum leading to the value for the $MTE=\frac{\hbar\omega -\Phi_{eff}}{3}\propto \sqrt{QE}$~\cite{dowell_quantum_2009}, clarifying the trade-off between larger $QE$ values and smaller transverse beam emittance. 
As an example, using longer laser wavelengths to decrease $E_{ex}$ is a clear path to smaller emittance values and larger brightness, but it also decreases rapidly the cathode QE, requiring more laser energy~\cite{hauri_intrinsic_2010}. Also, in the limit of $E_{ex}$ similar or smaller than the thermal energy $K_BT$ (where $K_B$ is the Boltzmann constant and $T$ is the temperature of the cathode) the approximations used in the calculation of electron transverse momentum spread fails as the tails of the Fermi-Dirac distribution dominate the spread, limiting the mininum achievable $MTE$ to $K_BT$~\cite{feng_thermal_2015}. The same behavior has been measured in semiconductor cathodes, as shown in Fig.~\ref{MTE}.

Given the low laser energy needed to obtain typical electron charges for UED setups (Sec.~\ref{section:QE}), it may seem natural to trade quantum efficiency for better beam quality. On the other hand, a large increase in laser beam energy compensating lower $QE$ approaching the work-function threshold may have detrimental effects. First, when coupled with a small focus at the cathode, it can lead to values of optical fluence approaching the material damage threshold. Furthermore, high intensity beams can increase the temperature of the transient electronic distribution within a material by orders of magnitude.  For sub-ps photoemission, electrons do not have time to thermalize with the lattice, since typical electron-phonon coupling constants are in the few-picosecond range. This can lead to photo-emission of hot electrons, contributing to the beam momentum spread and ultimately limit the achievable MTE~\cite{maxson_ultrafast_2017}. 

\subsubsection{Response time of a photoemitter}

Most photocathode materials have response times in the few femtoseconds to sub-picosecond range, dominated by the travel time of electrons from the bulk to the vacuum interface, which is determined either by penetration depth of the optical pulse and/or the photocathode film thickness. In certain materials (negative electron affinity semiconductor cathodes) the surface is chemically prepared to energetically boost the bottom of the conduction band above the vacuum level. Upon photon absorption electrons will reach the conduction band, and some of them will slowly relax to the bottom of the band via scattering with the lattice, while traveling toward the surface. Once there, they will escape into the vacuum thanks to the negative electron affinity, forming long temporal tails (up to 100 ps) with close-to-zero excess energy~\cite{bazarov_thermal_2008}. This effect is more visible when using small excess energies, while tends to disappear with increasing photon energies.

\subsubsection{Source size and spatial resolution in UED}
\label{sourcesize}

The size and shape of the photo-emitting area has an direct impact on scientific breadth of the instrument, influencing paramters as the transverse brightness and spatial resolution, and driving technological choices, such as the accelerating field, illumination geometry and repetition rate.

Single-shot ultrafast experiments require large peak currents $I_{peak}$, which are achieved using mm-scale source sizes. Such dimensions are only acceptable when the heterogeneity of the sample is not a concern.
When probing reversible dynamics in stroboscopic mode, the real space resolution of UED can be greatly increased by decreasing the spot size and optimizing for the current density $J_{peak}$ rather than for $I_{peak}$. At its limit, the stroboscopic modality could provide combined nanometer-femtosecond spatio-temporal resolution, ultimately enabling nano-UED~\cite{ji_nanoued_2019,feist_nanoscale_2018} and ultrafast STEM experiments. 
The sample illumination area for a fixed transverse coherence length scales linearly with the source size. Indeed, using similar considerations to the one that led to Eq. \ref{B4Dmin}, and assuming transverse emittance conservation along the beamline, the initial emitter diameter can be directly related to the spot size at the sample:

\begin{equation}
\frac{\sigma^{min}_{x,sample}}{\sigma_{x,cath}}\geq\frac{\pi \mathbb{R}_{min}}{\lambda_c s}\sqrt{\frac{MTE}{\ m_e c^2}}
\label{eq:spotsize}
\end{equation}
For typical MTE values of 0.5 eV and a resolving power above 10 at around 1~\AA$^{-1}$, we get $\sigma^{min}_{x,sample}>1.3 \sigma_{x,cath}$. This simple result puts in evidence the need of nanoscale emitters to reach nanometer-scale spatial resolution. Alternatively, transverse collimation downstream of the cathode can be performed, a common practice in static electron microscopes, with the result of decreasing the initial emittance at expenses of beam current~\cite{ji_nanoued_2019}(see Sec.~\ref{sec:beamsize}). 

In photo-emission,  two main factors limit the initial spot size: the numerical aperture (NA) of the optical delivery system, describing the laser beam convergence angle $\theta$ ($NA=sin(\theta)$ in vacuum), and  the wavelength $\lambda_{ph}$ of the laser used for photoemission. Even in the ideal case of $NA\approx1$, the laser beam waist $w_0$ is limited to $ w_0=\frac{\lambda_{ph}}{\pi}$. In practical circumstances the geometry of the setup may even prevent using large values of $NA$, due to physical constrains on the minimum distance of the last focusing lens from the cathode plane. Solutions to this issue have been investigated, for example by developing photocathodes operating in transmission geometry, allowing the last optical element to be positioned right to the back of the photoemission material~\cite{liu_csbr_2006}. 

In the last decade, laser-assisted electron emission from tips has been extensively explored to overcome the light diffraction limit. Selecting electrons emitted by the apex of the tip upon laser illumination provides nanoscale sources of femtosecond pulses. Laser triggered emission can be achieved via control of the temperature and the voltage applied to the tip, modulating the Fermi distribution tail and the potential barrier, exponentially suppressing electron emission in absence of laser. Linear photoemission from tips using near UV laser pulses can also be achieved upon coating of the tungsten apex with ZrO layer \cite{cook_improving_2009}~\cite{yang_scanning_2010} \cite{feist_ultrafast_2017}. 

The shape of the source has a strong impact on the amplitude and direction of externally applied field in the vicinity of the cathode plane ~\cite{TEM:book}, which in turn has an impact on the magnitude of electron emission and on the beam dynamics in the accelerating gap (see Sec.~\ref{section:cathodecurvature}).
Curved surfaces enhance the amplitude of the electric field at the surface, lowering the work function through the Schottky effect. On flat surfaces, such effect usually accounts for a decrease in the work function of no more than a few tenths of eV for all practically achievable accelerating fields. If a sharp tip with large aspect ratio $L/R$ is used instead, where $R$ is the tip radius and $L$ is the tip height, the field at the tip surface is greatly enhanced ($E_{enh}\propto E_{in}L/R$~\cite{podenok_electric_2006}), leading to a dramatic change in effective work function (up to more than 1 eV) and in photoemission yield. Note that in the limit of extreme electric fields (in excess of $10^9$ V/m), field emission rather than photoemission would dominantly contribute to the output current \cite{FN}. If the tip radius is comparable or smaller than the wavelength of the laser pulse used for photoemission, optical field enhancement takes place. Depending on the amplitude of the enhanced laser field, either weak or strong photoionization regimes can be achieved (measured by the Keldysh parameter $\gamma$~\cite{Keldysh}), leading to  
multiphoton photoemission~\cite{ropers_localized_2007} and/or optical field emission~\cite{hommelhoff_ultrafast_2006}.

At a first glance, the demands for high peak current and large current density may seem to be simultaneously achievable. As will discuss in the following, the two requirements instead drive the size of photoemission area in opposite directions. The number of electrons emitted from a flat cathode in a given time and from a given area is limited by cumulative image-charge fields at the cathode interface. As electrons get emitted from the material into the vacuum, charge at the surface promptly re-distribute to screen the bulk material from the external field. The total electric field $E_{tot}$ in the vacuum region between the emitted electrons and the cathode surface is therefore the sum of the externally applied electric field and the opposing image-charge field. In the limit of very short pulses the electron beam aspect ratio $A$ shows a "pancake-like" format, with $A=\frac{2 m R}{\Delta t^2eE_0}\gg 1$, $E_0$ is the external accelerating electric field, and $R$ and $\Delta t$ are the laser beam radius and pulse duration, respectively (considering for simplicity a uniformly charged cylinder). In this case the electron density in vacuum can be approximated as an infinitely wide sheet of charge, and the emission will stop when $E_{tot}=0$, leading to a maximum charge of $Q=\epsilon_0E_0\pi R^2$, where $\epsilon_0$ is the vacuum permittivity~\cite{BazarovPRL}. As an example, for an accelerating field of 20 MV/m, an emitter area larger than $17~\mu m$ in radius would be required to extract $10^6$ electrons.

Decreasing the source size to sub-micrometer changes the beam aspect ratio, eventually leading to cigar-like formats ($A<1$). In this case the finite transverse extension of the beam plays a dominant role in the extraction process, changing the functional form of the scaling laws for current density and brightness~\cite{MaxCurrent}. 
It is worth reporting the 4D brightness scaling for the case of large and small aspect ratio: 

\begin{equation}
 B_{4D}^max \propto
\begin{cases}
\frac{E_0}{\textrm{MTE}} & \textrm{for $A\gg1$ (pancake beam)} \\
\frac{E_0^{3/2}}{\textrm{MTE}}\frac{\Delta t}{R^{1/2}} & \textrm{for $A<1$ (cigar beam)}
\label{eqn.4dbrightness}
\end{cases}
\end{equation}

In the case of cigar aspect ratios, decreasing the source size will cause a smaller change in the maximum charge extracted than in the corresponding emittance (squared), with the important and often overlooked consequence of introducing a dependence between maximum 4D brightness and the source size. 
Therefore a possible scheme to achieve larger brightness values would include starting from a cigar-shaped electron beam, and then perform temporal compression downstream the electron gun. Indeed, as it will be discussed more in detail in Sec.~\ref{section:timecompression}, the electron beam can be temporally compressed with minimal implications on the transverse emittance (see for example~\cite{Filippetto2016}). This setup allows smaller initial spot sizes and disentangles spatial and temporal resolution. The drawback is an increased longitudinal emittance, that would ultimately limit the shortest pulse length achievable~\cite{maxson2017}.

\subsubsection{Towards brighter photoemission sources}
\label{section_advanced_cathodes}
Relevant research directions aim at increasing the brightness of electron sources by decreasing the cathode MTE or decreasing the photoemission source size and at the same time increasing the acceleration field. As shown in Fig.~\ref{MTE}, an effective way to reduce the MTE is to decrease the excess energy, up to the limit where the residual MTE of the emitted electrons is limited by the cathode temperature. Values of MTE as low as 26~meV have been measured at room temperature ~\cite{feng_thermal_2015}, while more recently measurements as low as 5 meV have been demonstrated by cooling single crystal Cu (100) surface to cryogenic temperatures~\cite{karkare_ultracold_2020}, about two orders of magnitude lower than typical MTE values from metal photocathodes. One of the drawbacks of working close to the work function threshold is the strong reduction in QE, which complicates the use of such cathodes, especially for applications targeting large peak currents (See Sec.~\ref{section:MTE}). Recently it has been shown that using ordered crystal surface structures can partially reverse the dependence between QE and MTE~\cite{karkare_reduction_2017}. Here the values of electron transverse energy can be constrained by a careful choice the energy band structure, decreasing the MTE of the emitted electrons even for relatively large excess energy values. 

Alternatively, semiconductor cathodes can provide low MTEs and very large QE, on the order of few to few tens of percent, mostly thanks to the suppression of electron-electron scattering leading to a much more efficient transport of excited electrons from the bulk to the vacuum interface. In such materials electrons occupy states up to the top of the valence band, while the conduction band is empty. The energy barrier to overcome in this case is the sum of the material band gap and the electron affinity, often enabling linear photoemission with visible or infrared photons~(see for example \cite{cultrera_thermal_2011,cultrera_alkali_2014,cultrera_ultra_2016}. 
The photoemission surface of such materials is often chemically very reactive, and contamination from the external environment rapidly lowers the QE by orders of magnitude \cite{dowell_cathode_2010,filippetto_cesium_2015}. Special high vacuum load-lock chambers are typically employed to transfer these photocathodes from the growth chambers to the electron guns.

A possibility to reduce the emission area to be much smaller than that achievable by direct lens focusing is offered by laser field impinging on nano-structured metallic surfaces that can excite traveling waves confined at the metal-dielectric interface, called surface plasmon polaritons (SPP). Mediated by SPP, whose wavelength can be much shorter than that of the excitation pulse, the optical field energy can be transported and concentrated in areas of sub-wavelength size, leading to large local field enhancement. This concept has been lately used to enhance absorption on metal tips~\cite{muller_nanofocused_2016}. More recently, the same idea has been studied to induce large enhancement factors on nanoscale flat surfaces ~\cite{durham_plasmonic_2019}, which could be extremely useful if the cathode is immersed in high field areas, where tips may not be ideal due to large amounts of field-emitted current and short lifetimes. 

\subsubsection{Laser systems}
\label{lasers}

A critical element in any UED setup is the ultrafast laser system which is used to provide pulses to the cathode and excite the sample. 
It should not come as a surprise that the chirped pulse amplification was a critical technological step in enabling the development of UED technique. Additional laser applications in UED setups include ponderomotive scattering for time-stamping the time-of-arrival of electrons with the photon beam \cite{hebeisen_grating_2008}(see Sec.~\ref{section:timestamping}), generating THz waves can be used to compress the beam or streak it \cite{Fabianska14_thzstreaking}. In most advanced cases, lasers provide the energy for actual electron acceleration (ACHIP, LPA) \cite{faure:UED}. 

Typical architectures for UED laser systems include a modelocked oscillator cavity followed by a chain of amplifiers to bring up the energy to the required levels. In cases in which RF is used to manipulate, control or diagnose the electron beam, it is important to choose the oscillator cavity length that can be easily synchronized with the RF frequency used in the experiment. Typically, a intra-cavity piezo-mirror is used to close a feedback loop to maintain phase-locking to an external signal. More on this is discussed in Sec.~\ref{section:clocking}. State-of-the-art systems are also able not only to lock the envelope of the laser pulse to an external signal, but also to lock the phase. CEP-phase locked phases so far have not been employed in UED setups, but this might change as attosecond electron pulses are starting to be used to probe attosecond dynamics \cite{Morimoto_atto_2018}. 

While most of the UED instruments up to now have operated using the Ti:Sa technology due to the large gain-bandwidth and clear advantage in the generation of ultrashort pulses of this crystal, the limitations associated with the poor efficiency and associated low average power as well as the progress in other competing laser technologies such as Yb:based lasers are increasing the diversity of the laser systems out in the field. As discussed in Sec.~\ref{SS}, one of the main characteristics of any setup is the targeted operation mode (ranging from single shot to stroboscopic). For the latter, being able to increase the repetition rate beyond 50 KHz greatly affects the laser technology choice. An important issue that requires a compromise in fact is the longer pulse length typical of the higher repetition rate and higher average power laser systems. Ti:Sa systems routinely generate $<$ 40 fs pulses, while the pulse length in Yb-based systems is 5-6 times longer. An open question is how to get ultrashort pulses at high repetition rates. Different technologies are being pursued ranging from OPCPA \cite{dubietis:opcpa} to employment of non linear compression techniques \cite{jocher:nlc}. 

Precise control of the laser distribution illuminating the cathode has been shown to improve the beam brightness especially in space charge dominated beamlines \cite{musumeci_blowout_2008}. Both transverse and longitudinal shaping of the laser pulse before photocathode illumination have been employed. In the transverse dimension, predetermined schemes like imaging an overfilled aperture, or refractive shapers, compete with adaptive computer-controlled approaches based on liquid crystal mask \cite{maxson:adaptive} or digital micro-mirror arrays \cite{li:DMD}. On the longitudinal size, the temporal profile can be controlled with dispersive crystals \cite{zhou:aBBO}, acousto-optic \cite{li:dazzler} and mechanical~\cite{cialdi:simple} spectral shaping. For oblique cathode illumination, the technique of pulse front tilt \cite{hebling:pft}, which is also used to velocity match the pump and the probe on the sample as discussed later (see Sec.~\ref{section_laserpump}), can be also applied.

The wavelength selectivity of the gain mediums does not cover all the possible wavelengths. For example in photocathode drivers it is useful to be able to tune the photon energy to the cathode work-function, and similarly when pumping a material one wants to excite certain optical modes and steer away from high reflectivity regions. Non linear frequency generation, both directly in crystals as well in optical parametric amplifier setups are usually added to the main laser system. While the price to pay in pulse energy is significant, the continuous wavelength tunability they offer allow exploration of new physics. For longer wavelengths, either difference frequency generation options in the OPA \cite{fischer:DFG} or optical rectification \cite{fulop:opticalrectification} can be used to generate THz which can be used for compression/diagnostics and also directly for pumping. 

\subsection{Electron dynamics}

\label{sectionII.c}

In the final step of the photoemission process, electrons escape the cathode surface and enter vacuum with a residual kinetic energy typically in the range of a few to 100s meV. Transport and control at these low energies is quite challenging, and electrons are therefore accelerated to higher kinetic energies, ranging from 100 eV\cite{ropers2014, Bainbridge16, Vogelgesang2018, muller14} for surface science and low-dimensional materials in reflection geometry, to keV and MeV levels, more typical for transmission modes.

As explained in Sec.~\ref{sourcesize} larger accelerating fields at the cathode surface allow to extract larger current densities, and thus enable higher beam brightness for a given cathode MTE. In this section we will review the electron beam dynamics downstream the cathode plane, including acceleration and compression, which allow tailoring of the beam phase space to the specific application, but may also lead to potential degradation of initial beam brightness due to nonlinear forces, time-varying fields and/or self-forces within the electron bunch.

    \subsubsection{The accelerating gap}
\label{Eacc}

\begin{figure}[ht]
\includegraphics[width=1\columnwidth]{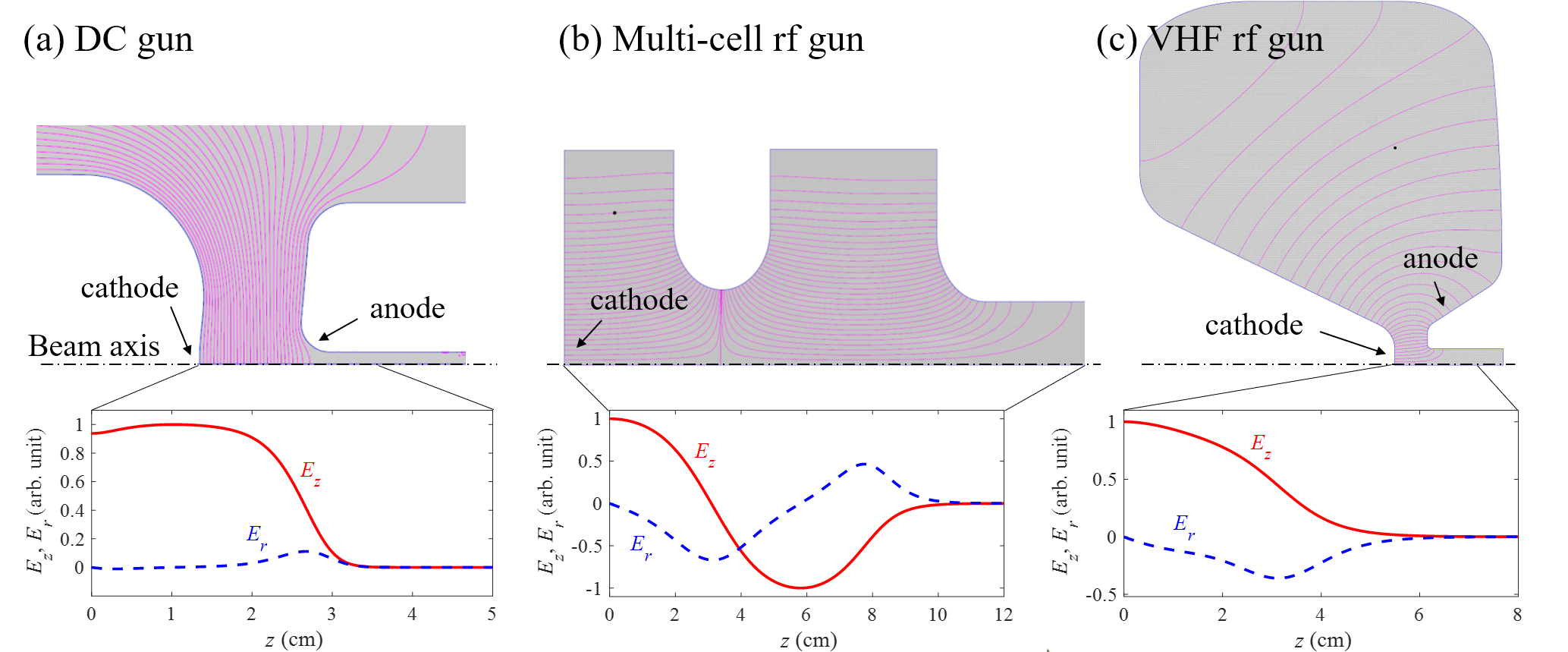}
\caption{For a (a) DC, (b) multi-cell RF gun, and (c) VHF quarter-wave resonator RF gun used for UED, the schematics of gun geometries are shown with field contours lines (equal-potential lines for DC gun and field lines for RF guns). The longitudinal ($E_z$, red solid) and transverse ($E_r$, blue dashed, small offset from the axis) field profiles are also shown.}
\label{fig:gungeometry}
\end{figure}

The most mature and widely used acceleration technologies use DC and RF fields~\cite{dowell_rao_book}. The schematics of the geometry and field profiles of a DC, multi-cell RF, and very-high-frequency (VHF) quarter-wave resonator RF gun are shown in Fig.~\ref{fig:gungeometry}. The geometry for these electron guns is essentially cylindrically symmetric, and the acceleration electric field can be written as:  
\begin{equation}
\label{eqn.accfield} E_z(z,r,t)=
\begin{cases}
E_0e_z(z,r) & \textrm{for DC fields} \\
E_0e_z(z,r)\sin(\omega t+\phi) & \textrm{for RF fields} 
\end{cases}
\end{equation}
where $z$ and $r$ are the axial and transverse coordinates, $z=0$ corresponds to the cathode position, $E_0$ is the peak electric field, $e_z$ is the normalized profile of the field distribution, and $\omega$ and $\phi$ are the RF angular frequency and phase, respectively. Maxwell equations relate the longitudinal field component $E_z$ with the transverse component $E_r=-\frac{r}{2}\frac{\partial E_z}{\partial z}$, which is important in the transverse evolution of the beam in the gun. In presence of time-varying fields, the ensuing magnetic field has also to be taken into account in the transverse dynamics, but bears no effect on the kinetic energy. 

In a static accelerating gap with a flat cathode (Fig.\ref{fig:gungeometry}(a)) electric field lines are normal to the surface and therefore only contribute to the increase of the longitudinal component of the particle momentum. Longitudinal single-particle dynamics is straightforward, and the final beam kinetic energy is simply the integral of the field $E_z$ over the longitudinal position $z$ multiplied by the electron charge. The kinetic energy of electrons accelerated by static fields is limited to $\approx$ 350 keV or lower by electrical breakdown, still in the non-relativistic regime. Here a variation of output energy has a strong effect on the final particle velocity, and hence the time of arrival of the beam at the sample (see Sec.~\ref{section:longitudinalDynamics}). 

The anode aperture is a perturbation from the ideal parallel plate geometry, which bends field lines outwards at the gap exit. The net effect on electron dynamics is transverse defocusing, typically requiring an optical element to re-capture the diverging beam after the gun. The strength of the electrostatic lens scales with the accelerating gradient, but at first order does not degrade the beam quality. As we will see later, in a time-dependent accelerating field the defocusing kick (visible by the magnitude of $E_r$ in Fig.~\ref{fig:gungeometry}(b) and (c) )) will also be time dependent, leading to an increase of the total emittance.

    \subsubsection{Electron acceleration via time-varying fields}
\label{RFdynamics}

Particle dynamics is more complicated in time-varying fields. In this section we describe the electron behavior in RF fields as an example. Most of the treatment can be extended to different frequency ranges. 
The longitudinal and transverse motion of electrons in an RF gun can be treated analytically ~\cite{kjkim89}, by modeling the fields as a standing wave of frequency $\omega$ with a given on axis amplitude profile $E_0e_z(z)$. Approximating $e_z$ as a sinusoidal function with a wave number $k=\omega/c$, the longitudinal acceleration field can be expressed as $E_z(z,t,\phi_0)=E_0\cos kz\sin(\omega t+\phi_0)$. The longitudinal equations of motion can then be rewritten, decomposing the standing wave into forward and backward traveling wave components, as 
\begin{equation}
\label{eqn.longieom} 
\frac{d\gamma}{dz}=\alpha k[\sin\phi+\sin(\phi+2kz)], 
\end{equation}
where $\phi(z,t)=\omega t-kz+\phi_0$ is the so-called synchronous phase. The use of the phase coordinate is particularly convenient for RF linacs, as particles reaching relativistic energies move along constant $\phi$ trajectories. The dimensionless parameter $\alpha=\frac{eE_0}{2m_ec^2 k}$ is a normalized measure of the strength of the accelerating field.  In order to capture electrons from rest $\alpha$ must be larger than 0.5 \cite{rosenzweig:book}, implying that higher frequencies require larger peak fields.

Particles are released from the photocathode with low speed, and quickly fall behind the synchronous phase until they reach relativistic energies. Due to the rapid acceleration of RF guns ($\alpha \approx$1), most of the phase slippage occurs in close vicinity to the cathode, where electrons are much slower than wave phase velocity. The final synchronous phase depends on the launch phase $\phi_0$ for a given gun geometry and operation field strength. This dynamics is an intrinsic feature of particle acceleration with time-varying fields. If $\alpha$ is too small, the electrons do not gain enough energy during the accelerating phase and keep slipping back in phase until they start experiencing a decelerating field, like a surfer with not enough initial speed to catch the incoming wave. The implications of such dynamics on the bunch length and time of flight of electrons will be discussed more in depth in Sec.~\ref{section:timecompression}.  

\begin{figure}[ht]
\includegraphics[width=1\columnwidth]{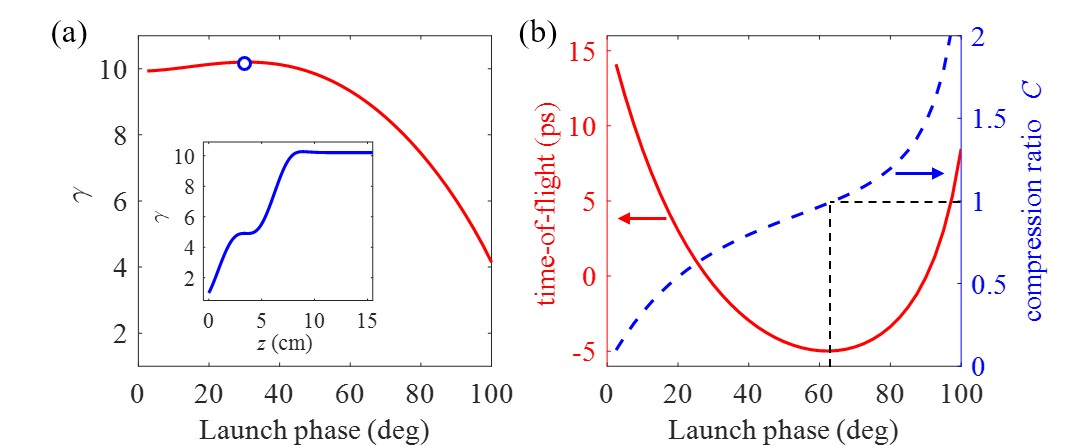}
\caption{For a 1.6 cell S-band RF gun operating at 100 MV/m, (a) the beam energy $\gamma$ at gun exit as a function of the launch phase. (inset) the evolution of $\gamma$ in the gun at the maximum energy launch phase, indicated by the blue circle. (b) Relative time-of-flight from the photocathode to $z=15$ cm, and bunch length compression ratio $C$ as a function of launch phase.}
\label{fig:gundyn}
\end{figure}

The kinetic energy at the gun exit is a function of  $\phi_0$. An example of $\gamma$-$\phi_0$ correlation is shown in Fig.~\ref{fig:gundyn}(a). The example corresponds to the SLAC-UCLA-BNL 1.6 cell S-band RF gun, one of the most widely used sources for relativistic UED applications, operating at a peak field of $E_0$ = 100 MV/m ($e_z$ profile shown in Fig.~\ref{fig:gungeometry}(b)). Such correlation translates to RF-induced energy spread in an electron beam. For a finite laser pulse length illuminating the cathode, electrons emitted at different times will experience a different instantaneous accelerating field amplitude $E_0e_z\sin(\phi_0)$. Neglecting space charge effects (see Sec.~\ref{section:spacecharge}), the launch phase providing maximum energy gain $\phi_0=\phi_m$ is also the phase which minimizes the total energy spread. For the S-band gun example above, $\phi_m \sim$ 30$^\circ$ (blue circle in Fig.~\ref{fig:gundyn}(a)). The accelerating field experienced by the particles at photo-emission in this case is roughly $50\%$ ($\sin30^\circ$) of the peak acceleration field. The inset shows the evolution of the beam energy inside the gun. Higher peak fields or different gun geometries can be exploited to obtain larger values of optimal injection phase, increasing the accelerating field at emission. For example a 1.4 cell S-band RF gun can shift $\phi_m$ to 70$^\circ$ or higher, increasing the field at photo-emission to $\sim$95\% of the maximum, and leading to higher beam brightness~\cite{limusumeci:prapplied}.

In the VHF range (30-300 MHz)~\cite{APEXgun},the phase slippage become negligible and the launch phase is therefore much closer to 90$^\circ$, allowing to take full advantage of the maximum accelerating field.

Transverse RF fields  (see Fig.~\ref{fig:gungeometry})act as time-dependent focusing/defocusing lenses. The variation in focusing strength experienced by different longitudinal beam slices, (i.e. the head, center and tail of the beam), causes the transverse phase space distribution to fan out in correlation with the longitudinal beam coordinate. This increases the area of the beam transverse phase space and induces RF-emittance growth. This RF-induced effect is minimized by choosing the initial launching phase so that the beam exits the gun at the maximum energy \cite{kjkim89}. In this case the contribution to the transverse emittance is $\epsilon_x^{\textrm{rf}}=\frac{1}{4mc^2}eE_0 \sigma^2_x \sigma^2_{\phi}$, where $\sigma_x$ is the rms transverse beam size and $\sigma_{\phi}$ is the rms longitudinal beam size in radians of RF phase. In UED applications where spot sizes are less than 100 $\mu$m and $\sigma_\phi$ is 0.1 degrees or smaller, this effect can be often neglected.

    \subsubsection{The effect of the cathode curvature}
   \label{section:cathodecurvature}

The profile of the photocathode surface has an impact on the output electron beam parameters and dynamics in the gap. In general, the area can either have a flat or curved profile. In the case of the flat profile, the field lines will be normal to the surface and all the acceleration will be in the longitudinal direction with no effects on the transverse plane. 
In the case of a curved surface, three different cases can be distinguished, comparing the radius of curvature $R$ with the laser spot size $r$ used for photo-emission. If the surface radius of curvature is large, the main effect is a distortion of the field in the cathode vicinity, adding transverse components and leading to transverse focusing (concave) or defocusing (convex) effects. The cathode is an equipotential surface, with $\Phi(\mathbf{r})=0$. Expanding the electric potential in $r$ and $z$ to the second order, under the assumption of $R\gg r$ one finds aberration components due to curvature to be proportional to $1/R$~\cite{HAWKES20181}.

As the cathode radius of curvature gets smaller and becomes comparable to the laser spot size, both transverse and longitudinal effects need to be considered. The electric field enhancement along a curved surface discussed earlier, can be used to increase the accelerating field in the cathode area while keeping a large transverse emission size, obtaining at the same time extraction of multi-electron beams and ultrashort pulses from setups with otherwise modest accelerating gradients, typically DC guns \cite{petruk_shaped_2017}. A similar approach can be taken in cathodes for RF guns. In this case the time-varying nature of the field can be used at one's advantage for beam temporal compression. By fabricating curved cathodes, a radial-temporal correlation is established by means of two related effects: the delay of the outer region of the laser pulse in reaching the cathode surface with respect to the central area, and the different accelerating field amplitude experienced by the particle at birth. Such concave shape can be optimized to pre-compensate for the non-isochronicity of the following focusing elements, leading to shorter final electron pulse \cite{de_loos_radial_2006}.

For tip-like cathodes, the radius of curvature is orders of magnitude smaller than the illuminating laser, and in most of cases even smaller than the laser wavelength. The main advantage of a tip is that the source size is now determined by the physical extension of the tip and not by the laser spot size.
Accelerating field at the tip apex can be enhanced by factors exceeding 100, with a longitudinal extensions comparable to the tip radius. While this may locally increase the maximum brightness achievable, it also increases field non linearities and, in order to obtain a high brightness beam, heavy collimation is needed downstream the accelerating gap, selecting only electrons emitted from the tip apex (see Sec.~\ref{sec:beamsize}).

\subsubsection{Temporal beam evolution in simple systems: Vacuum dispersion}
\label{section:longitudinalDynamics}

In this section we will review the role of key instrument parameters on the beam longitudinal dynamics in absence of space charge. The evolution of the beam center of mass is unaffected by self-fields, and we are therefore able to provide approximate analytical equations that can be used to accurately predict the final energy and arrival time of the electron(s) at the sample. At the same time, for an accurate prediction of the final pulse length at the specimen both longitudinal emittance and the eventual contribution of space-charge forces to the dynamics need to be accounted for. This requires solving self-consistently the Maxwell equations coupled to the equations of motion for the beam, and it is generally achieved through the use of sophisticated simulation codes (see for example Fig.~\ref{fig:velocitybunching}). 

In the simplest setup, which includes a static accelerating field within a gap and a downstream drift to the sample, accelerating field fluctuations and beam energy spread at emission (the electron excess energy) contribute to shot-to-shot energy and time-of-flight variations.  Variation in electron energy translates in time of flight fluctuations through vacuum dispersion. To quantify the impact of such effect on the instrument performance we first consider only the accelerating gap starting from the photocathode surface, and then we include the transport from the output of the gun to the sample. 

\begin{figure}[ht]
\includegraphics[width=1\columnwidth]{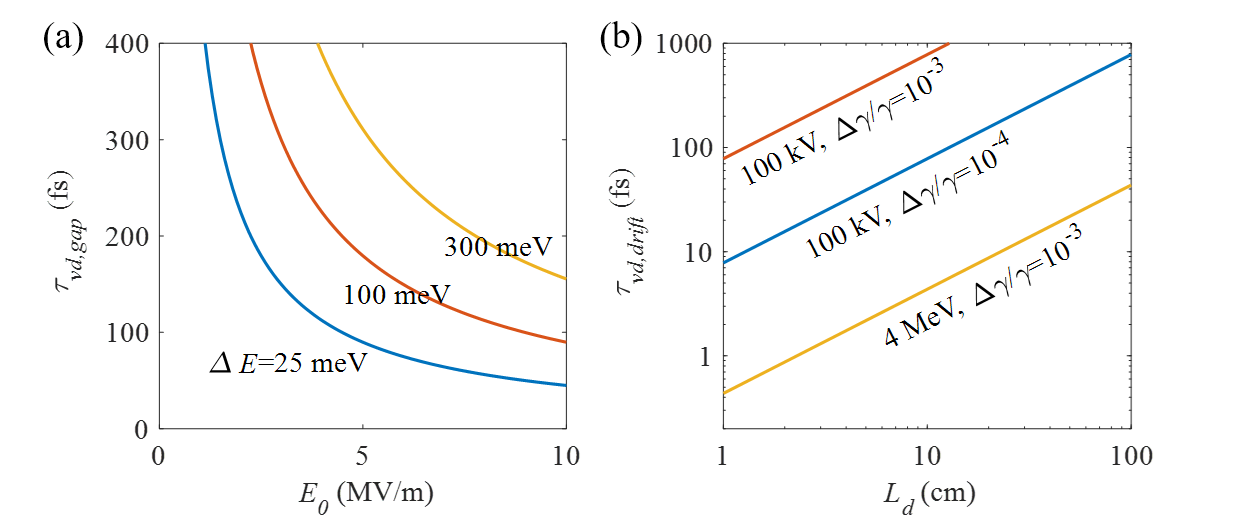}
\caption{(a) Dependence of vacuum dispersion broadening $\tau_{vd,gap}$ on the acceleration field $E_0$ inside an DC accelerating gap for several different initial energy bandwidth of photoelectrons $\Delta 
E$. (b) Broadening of the bunch length in a drift transport channel, for different beam energy and energy spread values.}
\label{fig:vd}
\end{figure}

We define $\tau_{vd,gap}$ as the temporal distance at the output of the accelerating gap between two electrons photo-emitted from the cathode at the same time with respect with the laser arrival time. Depending on the operation mode, $\tau_{vd,gap}$ represents the shot-to-shot TOF fluctuations (single-electron mode) or the final temporal spread of the beam (assuming negligible space charge effects).
The value of $\tau_{vd,gap}$ at the output of the static gap is mainly affected by the energy bandwidth of the photo-electrons at emission $\Delta E$, which depends on the detail of the photocathode material and driving laser.  Fluctuations in the amplitude of the accelerating fields are generally below $1e-4$ for state-of-the-art high-voltage power supplies, and can be ignored inside the short accelerating gap, while they will need to be included when calculating the TOF to the sample. For an accelerating electric field amplitude $E_0$ we find $\tau_{vd,gap}=(\sqrt{2}m\Delta E)^{1/2}/eE_0$ ~\cite{baum_2010_singleelectron}. Figure~\ref{fig:vd}(a) shows the dependence of $\tau_{vd}$ on $E_0$ for several different values of $\Delta E$, providing a lower limit for the bunch length achievable~\cite{li2017femtosecond, duncan_lossless_20}. The equation for $\tau_{vd,gap}$ reported above uses a simple non relativistic model which is approximately valid also for higher energies. Indeed, vacuum dispersion-induced broadening is quickly suppressed through rapid acceleration and energy gain, so the main contribution to $\tau_{vd}$ is at low energies. The final bunch length at the gun exit is the convolution between $\tau_{vd}$ and the initial electron pulse length just outside the cathode. Lastly, it is worth noting the inverse linear scaling between of $\tau_{vd}$ and $E_0$, which highlights the importance of high accelerating fields.

Vacuum dispersion in the gap defines the minimum pulse length achievable in a UED setup in absence of temporal compression, with the sample ideally placed right at the output of the accelerating region.

In a drift transport channel, the TOF and temporal broadening are fully determined by the particles kinetic energy and energy spread. We can express this dependence as:  
\begin{equation}
\tau_{vd,drift}=\frac{\Delta\gamma}{\gamma}\frac{R^{drift}_{56}}{\beta c}=\frac{L_d}{c}\frac{\Delta\gamma}{\beta^3\gamma^3}
\label{eq:tof_drift}
\end{equation}
where $L_d$ is the drift distance, and $R^{drift}_{56}=\frac{L_{d}}{\beta^2\gamma^2}$ is the longitudinal dispersion function in a drift section~\cite{england_sextupole_2005}. Figure~\ref{fig:vd}(b) shows the value of $\tau_{vd,drift}$ for different kinetic energies. For a given a target pulse length at the sample, higher beam energies allow for larger energy spreads. In simple UED setups (i.e. with no compression), particles at the beam head remain at the head during propagation and in order to calculate the total contribution to TOF fluctuations or, in absence of space charge, the bunch temporal spread, it is possible to simply sum up the dispersion terms in the gun and in the following drift as $\tau_{\Delta pp}=\tau_{vd,gap}+\tau_{vd,drift}$.

    \subsubsection{Space charge effects}
\label{section:spacecharge}

Space charge forces, i.e. interaction between electrons, play a significant role in the dynamics of high brightness, ultrashort electron beams. In particular, space charge forces act as defocusing forces in transverse and longitudinal planes, limiting the beam charge density in real space, correlating pulse length and beam charge, and potentially degrading beam brightness and the spatio-temporal resolution in UED experiments.

It is instructive to first look at the scaling of the Coulomb interaction between two electrons to understand how the space charge forces scale with their kinetic energy. Consider the case of two particles moving with parallel and constant velocity $v=\beta c$, and with longitudinal and transverse separation $s$ and $x$, respectively. The total space charge force (electric and magnetic fields) experienced by the trailing particles is \cite{zangwill}: 
\begin{eqnarray}
F_l & = & -\frac{1}{4\pi\epsilon_0}\frac{e^2s}{\gamma^2(s^2+x^2/\gamma^2)^{3/2}} \\
F_t & = & \frac{1}{4\pi\epsilon_0}\frac{e^2x}{\gamma^4(s^2+x^2/\gamma^2)^{3/2}}
\label{eqn:scforces} 
\end{eqnarray}
If the electrons are purely transversely separated, then $s=0$ and $F_l$ vanishes, while $F_t\propto\frac{ 1}{\gamma x^{2}}$. Another way to understand this scaling is by performing the Lorentz transformation between the lab frame $K$ and the particles rest frame $K'$ moving at $v$ with respect to $K$ in the longitudinal direction. As the two particles move, the electric field experienced by one of the particle, as seen in the laboratory frame is $E_t=\gamma E'_t=\frac{1}{4\pi\epsilon_0}\frac{e \gamma}{x^2}$.
The increase of $E_t$ with $\gamma$ is a result of the growing anisotropy of electric field lines with increasing particle speed, which concentrate to within a transverse cone of opening angle on the order of $1/\gamma$. Calculating the Lorentz force on the electron we then find that $F_t=eE_t/\gamma^2$, where the factor $\gamma^{-2}$ accounts for the opposite signs of electric and magnetic force components, retrieving the initial scaling $F_t\propto\frac{1}{\gamma}$.

A similar reasoning can be carried out for the longitudinal component of the force. From Eq.\ref{eqn:scforces} we find that, for $x=0$ or $s\ll x/\gamma$,  $F_l\propto\frac{1}{\gamma^{2}s^{2}}$. Note that $s$ is proportional to $\beta$ for a fixed temporal separation, and hence $F_l\propto\frac{1}{\gamma^{2}\beta^{2}}$. The acceleration of the electron is $a_l=\frac{F_l}{\gamma^3m}$, where the $\gamma^3$ dependence accounts for the increasing difficulty in changing the speed of the electrons when approaching the speed of light, and $m$ is the electron's rest mass. The space charge-driven particle separation $l$ at a downstream position $L$ is given by $l=\frac{1}{2}a_lt^2$, where $t=\frac{L}{\beta c}$ is the average time of flight. Putting all together, $l\propto\frac{L^2}{\gamma^5\beta^4}$, highlighting the benefit of increasing the beam kinetic energy to counteract space charge effects.

For many-electron beams, the space charge force acted upon each electron is generally calculated by integrating over a smooth charge density distribution, rather than summing up the pair-wise Coulomb forces between the target electron and each and every other electron. The smooth field approach is valid when the field from each particle is screened by surrounding electrons within a distance equal to the Debye length $\lambda_D= \left( \frac{\epsilon_0 \gamma k_B T_b }{n e^2)} \right)^{1/2}$, where $n$ is the electron number density and $T_b$ is the effective temperature in the beam rest frame \cite{reiser:book}. Since the Debye length for UED beams is usually much larger than the average spacing between electrons (i.e. $n^{-1/3}$), the large number of particles inside a Debye sphere has the effect of smoothing out the space charge field. In this case a collective description of the beam distribution is more useful, and the particular shape of the distribution plays an important role in the space charge model. Nevertheless, a first order description of the dynamics can be obtained using the envelope equations, i.e. the equations that determine the evolution of the second order moments of the beam distribution. 

A first example of envelope equations is found in one of the first quantitative studies of non-relativistic space-charge driven bunch lengthening using simple analytical models~\cite{Siwick:2002hiba, Reed:2006jbbaca, Ischenko_sc_2019}. In this case the radial beam envelope was assumed constant and there is only one equation to be solved for the longitudinal beam size. For a non-relativistic pancake-shaped electron bunch with the radius $r$ much larger than the total length $l$, the evolution of the bunch length can be written as
\begin{equation}
\frac{d^2l}{dt^2}=\frac{Ne^2}{m\epsilon_0\pi r^2}\left(1-\frac{l}{\sqrt{l^2+4r^2}}\right),
\label{eqn:meanfieldmodel} 
\end{equation}

If initially $l$ is much smaller than $r$, the first term on the right-hand-side of Eq.~\ref{eqn:meanfieldmodel} dominates the beam expansion. As the bunch becomes longer, $d^2l/dt^2\rightarrow0$, which implies that the lengthening rate $dl/dt$ reaches a constant value after the potential energy of the electron bunch is converted to kinetic energy. This quantity represents the velocity spread of the bunch, since electrons at the head and tail of the bunch are driven by the space charge forces towards opposite directions. Although it is based on a simple model, Eq.~\ref{eqn:meanfieldmodel} gives results in good agreement with particle-tracking simulation tools, which are capable of more accurately dealing with realistic beam profiles. 

The Eq. \ref{eqn:envelope1} and \ref{eqn:envelope2} are the non-relativistic simplified case of coupled envelope equations \cite{reiser:book}, which for a constant beam energy can be written as:
\begin{eqnarray}
\label{eqn:envelope1} 
    \sigma_z''+k_{0z}\sigma_z-\frac{K_l}{\sigma_z^2}-\frac{\epsilon_z^2}{\sigma_z^3}=0 \\
    \sigma_r''+k_{0r}\sigma_r-\frac{K}{\sigma_r}-\frac{\epsilon_r^2}{\sigma_r^3}=0
\label{eqn:envelope2} 
\end{eqnarray}
where the evolution is followed along the longitudinal coordinate $s$, and $k_{0z,r}$ represent the transverse and longitudinal focusing (various techniques to implement longitudinal focusing are discussed in the next sections) and the last terms can be interpreted as pressure forces preventing the beam sizes to become infinitely small for finite beam emittances. These equations are coupled by the perveance terms $K,K_l$ which represent the smooth space charge fields contributions to the envelope evolution, and naturally depend on the beam aspect ratio. In the limit of very low charge beams, these terms can be neglected. 

The transverse perveance is $K \approx \frac{I}{I_A \beta \gamma}$, where $I$ is the beam current and $I_A=17.04$ kA is the Alfven current. The energy dependence of $K$ shows the $\frac{1}{\gamma}$ scaling discussed above. For an infinitely long beam of current, only the second equation is relevant. 
For bunched beams, $K_l = \frac{Qr_c g}{\beta^2 \gamma^5 }$, where  $Q$ is the bunch charge and $r_c$ is the classical electron radius. When the bunch is long $g \rightarrow 1$ and this is essentially the relativistic generalization of Eq. \ref{eqn:meanfieldmodel} in the limit $l \rightarrow \infty$. For shorter bunches, $g$ is a more complicate function of the aspect ratio of the beam in its own rest-frame. 

Regardless of the specific functional form of $g$, the strong $\gamma$ dependence of $K_l$ illustrates the scaling of the space charge-induced bunch lengthening with energy.
Larger $\gamma$ values allow for higher charge density and bunch charge for single-shot experiments, and help maintaining ultrashort bunch lengths over a longer distances $L$, to accommodate for sophisticated sample delivery systems and other complex setups including front sample illumination, gas phase and liquid phase samples, etc.

\begin{figure}[ht]
\includegraphics[width=1\columnwidth]{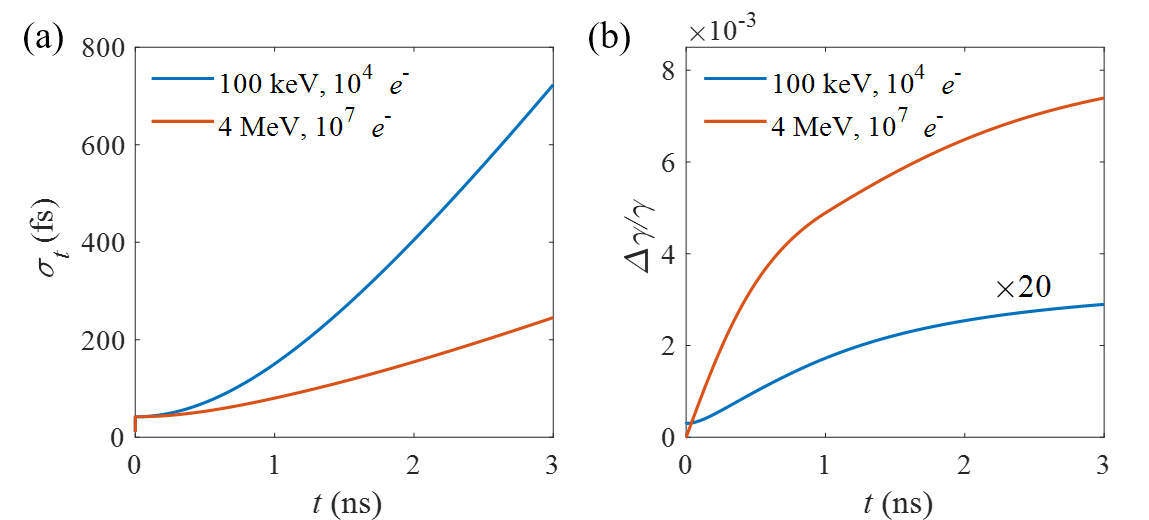}
\caption{Comparison of space-charge driven evolution of the (a) bunch length and (b) energy spread of a 100 keV and 4 MeV electron beams. The two beams start with otherwise identical initial conditions, including 100 $\mu$m radius, 10 nm-rad normalized emittance, and 50 fs rms bunch length.}
\label{fig:pulsebroadening}
\end{figure}

To quantitatively evaluate space charge effects on the beam evolution, particle tracking codes are heavily employed. Figure~\ref{fig:pulsebroadening} shows bunch lengthening and energy spread evolution for a 100 keV electron bunch containing $10^4$ electrons. In comparison, for a 4 MeV electron bunch even with 1000 times higher bunch charge the space-charge driven broadening is much less evident. It is interesting to notice that, however, space charge induced increase of energy spread is lower in the case of the 100 keV beam, due to the rapid decrease in beam charge density. Besides the different energy, the two beams start with identical initial conditions, 100 $\mu$m radius, 10 nm-rad normalized emittance, and 50 fs rms bunch length. The initial energy spread is at a level that the vacuum dispersion has negligible effects on the final bunch length, and the final energy spread is dominated by the space charge forces rather than initial conditions in both cases.  
 
Transverse space charge forces act as defocusing forces to electrons which, to the first order, can be counter-balanced by external focusing optics. If non-linear space charge forces are present, however, they can lead to distortion and even filamentation of beam transverse phase space, which leads to an increase of the rms emittance. shaping of the electron bunch can be used to control the charge density distribution and mitigate this issue. In particular, uniformly filled ellipsoidal distribution have linear self-fields in all three dimensions and can be used to preserve the brightness from the photocathode to the sample. 
Various beam shaping techniques, mostly by tailoring the spatial, temporal and spectral profile of cathode driving laser pulses, have been proposed and experimentally explored. 

One appealing approach, inspired by the similarity with the gravitational potential fields of galaxies \cite{chandrasekhar:ellipsoidal}, is to take advantage of the self-expansion (blow-out regime) of an ultrashort, transversely spherical electron beam. The main advantage for UED experiments of this beam regime which has been simulated and experimentally verified ~\cite{luiten2004,musumeci_blowout_2008}  is the possibility of using downstream temporal compression (see Sec.~\ref{section:timecompression}), to obtain ultrashort pulses, limited only by the initial longitudinal emittance. Nevertheless, tight constraint associated with the transverse beam size at the cathode and the image charge distortions, limit the initial 4D brightness and decrease the obtainable transverse coherence length.  

Alternatively, uniform ellipsoid beams can be formed by illuminating the photocathode with a very small transverse size and longitudinally parabolic laser pulse, and then the electron beam will expand transversely under its self-field~\cite{Claessens05, li_nanometer_2012}. This regime is particularly relevant for UED as they are characterized by tiny emission areas and ultralow emittances and bunch compression can be used to shorten the relatively long initial bunch length. 

\subsubsection{Temporal compression}
\label{section:timecompression}
    
One can take advantage of the time-dependent nature of oscillatory RF accelerating fields to manipulate the longitudinal phase space (LPS) of ultrafast electron beams. Typically the RF fields (depending on the injection phase) induce an energy modulation or chirp in the beam that through the effect of dispersion results in a temporal compression (or stretching) after some propagation distance. 

This dynamics occurs first in RF guns in the vicinity of the cathode, where electron energy is low. In this region the acceleration depends linearly on the experienced field amplitude, determining the rate of change of the electron velocity, its final energy and the TOF of electrons through the gap. As the electrons become relativistic and their velocities approach the speed of light, the dispersion in straight region of beamlines becomes negligible.

In order to quantify this effect we can consider two electrons injected into one cavity at different times $\Delta t_{inj}$, and define a compression factor $C$ of the cavity as the ratio between $\Delta t_{inj}$ and the difference in time of arrival at the cavity output $\Delta TOA$~\cite{Filippetto2016}:

\begin{equation}
\label{eq:compressionFactor}
C(\Phi_0)=\frac{\Delta TOA}{\Delta t_{inj}}=\frac{\Delta t_{inj} + \Delta eTOF}{\Delta t_{inj}}=1+ \frac{\Delta eTOF}{\Delta t_{inj}}
\end{equation}

Figure \ref{fig:gundyn}(b) reports an example of the simulated TOF from the cathode to the exit of an S-band RF gun ($z=15~\textrm{cm}$) as function of launch phase (red curve). By selecting the launch phase appropriately, this correlation can be exploited for temporal manipulation, as shown by the blue dashed curve, resulting in a  compression of the temporal distance between the two input electrons (C$<$1) ~\cite{Wang:1996up, li_temporal_2009}. It is important to note here that a larger sensitivity of the beam TOF to the injection phase $\phi_0$ poses stringent requirements on the laser-to-RF phase locking stability (see Sec.~\ref{section:clocking}). Such sensitivity is minimized for values of $C$ close to 1 ($\phi=62^\circ$ in Fig.~\ref{fig:gundyn}(b)), which is naturally also the point in which the correlation between phase and TOF vanishes.

When a bunching cavity is present (see e.g. the scheme in Figure~\ref{fig:setup} ), the simultaneous presence of two RF cavities, electron gun and bunching cavity, complicates the analytical derivation, introducing correlations between otherwise uncorrelated variables, such as the amplitude of the first cavity and the injection phase into the second one. Indeed, amplitude and phase fluctuations of the gun fields modulate the output energy, which causes TOF fluctuations in the subsequent drifts following Eq.\ref{eq:tof_drift}, resulting in fluctuation of injection phase into the buncher. For a detailed derivation of the general beamline see \citet{Filippetto2016}.

\begin{figure}[ht]
\includegraphics[width=1\columnwidth]{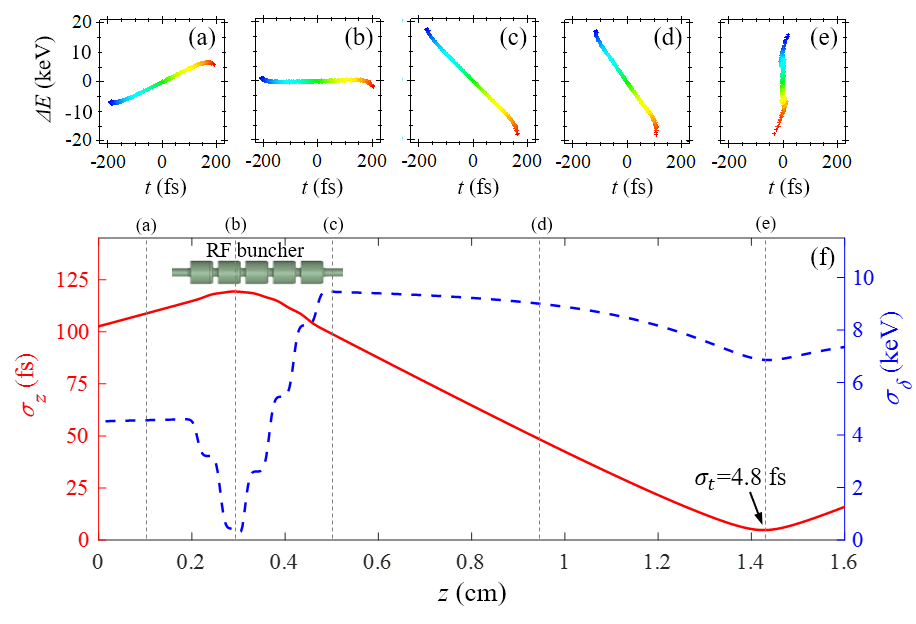}
\caption{(a-e) LPS distributions at various locations during the temporal compression process and (f) evolution of the bunch length and energy spread. The electron bunch is positively chirped (a) before the RF buncher due to space charge forces. The chirp is then minimized (b) and reversed (c) by the RF buncher. In the drift space after the buncher the electron beams undergo shear motion (d) in LPS torwards vertical orientation and reaches minimal bunch length (e).}
\label{fig:velocitybunching}
\end{figure}

In order to understand the dynamics in bunching cavities, we consider a beam of particles traveling in vacuum with a certain average energy $\gamma mc^2$ and spread in time $dt$. To achieve temporal compression one first need to obtain the right correlation coefficient in the $\gamma$-$t$ LPS. When used at the so-called zero-crossing phase, the bunching cavity provides zero net acceleration, but imparts an energy chirp $h=\frac{d\gamma}{dt}$ on the beam, with a negative slope in the $\gamma$-$t$ distribution of magnitude $h=e\omega_0V_0/mc^2$, where $\omega_0$ and $V_0$ are the angular frequency and total integrated voltage of the structure, respectively. 
As the beam travels through the downstream transport line, the chirp leads to temporal compression via the longitudinal dispersion $R^{drift}_{56}/\beta c$. In the LPS, this process can be seen as a shear motion of the $\gamma$-$t$ distribution, i.e. electron trajectories in the plane move horizontally (maintaining a constant $\gamma$) until the projection of the distribution $t$ is minimized and the beam reaches the shortest bunch length as depicted in Fig.~\ref{fig:velocitybunching}. In the case of a straight drift channel (and neglecting space-charge defocusing forces), the chirped beam reaches the longitudinal focus when $hR_{56}\beta c=-1$, after a distance $L_f$ equal to $L_f=m(\beta c)^3 \gamma^2/e\omega_0V_0$. The time-dependent electric fields used at the scope have frequencies spanning from the RF to the THz range, and amplitudes capable of generating 1/few keV/ps correlations or larger, required to efficiently compress electron beams with kinetic energies above 100 keV. The wavelength of the field should be chosen much longer than the electron beam duration, in order to produce a (quasi-)linear energy-time correlation. 

It is worth pointing out a distinction between the minimization of $\Delta eTOF$, and electron beam compression. While Eq.\ref{eq:compressionFactor} provides a direct link between $C$ and $\Delta eTOF$, the former parameter can only be quantitatively associated with actual electron beam temporal compression in the case of negligible longitudinal space charge effects, as in single-electron mode operations. When dealing with a beam of multiple electrons, space charge fields will increase during compression and may eventually become important. To obtain the shortest pulse length at the sample, the bunching cavity field will then need to be set to higher values, in order to pre-compensate for the downstream space-charge de-bunching. This will lead to negative values of C, possibly even smaller than -1, with a consequent amplification of the input temporal jitter. For such reason temporal compression needs to be designed carefully. Despite providing shorter electron pulses at the target, it may be detrimental to the overall temporal resolution.

The use of RF fields for energy modulation and temporal compression described above has long been used in vacuum electronic devices as well as in electron photoinjectors driven by DC or RF guns. In UED, such technique was first introduced for 100 keV electron beams \cite{Oudheusden:2007foba}, demonstrating 100 fs short beams with up to $10^6$ electrons, via the use of a single-cell 3-GHz cavity with sub-kW RF power~\cite{Oudheusden:2010cfbaca}. For MeV electron beams the required buncher voltage is much larger due to the unfavorable scaling of the vacuum dispersion with beam energy (Eq.~\ref{eq:tof_drift}). Nevertheless MeV electron beams have been successfully compressed to below 10 fs rms~\cite{Li:2011ivba, maxson_ultrafast_2017}.

THz radiation can be very efficient in compressing electron beams, due to the 2-3 orders of magnitude larger $\omega_0$ compared to RF fields. Recent experiments have shown laser-generated THz radiation combined with interaction structures for coupling and enhancement can effectively compress keV-scale beams to bunch lengths below 100 fs~\cite{kealhofer_all-optical_2016, zhang_steam_2018}, reaching below 30 fs with MeV-scale beams~\cite{snively_femtosecond_2020, zhao_femtosecond_2020}. Further developments along this line of research are rapidly advancing, including e.g. increasing the electron beam-THz interaction length and improving the symmetry of THz structures and fields to optimize the electron beams qualities. THz compression simplifies the apparatus by removing the RF power source and RF-to-laser synchronization system. With laser-generated THz radiation, which is intrinsically synchronized with the pump laser, the time-of-flight of compressed electron beams may actually be stabilized, improving the temporal resolution.     

Many factors contribute to the the minimum bunch length achievable. Due to the non-linear relationship between $\gamma$ and $\beta$, the LPS will develop nonlinear correlations even for an ideal linear chirp~\cite{Zeitler:2015fsba}. Also, depending on the ratio between input beam duration and bunching field oscillating period, the induced energy chirp would include some amount of nonlinear $\gamma$-$t$ correlations (i.e. 3rd order of RF fields due to the sinusoidal potential). For THz fields the full period of the wave is comparable with the electron beam pulse length (ps-scale), and the particular temporal profile depends on the spectral content, but usually contains higher degrees of nonlinear $\gamma$-$t$ correlations. 

Another limitation is represented by space charge effects. The charge density increases during transport and compression, and the space charge field may develop nonlinear components associated with the particular charge density profile, including curvatures in the beam core and tails at the beam edges (see e.g. the tails of the distribution in Fig.~\ref{fig:velocitybunching}d)). The curvature of the RF field could be exploited in this case to equalize the space charge driven non-linearities and achieve shorter bunch lengths~\cite{Zeitler:2015fsba}. Precise control of LPS and electron beam compression beyond the femtosecond-scale is an active research topics in beam physics that will directly benefit UED applications. 

The total longitudinal dispersion $R_{56}$ along a beamline can be tuned using specifically designed electron optics to provide compression without the need of an active cavity relying on the energy chirp induced by the space charge forces. For example, while in a drift high energy particles arrive earlier (positive dispersion), in a dipole magnet higher energy particles will arrive later than low-energy particles at the magnet output (negative dispersion). Therefore it is possible to design beamlines where a combination of magnets and drift sections lead to the isochronous condition ($R_{56}=0$) i.e. the particle TOF is independent from its energy, or to bunch compression with a non-zero $R_{56}$ and properly tuned beam chirp~\cite{Smirnov_subTHz_2015, Mankos2017}.

\begin{figure}[ht]
\includegraphics[width=1\columnwidth]{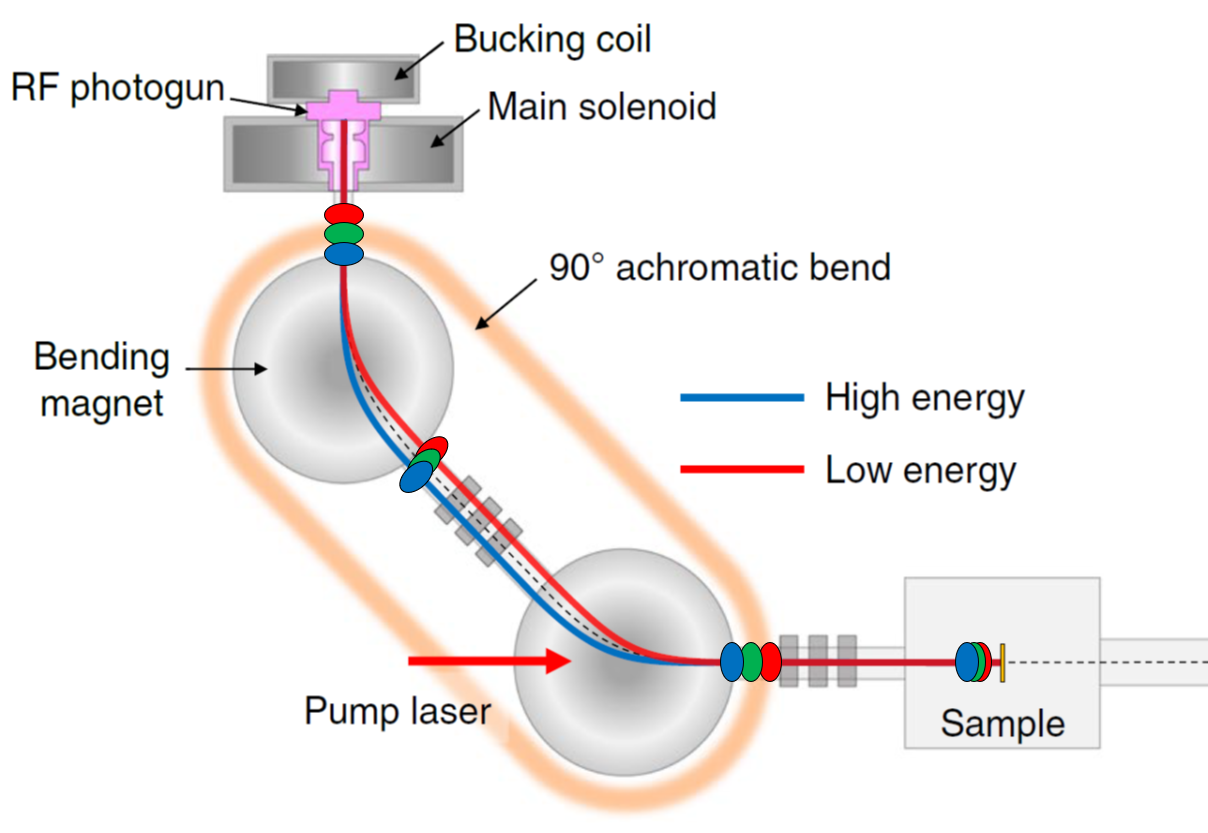}
\caption{Schematic of a DBA transport line following an RF gun for bunch length and time-of-arrival manipulation. Adapted from Ref.~\cite{kim:isochronous}}
\label{fig:dba}
\end{figure}

Symmetric and asymmetric double bend achromatic (DBA) transport lines with tunable $R_{56}$ have recently been demonstrated in MeV UED setups to improve the temporal resolution~\cite{kim:isochronous, Qi2020}. A DBA layout is shown in Fig.~\ref{fig:dba}. The positive chirp at the entrance of the first bending magnet is mostly induced by space charge effects. The DBA transport line is configured to a proper $R_{56}$, to compress the beam to its minimum at the sample location. The DBA approach is usually referred to as a 'passive' scheme, given the absence of active RF or THz bunching structures, which eliminates instability sources including RF amplitude and phase fluctuations and THz amplitude and waveform fluctuations. On the other hand, the performance of passive schemes is subject to the fluctuations of the initial energy chirp and higher-order effects in the transport line.   

Finally, manipulation of electron beams using optical laser is an attractive technique for generating sub-fs temporal structures in the beam. Ultrashort laser gating have generated isolated 30-fs temporal structure in the electron-energy spectrum from a 500 fs long electron pulse through photon-electron-nanostructure interactions~\cite{Hassan17}. 

Taking advantage of the significantly shorter wavelength of optical lasers compared to RF or THz radiation, laser fields can create extremely fine structures in the phase space of electron beams, generating  trains of attosecond-long pulses ~\cite{kozak_atto_2018,Ropers_Ramsey_2016, kazak_atto_2017, Ropers_atto_2017, Morimoto_atto_2018}. Such an attosecond bunch train provides a powerful tool for studying cycle-reversible structure dynamics under optical excitations.

One of the most exciting research frontiers on electron beam manipulation is to further pushing the limit in time towards generation of isolated attosecond electron pulses \cite{Morimoto_atto_2018, Priebe17_attosecond, Vanacore18_attosecond, Yalunin21_attosecond}.

 \subsubsection{Evolution of the beam energy spread}
\label{sectionII.c.5}

The energy spread of an electron beam contributes to the blurring of the diffraction pattern as it effectively induces a spread in the electron wavelength and therefore of the diffraction features. Mathematically one has $\Delta E / E = \Delta \lambda / \lambda$. Common energy spread values, on the order of $10^{-3}$ or smaller, have negligible contribution when compared with emittance-induced blurring (one order of magnitude lower or more). Nevertheless, there are cases where such effect becomes important, for example in laser-plasma based electron sources \cite{faure:UED}), where energy spread values can be in excess of 1\%.

Contributions to the beam energy spread include the excess energy in the photoemission process, the variation of the accelerating field instantaneous amplitude over the beam duration, and the work done by space charge forces. Values of the energy spread at emission depend on the particular setup and illumination characteristics,  and range from a few meV to a few eV (see Sec.~\ref{sectionII.b}).Time dependent fields used for compression cause correlated increase of beam energy spread, as already explained above. Furthermore,transverse variations of the accelerating electric field within the beam result in additional energy spread. The characteristic spatial scale over which the longitudinal field changes is related to its wavelength \cite{deloos:radialcompression}, and $\Delta E/E \propto \sigma_r^2 / 2 \lambda_{RF}$. Note that this energy spread can in principle be compensated by removing the transverse-longitudinal phase space correlations with a proper beam transport \cite{duncan_lossless_20}. Lastly space charge forces also contribute to additional correlated (chirp) and uncorrelated (Boersch effect \cite{hansen:boersch}) energy spread.

\begin{figure}[ht]
\includegraphics[width=0.7\columnwidth]{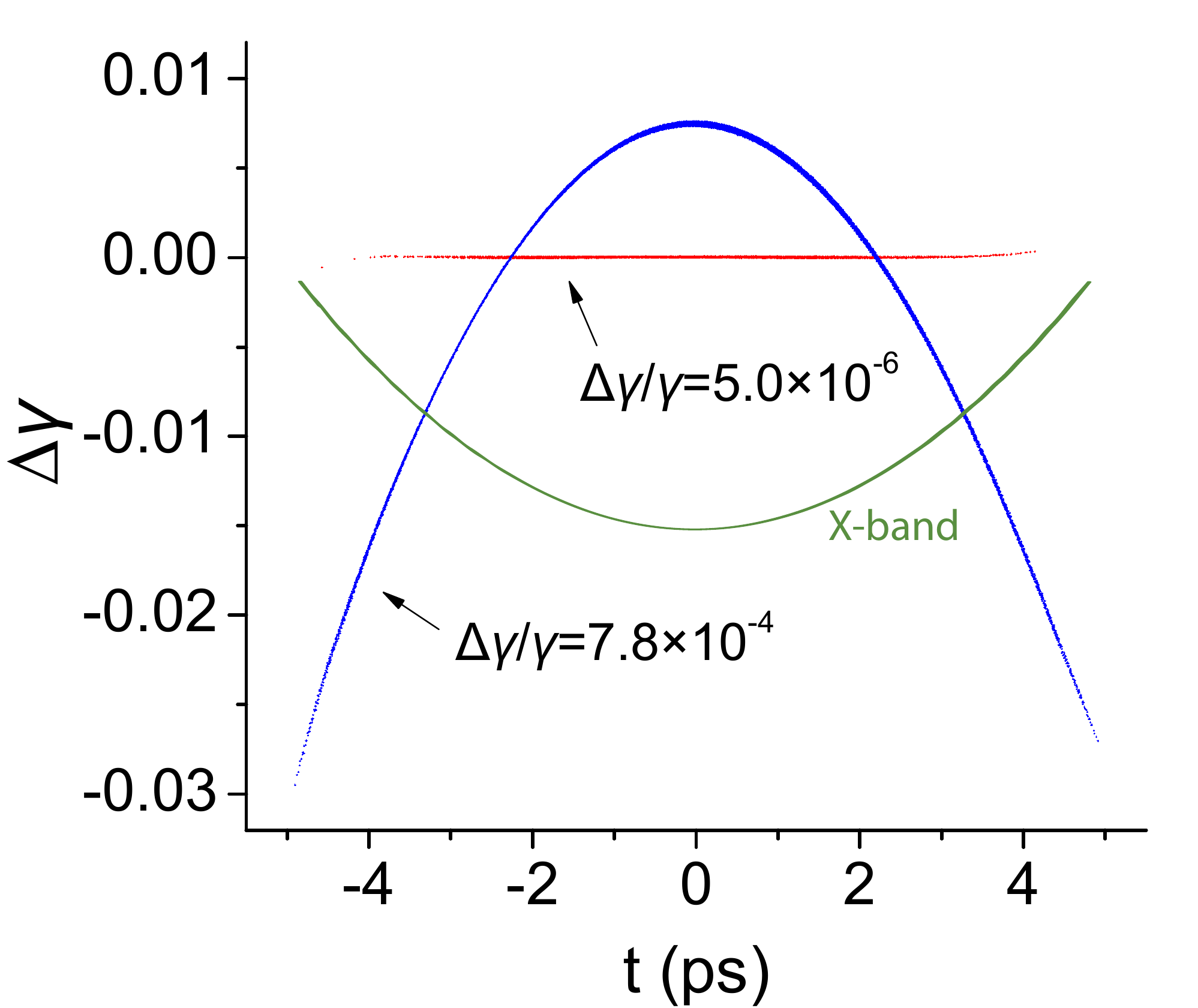}
\caption{Schematic cartoon of RF linearization process. Non-linear correlations in the beam longitudinal phase space are removed using a higher harmonic X-band cavity.}
\label{Fig:RF_linearization}
\end{figure}

Linearization cavities can be used in the process of minimizing linear and non linear correlations in the longitudinal phase space \cite{limusumeci:prapplied}, decreasing the overall energy spread and enabling much shorter bunch lengths (see Fig.~\ref{Fig:RF_linearization}).

The energy spread can also be filtered out  by collimation in a dispersive section \cite{Filippetto2016} or in non-relativistic beamline using Wien filters \cite{wienfilter}. This processes remove charge and do not change the beam peak brightness but could be useful if a truly monochromatic illumination of the sample is desired. In transmission electron microscopy this is common as an important effect of the energy spread is related to the chromatic aberrations in the lenses in the column thus affect the final spot size and spatial resolution.

Shot-to-shot energy fluctuations also appear as source of energy spread in experiments requiring accumulation. In DC-based electron guns the stability of the high voltage power supply is typically at the 1 ppm level. High power RF amplification needed for relativistic beam acceleration has energy stability on the order of 100 ppm at the state-of-the-art.

\subsection{Technologies for electron acceleration}
\label{section:acc_tech}

\subsubsection{DC sources}
The first UED apparatus was realized in 1982 using a modified streak camera with DC accelerating fields~\cite{Mourou1982}. The importance of high accelerating fields was well recognized early on and continuous efforts have been directed over the years towards optimizing the design and surface processing aimed at reaching higher breakdown thresholds. Today, acceleration via DC fields provides highly reliable sources for diffraction and microscopy in a compact setups with typical energies up to 200-300 keV.

\begin{figure}[ht]
\includegraphics[width=1\columnwidth]{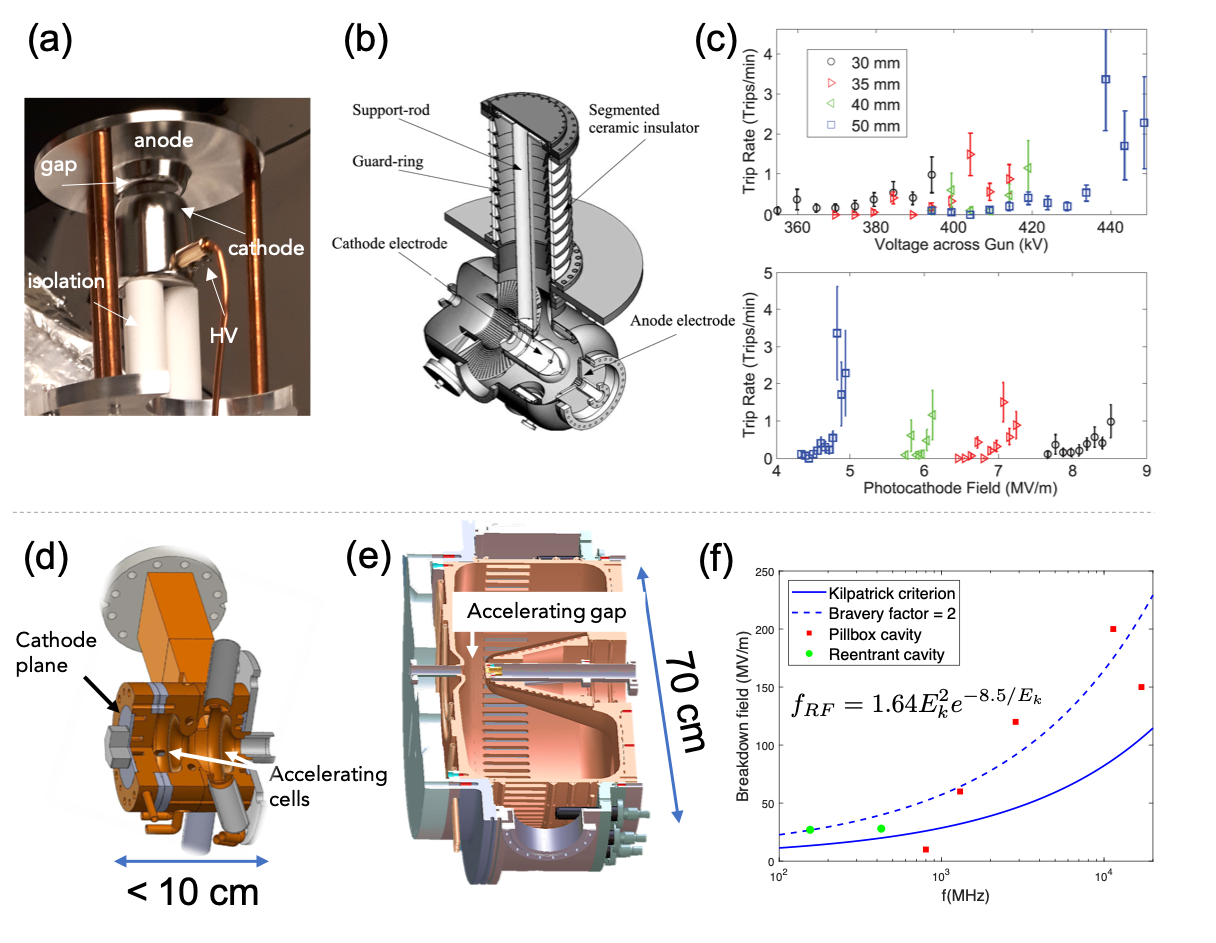}
\caption{(a) Example of a 30 kV electron gun. The accelerating gap is 3mm. (b) Detailed schematics of a 500~kV DC electron gun, from~\citet{nagai_high-voltage_2010}; (c) Maximum accelerating field and output energy for different gap sizes from an electron gun with variable gap, from~\citet{maxson_design_2014};(d)the SLAC/UCLA/BNL high gradient pulsed 1.6 cell S-band RF gun;(e) the APEX gun at LBNL, an example of CW, normal conducting RF electron gun; (f) the frequency dependence of the breakdown field (adapted from \citet{dowell_rao_book}).
}
\label{fig:Eguns}
\end{figure}

The complexity and cost of this technology (in particular for power supplies and the insulating stages) increase very fast with the applied gap voltage (see Fig.\ref{fig:Eguns}(a),(b)) ~\cite{osaka_3MV_1997, japan_1p2MV}, and the design of a DC gun producing beam energies larger than 350 keV becomes a dedicated research effort.  
Indeed, as shown from the experimental data reported in Fig.\ref{fig:Eguns}(c), the accelerating field and the output beam energy can not be simultaneously optimized. Fields up to 10 MV/m can be routinely achieved with short ($<$ 1~cm) gaps~\cite{maxson_thesis_2015, dowell_rao_book, bazarov_comparison_2011} producing electron energies up to 100 keV, but maintaining similar field level for larger cathode-anode distances has turned out particularly hard to achieve. 
The weakest point of the high-voltage system is the insulator separating the two electrodes. For large voltage across the electrodes, the insulator size must be increased to keep the electric field in the ceramic under control. The larger surface also increases the probability of field emitted electrons (for example from the support rod itself) to strike it, causing heating and local increase of voltage which may lead to punctures and damage. Mitigation strategies include the use of segmented ceramic parts alternate to metallic shields, as shown by the structure in Fig.~\ref{fig:Eguns}(b)), which allowed to reach 500~kV voltage across the gap~\cite{nagai_high-voltage_2010}.

Another technological challenge in the quest for high electric fields in a DC gun is represented by arcing along the ceramic insulator in the so called triple-point junction, i.e. the area where the electrode meets the insulator and the vacuum. Electric field in this area may become locally very large, due to the presence of unwanted small voids between the electrode and the ceramic~\cite{miller_surface_1989}. 

DC-gun designs can also include the possibility for cathode back-illumination. Such geometry is optimal for ultra-compact systems, where the space-charge driven electron beam bunch lengthening is kept under control by the minimization of the distance between the gun and the sample ~\cite{Sciaini2011, Waldecker_2015}.
The extremely good vacuum performance achieved (in the $10^{-12}$ Torr range) can be exploited to test very sensitive photocathode materials with enhanced performance, such as high QE, low intrinsic emittance, low work function, emission of polarized electrons, etc., while higher energies could be achieved by adding linac boosters downstream of the gun~\cite{Feng_2015, zhou2020velocity}. DC guns have also been optimized at low voltages for targeted applications ~\cite{Badali_10keV_2016}, or cooled down to cryogenic temperatures to obtain brighter beams via lower cathode MTE~\cite{Luca:2015cooled, karkare_ultracold_2020}.

Transmission electron microscopes use DC acceleration. The microscope column can be modified to accommodate the input of a laser pulse for both photoemission and sample excitation, therefore adding ultrafast temporal resolution to the device \cite{piazza_design_2013, Houdellier18, ruan:TEM, plemmons2015, Lobastov_Weissenrieder}, and used both in microscopy and diffraction mode. 
While the electron gun is not designed for optimal beam brightness (accelerating fields are usually of the order of 1 MV/m), such devices are very attractive as the microscope column provides outstanding control of the spatial beam properties. 
Ultrafast TEMs can photo-emit from flat cathodes \cite{ji:flat}, achieving large currents from large photoemission areas, in analogy with typical custom UED setups, or use field-assisted photoemission from tips ~\cite{feist_nanoscale_2018}. In both cases multiple apertures are used to select the core of the beam and obtain small spot sizes.
Typically, an additional condenser lens is added to the column for flexibility (sometimes called C0, see for example ~\cite{piazza_design_2013}). Two viewports and two in-vacuum mirrors are also added to the instrument, to deliver laser pulses to the photocathode and sample respectively. 
Convergent electron beam diffraction using modified TEMs has been shown to achieve spot size at the sample of few nm, with sub-picosecond temporal resolution \cite{feist_ultrafast_2017}. Electron flux is the price to pay for the high spatio-temporal resolution ($<<$1 electron per shot), leading to very long acquisition times. On the other hand TEM columns can today reach very high long-term stability, thanks to continued decades-long engineering development.

    \subsubsection{RF-based pulsed sources}

Radiofrequency electron guns \cite{sheffield1988alamos} operate with accelerating fields larger than $\sim$100 MV/m ~\cite{highgradient18} and multi MeV-level output beam energy, owing to the favorable scaling of the breakdown field with RF frequency (Kilpatrick criterion, reported in Fig.~\ref{fig:Eguns}(f),~\cite{kilpatrick_criterion_1957}). Such technology allows the generation of low emittance, high bunch charge beams (up to $10^9$ electrons per pulse), and its potential to generate beams suitable for UED applications was already recognized during the early stages of development~\cite{Wang:1996up, pac03_ued, JKoreanPhys06}. 
On the other hand the use of RF guns complicates the UED setup, requiring high-power RF sources stability at the edge of the present technology, and femtosecond phase synchronization between laser and RF. 

One potential drawback of the high fields in the cavity, is the generation of unwanted electrons field-emitted from the walls of the cavity every RF cycle and accelerated into the beamline (dark current). In UED applications the dark-current degrades the SNR at the detector, requiring  filtering schemes along the line, as for example transverse collimation or time-gated acquisition. In order to minimize this issue, short RF pulses are sought, but the minimum duration is set by the cavity filling time$\tau_{RF}$, ranging from a few to few hundreds $\mu$s. Some RF designs utilize overcoupling to shorten $\tau_{RF}$, at the expenses of reduced power delivery to the cavity due to the consequent impedance mismatch. Typical cavities require multi-MW peak RF power to establish 50-100 MV/m acceleration fields. Such high peak power bears important consequences on the maximum attainable repetition rate of both the guns and the RF power sources (typically high-power klystrons amplifiers). Indeed the maximum duty cycle of such a high power source is of the order of $1E-3$, while the gun operations are limited to around 1000 Hz, due to the RF-induced heat load on the structure surfaces. 

A final consideration on the RF design is related to the presence of high-order cavity modes which can affect the beam dynamics. Quadrupole components in the RF fields arise due to the asymmetries in the cavity geometry (vacuum pumping holes, couplers, laser ports), and can severely affect the beam dynamics. 
Designs with symmetric coupling or racetrack cavity geometry are employed to minimize these effects \cite{dowell:skew}. 

The main R\&D efforts to further improve pulsed RF guns performance include increasing the acceleration fields, the duty cycle and rep-rate, and the integration of advanced photocathodes in the RF cavity. 
Cryogenic pulsed RF guns are a promising research direction to push the limits of beam brightness ~\cite{Rosenzweig_cryogun_19}, as copper at cryogenic temperatures has significantly lower resistivity loss and can withstand much higher surface fields~\cite{Cahill_cryoxband_18, Cahill_rfloss_18}. Increasing the frequency to the X-band region has been another main R\&D thrust, with the potential to roughly double the acceleration fields of those of S- and L-band guns~\cite{Limborg_xband_16, Marsh_xband_18}.
Finally, recent implementation of advanced photocatode replacement systems coupled to high frequency RF guns will soon open the doors to testing a much wider range of materials, well beyond what has already been done with Cu, Mg and Cs$_2$Te
~\cite{qian_mg_2010,ttfgun2000, Terunuma2010,filippetto_cesium_2015}. The combination of low MTE cathodes and high acceleration fields will create unprecedented peak beam brightness, is ideal for single-shot UED measurements. 

\subsubsection{Continuous-Wave RF sources}

Pushing the repetition rate of RF guns is a challenging endeavour. RF currents on the cavity walls cause ohmic losses, and eventually the power density dissipated on the cavity walls can not be efficiently removed anymore. For a given energy gain, the power density is a steep function of the RF frequency, proportional to $f_{RF}^{5/2}$ ~\cite{Wangler}, making such a problem more important for higher frequencies, and effectively setting peak beam brightness (higher frequencies higher fields) against repetition rate.

Continuous-wave room-temperature normal-conducting RF guns have been developed in the context of high rep-rate X-FELs, and usually operate at lower frequencies in order to sustain the continuous field. As an example, the APEX gun~\cite{APEXgun}  operates with frequencies in the very-high-frequency (VHF) range. Long-term stability at $>$20 MV/m acceleration fields, with a kinetic energy of up to 800 keV has been demonstrated, with input power of the order of 100~kW. The chosen resonant frequency ($\sim$186 MHz), aims at balancing high accelerating fields and thermal load. 
While the prototype VHF gun is presently part of the HiRES beamline, for high rep-rate UED experiments~\cite{Filippetto2016}, an improved version is used to drive the LCLS-II XFEL ~\cite{lclsiigun2014}. There is ongoing optimizing effort to further improve the acceleration field to 30 MV/m, approaching the limit of the allowable surface heat density~\cite{qian_cwgun_2019, pitz_cwgun_2019, apexii2017}. 

An alternative solution which would greatly reduce the thermal management issue relates to the use of superconducting RF (SRF) technology~\cite{SRF_BNL_2020}. SRF accelerating structures are characterized by extremely low surface resistivity and thus can support high RF fields with minimal power consumption. A CW SRF gun has the potential to operate with higher acceleration field and higher kinetic energy than a normal-conducting CW gun. The underlying physics and fabrication technologies for SRF cavities have been under intense R\&D in the past decade, and are now used in large-scale in many facilities~\cite{Grassellino_2013, Grassellino_2017}. This technology however, still faces various challenges to be able to stably operate at high field and high energy~\cite{fes_report_2016}, especially when used in electron guns. The main technical difficulties include handling of RF and thermal junctions between the SRF gun body and the cathode substrate, and contamination of the gun surface by cathode particulates. Quarter-wave resonator type VHF SRF guns at $\sim$200 MHz operate at 4 Kelvin and have rather large characteristic dimensions, and thus could be more likely to overcome the two challenges mentioned above~\cite{wifelgun2012}. Other promising approaches are the multi-cell L-band SRF guns developed at DESY and HZDR, using respectively a superconducting Pb cathode welded to the Nb gun body~\cite{desysrfgun18} and a Mg cathode~\cite{xiangmg2018}. The Pb and Mg cathodes are both suitable for low charge operation for UED. Ongoing R\&D efforts aim at bringing SRF guns to reliable operations at $\geqslant$40 MV/m field and multi-MeV kinetic energies. 

When using CW-RF guns, each RF bucket can be filled with one electron pulse, so the maximum attainable repetition rate is equal to the RF frequency. In UED experiments, considerations on the available laser energy  and sample relaxation times can limit the repetition rate further. Due to the CW operation, system noise can be characterized and potentially suppressed over a much wider bandwidth, thanks to fast electronic feedback. Therefore the amplitude and phase in a CW gun can in principle be controlled to high precision, obtaining higher energy stability than in the case of pulsed systems. High rep-rate detectors and beam instrumentation are an active area of development with many commonalities between UED and FEL requirements and similar rewards. 

    \subsubsection{Advanced electron sources}
 \label{sectionII.b.3}
 
In the following we provide an overview of the main research directions aimed at the development of new electron sources at the time of this review.

\paragraph{THz gun and acceleration}

Extending electron beam acceleration devices to THz-scale frequencies could potentially allow to reach GV/m gradients, leading to a leap in beam brightness. Recent progress in this direction led to increased energy gain from a few keV to hundreds of keVs~\cite{Huang_thzgun_16, nanni_thz_15, zhang_steam_2018, othman_2020_thzgun}, and promising potential to reach the MeV level~\cite{Fallahi_thz_2016}. The dimensionless parameter $\alpha\propto E_0/\omega$ (see Sec.~\ref{RFdynamics}) presently achieved in THz-based electron guns is significantly smaller than unity.
Therefore, severe phase slippage occurs between the electron beam and the THz field, limiting the effective interaction distance and energy gain.
Other active research areas in this field include both the fine control of field amplitude and phase, and THz-gated photoemission ~\cite{carbajo_thzcathode_20}. Geometric apertures of THz guns are comparable in size to the wavelength of the field and thus can accommodate micrometer-sized beams for UED setup\cite{DZhang21_thzued}. A distinct advantage of THz acceleration over RF sources is the intrinsically jitter-free acceleration: the THz pulse can be derived from the pump laser system. At the same time, THz production is based on a nonlinear process, and a stable accelerating field requires exquisite control on the laser amplitude. 

\paragraph{Laser-acceleration based electron sources}

Laser-driven acceleration is based on ultrashort and ultraintense lasers to achieve acceleration gradients up to three orders of magnitude higher than that of conventional RF accelerators. The main challenge is to identify suitable coupling mechanisms between the transverse electromagnetic waves and the longitudinal electron motion. In laser-plasma accelerators (LPA), this coupling is performed via excitation of a longitudinal plasma wave in a gas using intense laser pulses, producing gradients up to 10 GV/m.  

LPA-based electron sources share with THz-based acceleration the advantage of obtaining  electron bunches intrinsically synchronized with the drive laser. In addition, the temporal duration of the accelerated bunches is inversely proportional to the plasma frequency, which can be controlled by the plasma charge density, naturally producing  few fs electron bunches. 
The use of sub-MeV electron beams generated by laser driven acceleration for diffraction measurements has been demonstrated~\cite{Tokita:2009by, faure:UED, he_capturing_2016}.
One important challenge for UED applications is to be able to  preserve the short bunch length during propagation to the sample, and obtain repeatable beam parameters.
One strategy to improve stability (at the cost of beam current) is to use a magnetic beam transport line with collimators to select a predefined region in phase space and then maximize the LPA accelerator overlap with the acceptance window of the system. The use of a collimator in a dispersion region was demonstrated to be beneficial in improving the transverse quality of the beams and select a fixed energy band  ~\cite{Tokita10, Faure16concept}. Since the time-of-flight of electrons depends on their energy, monochromatization of the beam also stabilizes the time-of-arrival, improving the temporal resolution to sub-10 fs levels. LPAs provide a promising route to realize an all optical, jitter-free approach for UED, with ongoing efforts to improve the quality, stability and rep-rate of the electron beams.  

\begin{figure}[ht]
\includegraphics[width=1\columnwidth]{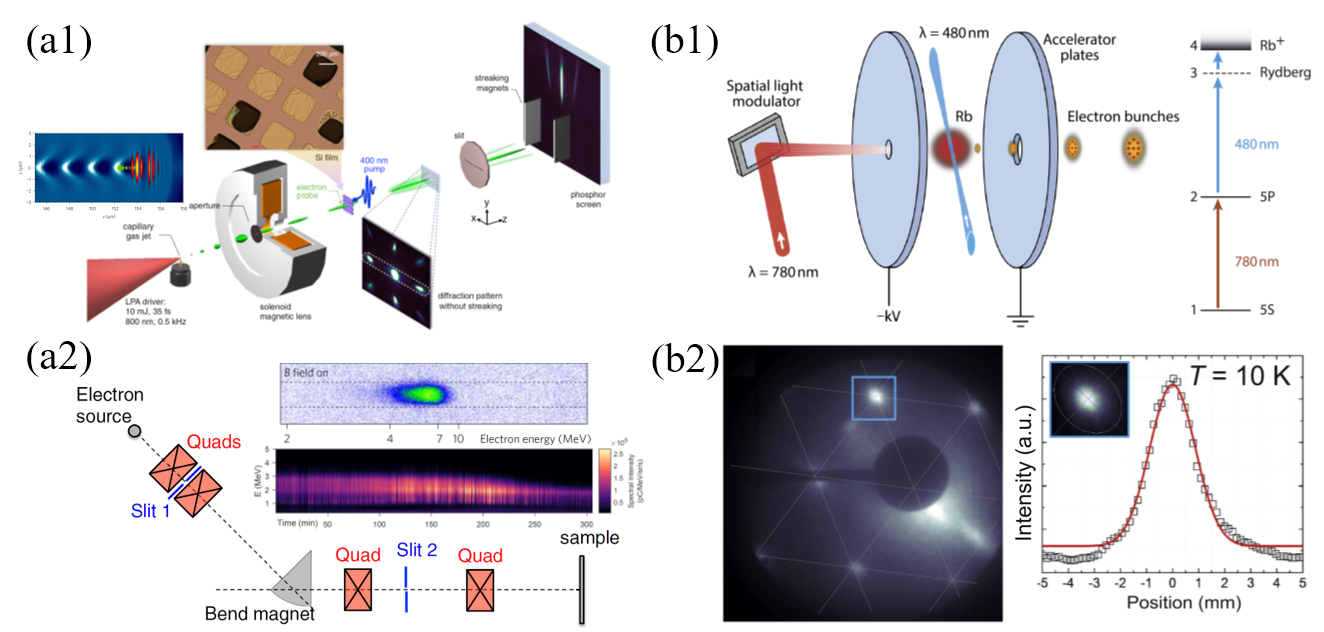}
\caption{Examples of advanced sources for UED. (a1) Schematic of a LPA electron beamline for UED in which the Silicon diffraction patterns consists of Bragg strips due to the relatively large energy spread \cite{he_capturing_2016}, and (a2) a transport beamline with collimation utilizing LPA electron bunches to reach 10 fs level temporal resolution \cite{Faure16concept}. The inset shows the LPA energy spectrum with long-term stability optimized and suitable for MeV UED purpose \cite{Rovige_longterm_2020}. (b1) Schematic of an ultracold MOT sources and the trapping and ionization energy levels of Rb atoms \cite{McCulloch_2016}, and (b2) the graphene diffraction pattern obtained with a MOT source and the source temperature is retrieved to be 10 K \cite{Mourik_MOT_UED14}.}
\label{fig:advancedsource}
\end{figure}

\paragraph{Ultracold sources}
Near-threshold photo-ionization of magneto-optically trapped (MOT) atoms is another novel approach for generating low emittance, high coherence electron beams~\cite{Claessens_cold_2007, Scholten2011, Luiten_cold_2014}. This approach takes advantage of the progress in atomic cooling techniques over the last two decades~\cite{Killian_mot_1999, Robinson_mot_2000}. The schematic of a MOT electron source is shown in Fig.~\ref{fig:advancedsource}(c). For example, for the commonly used Rubidium sources, a cloud of $^{85}Rb$ atoms are first excited from $5s$ to $5p$ state and then ionized by a second laser pulse to release photoelectrons, which are then immediately accelerated by an electric field. Laser pulses used for excitation and ionization usually propagate in perpendicular directions and form a source volume of hundreds of $\mu$m in all three dimensions in order to extract at least 10$^6$ electrons as the maximum density of the MOT is limited to below 10$^{12}$ cm$^{-3}$. The excess energy of the photoelectrons can be tuned by the central wavelength and bandwidth of the ionization laser, with the latter constrained by the choice of the laser pulse duration (through the Fourier transform limit). An interesting phenomena is that the excess energy of the extracted electron beams has been shown to remain well below the bandwidth of the ionization laser due to the complex interplay of the laser field and the potential of Rb$^+$ ions. The effective temperatures of MOT sources were shown to be as low as 10 K, significantly lower than that from common solid state photocathodes~\cite{Engelen13}. There are ongoing R\&D efforts to further increase the density in the MOT and hence the brightness of the source for UED applications.  

\paragraph{RF-streaked ultrashort bunch train}
Generation of picosecond to sub-ps electron beams usually relies on photoemission sources using ultrafast lasers. A new concept for producing a train of ultrashort bunches without laser has been proposed and experimentally demonstrated~\cite{verhoeven2018high, VANRENS2018, qiu2016, Lau_stroboscopic_2020}. In this scheme, an RF deflecting cavity and a collimation slit are inserted between the electron source and sample of a conventional TEM. The cavity imparts a time-dependent angular kick to the DC electron beams, causing electrons to be deflected transversely depending on their arrival time. Only electrons arriving close to the zero-crossing phase will experience weak enough deflection and propagate through the slit. This scheme therefore imparts a temporal structure to a continuous stream of electron at expenses of beam current, with a fixed repetition rate equal to two times the RF frequency. Controlling the parameters of the setup, including the deflection strength, the location and width of the slit etc., the temporal duration of the pulses can be adjusted, together with the average number of electrons in each pulse, while maintaining the beam quality to reach high spatial resolution \cite{zhangkruit}. The rep-rate of the pulses can be GHz using a single cavity, tens of MHz relying on the beating of two GHz cavities~\cite{dualmode2018}, or tunable from 0.1 to 12 GHz using RF-driven traveling wave stripline elements~\cite{Jin2019}. A similar method for generating short electron pulse trains at high rep-rate from an originally DC electron beams is to utilize a photoswitch as a beam blanker~\cite{Kruit18}. The GHz electron pulse trains instruments are suitable for studying ferromagnetic resonance in magnetic materials, magnons in spintronics, and electromagnetic fields \cite{fu2020direct} and atomic structures in MEMS/NEMS systems etc. under synchronized GHz RF excitations. Pulsed electron beams alone have also been explored to potentially relax radiation damage to samples~\cite{Kisielowski19, Choe2020}. 

\subsection{Control and measurement of ultrafast pulses of electrons}
\label{sectionII.d}

Measuring and controlling femstosecond electron beams is a challenging endeavour shared among many techniques for ultrafast science, such as free-electron lasers and ultrafast electron diffraction and microscopy setups. In UED, given the small number of electrons per pulse, accurate measurements of arrival time and pulse duration suffer from low signal-to-noise ratio and long acquisition times. Strong lateral focusing of electron pulses into nanoscale dimensions is complicated by the action of space charge forces, inducing large energy spread and non-linearities in the beam phase space, by the large beam emittance produced by flat cathodes and by lens aberrations. 
To further complicate the matter, beam properties are most useful if measured in \textit{real-time}, i.e. contextually with the experiment. 

In what follow we provide an overview of the state-of-the-art techniques for measuring and control of electron beams in a UED beamline. 

    \subsubsection{Measuring the duration of ultrashort electron pulses}
\label{section:temporaldiagnostic}

\begin{figure}[ht]
\includegraphics[width=0.9\columnwidth]{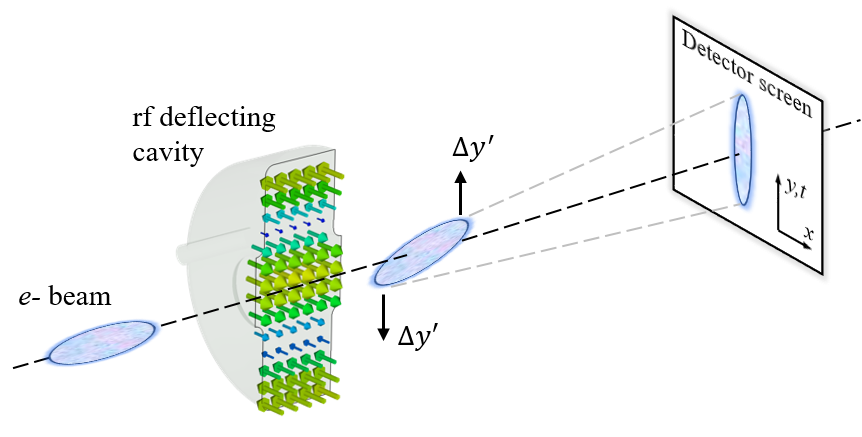}
\caption{Principle of bunch length characterization using an RF deflecting cavity. Mapping the electron beam temporal distribution into the transverse density profile.}
\label{fig:rfdeflector}
\end{figure}

Information on the electron beam temporal distribution can be encoded into one of the transverse directions through streaking technique, which uses time varying fields to introduce transverse-to-longitudinal correlations. A time-dependent kick in transverse momentum is applied (streaking), and then mapped into a transverse profile via a drift section or electron optical transport line. The necessary fields for beam streaking include quasi-DC, RF, THz, as well as optical fields. 
DC-like streaking fields are generated by ramping a DC field perpendicular to the beam trajectory, between two electrode plates, and have been used for long time in streak cameras to characterize the bunch length of low energy photoelectron beams. Optically triggered streak cameras can provide enough electric field amplitude for obtaining sub-picosecond resolution in non-relativistic setups. Photoswitch-based devices encode information related to the electron beam TOA at the sample within the diffraction pattern image (centroid motion of the peak along the streaking direction), obtaining 150 fs resolution after temporal binning \cite{Gao2013}. More recently the same technique has been demonstrated adequate to measure the bunch length of tens of keV electron beams with $\sim$100 fs resolution ~\cite{Kassier_2010}. The extension of this technology to higher temporal resolutions, higher repetition rates and higher energy beams is hindered by electric breakdown of the photoswitch material in vacuum.  
Beam transverse deflection with an RF cavity was first demonstrated with the Lola cavity~\cite{lola}. The principle of use of a deflecting cavity is shown in Fig.~\ref{fig:rfdeflector}. For a detailed beam dynamics treatment in presence of RF deflecting cavities see~\cite{Floettmann:2014jfba}.  The resonating structure usually operates with an HEM11 mode, imposing a strong time-dependent transverse momentum kick to electrons. Assuming no deflection for the longitudinal beam center, the streaking strength is $K=\frac{e\omega V_0}{mc^2\gamma}R_{12}$, where $\omega$ is the angular RF frequency, $V_0$ is the maximum deflecting voltage, and $R_{12}$ is the transfer matrix coefficient for mapping the transverse angular coordinate from the deflecting cavity to position on a downstream transverse detector ($R_{12}=L$ for a drift space of length $L$). Using RF deflectors with appropriate $V_0$ and $\omega$, femtosecond resolution has been demonstrated on ultrarelativisitc beams ~\cite{maxson_ultrafast_2017, lcls_xtcav_2015}. The ultimate resolving power of the instrument is limited by both the beam uncorrelated divergence(see~\ref{brightness}), and the maximum voltage achievable. Indeed, a first requirement constrains the transverse angular spread of the beam $\sigma_{r'}$, to be much smaller than the difference in RF streaking kick between two time-points to be distinguished, i.e. $K\sigma_t\gg\sigma_{r'}$. At the same time, small beam sizes are needed inside the RF structure to avoid off-axis field distortions, and at the final detector to contain the beam inside the total screen size and avoid spreading the signal over too many pixels, which would limit the SNR. 

To obtain larger streaking field, higher frequencies in the optical and THz range could be pursued. THz streaking of electron beams was first introduced in~\citet{fabianska_split_2015}. To increase the field amplitude, the authors propose and design a split ring resonator geometry that enhances the field in the gap. More in general, nano and micro-structured surfaces can be used to locally enhance the THz field and introduce amplitude and phase differences between the $E$ and $B$ components, with physical geometries ranging from butterfly triangles to  parallel-plate waveguides. Upon illumination with THz radiation, such structures have demonstrated sub-femtosecond temporal resolution on non-relativisitc (30 kV) beams~\cite{kealhofer_all-optical_2016}, and $\approx$10 fs for relativistic, MeV-class electron pulses ~\cite{zhao_terahertz_2018,li_terahertz-based_2019}. Dieletric-line waveguides driven by THz offer highly linear fields with reasonable transverse dimensions, which are also suitable for streaking measurement \cite{zhao_terahertz_2019,Lemery:IPAC2017}. Furthermore, THz fields have been used for temporal compression of beams, with simultaneous suppression of the relative time jitter, leading to a sub-50 fs overall temporal resolution \cite{zhao_femtosecond_2020,snively_femtosecond_2020}.

Electron energy modulation via direct interaction with optical near-field from laser pulses can be used for retrieving pulse length and relative electron beam-laser time jitter. If the electrons are suddenly launched into a high field region, with boundary conditions allowing electric field in the longitudinal direction, electrons will be accelerated or decelerated depending on the phase, and energy sidebands will appear in the spectrum, showing higher-order periodic modulations separated by the laser photon energy reaching tens of eV. Analysis of the side bands reveals information on the electron beam duration and time jitter ~\cite{kirchner_laser_2014}. Narrow beam energy spread is required to resolve the modulations, limiting the operation mode to single electron emission. On the other hand, utilizing carrier-envelope-stabilized pulses, sub-femtosecond resolution can be achieved.

Direct electron-laser interaction in vacuum, i.e. ponderomotive scattering of electrons by laser fields, have also been used to characterize the bunch length of electron beams. The present laser technology provides access to high-peak laser intensities from commercial table top systems, in the region of $10^{17}$ W/cm$^2$, which can be used to drive non-linear processes and enable energy exchange with free electrons in vacuum. The ponderomotive force acting on an electron beam upon interaction with a laser field depends on the spatial gradient of the field envelope and adds an outward drift component to the motion, superimposed to the quiver oscillations driven by field  oscillations in time ~\cite{kibble_mutual_1966,Gao_2012}, providing a direct mean for obtaining the longitudinal convolution between laser and electrons.

The technique has been demonstrated in accumulation mode with non-relativistic UED setups~\cite{siwick_characterization_2005}, and subsequently improved via laser local intensity enhancement using optical interference, obtaining higher resolutions with lower laser energies~\cite{hebeisen_grating_2008}. 

    \subsubsection{Time-stamping}
        \label{section:timestamping}

Online, single-sot measurements of the relative time delay between pump and probe pulses provide a clear route towards higher temporal resolution.
Such development needs to be carried out in conjunction with novel signal detection methods enabling high frame rate acquisition of single-shot UED patterns, which would then enable tagging of each frame with a specific measured pump-probe delay.  
The first demonstration of electron beam time stamping was performed in 2005 \cite{cavalieri_clocking_2005}, via an electro-optical sampling of the electron beam electric field (EOS, see~\cite{valdmanis_subpicosecond_1986}). The THz-components of the electric field co-propagating with the beam induce transient birefringence in an off-axis anisotropic crystal.The change in index of refraction is sensed by a probing laser beam, encoding beam temporal information in the spatial, temporal or spectral distribution depending on the particular setup. Alternatively, the electro-optical  conversion can be performed outside the vacuum chamber ~\cite{lohl_electron_2010}, achieving sub-10 femtosecond resolution. As the signal strength decreases strongly with the charge, so does the the measurement accuracy. At 10 pC, the single-shot temporal resolution has been measured to be ~200 fs ~\cite{scoby_electro-optic_2010}. The use of nanostructured surfaces would allow greater THz detection efficiency, thanks to plasmonic enhancement. Recently photo-conductive antennas have been used to detect the beam arrival time of a 1 pC beam~\cite{snively_non-invasive_2018}.  

Temporal streaking of electron beams can provide sub-femtosecond resolution in time of arrival. The technique is mostly used for measurement of longitudinal beam distribution (Sec.~\ref{section:temporaldiagnostic}), but it can also be applied to the measurement of beam shot-to-shot temporal jitter. The information obtained in RF streaking corresponds to the jitter between the electron beam arrival time and the phase of the RF wave, not of the optical excitation pulse. If THz or optical frequencies are used, the streaking field can be derived directly from the pump laser, maintaining phase-coherence and providing direct pump-probe time-stamping information. 

Although beam streaking is a destructive measurement, it could be in principle applied to the undiffracted beam downstream the detector, if let through.
Linear correlation between electron beam energy and time of flight has been experimentally demonstrated over a broad range of energies for a system without a bunching cavity \cite{zhao_terahertz_2018}, implying that a simple spectrometer system could be used as non invasive time-stamping tool. Going to even shorter wavelengths holds the potential for attosecond-scale control. 
Laser-electron interaction, such as the energy modulation or the ponderomotive scattering described above to measure the pulse length, could be used in place of an RF cavity, for directly retrieving relative electron beam-laser time jitter. 

   \subsubsection{Measuring time-zero}

All methods described in Sec.~\ref{section:timestamping} for time-stamping can also be used to find temporal overlap between the optical pump and the electron probe at the exact sample location, also called time-zero. On the other hand, performing time stamping measurements requires additional tools not necessarily included in every UED setup. Given the primary importance of establishing time-zero in ultrafast experiments, alternative methods have been developed, mostly based on destructive interaction between electron beam and laser mediated processes. The main challenge is to develop a simple and robust procedure to retrieve time-zero with sub-picosecond precision, which could be implemented rapidly during an experiment, ans possibly repeated multiple times during a data-acquisition run. 

For low density targets, like gaseous materials, hours of integration may be needed to obtain diffraction images with good SNR, making such experiments particularly sensitive to slow drifts of time-zero, due to variations of the system, such as  environmental conditions or calibration constants. 

For example, the thermal coefficient of delay  (TCD) of a typical optical fiber or coaxial cable is in the range of $10~ps/km/^{\circ}C$ or larger. As a consequence the time delay of signals traveling through a fiber/cable to reach the receiver depends on the ambient temperature. A change in temperature will then result in a phase shift at the receiver. As experiments with long accumulation times can run overnight, large temperature variations are expected if not compensated by adequate temperature control, and time-zero measurements will likely need to be repeated periodically throughout the experimental scan, justifying the need of a technique readily available contextually to the experiment.

Electron beam shadowgraphy of transient electric fields in a laser-induced plasma has been extensively used as time-zero tool in UED experiments \cite{park_synchronization_2005}, but also as a scientific technique for the study of  laser-induced ablation in solids \cite{hebeisen_grating_2008,zhu_four-dimensional_2010}
and optical-field-ionization in plasmas~\cite{centurion_picosecond_2008}.
Here an intense ultrafast laser pulse illuminates a target material, triggering the injection of a plum of electrons in vacuum. The UED electron pulse acts as a sensitive probe for the the transient electric field associated with the expansion of the electron cloud in vacuum. Temporal pump-probe scans reveal the evolution of the fields in the vicinity of the interaction region. For the purpose of time-zero measurements, the exact mechanism of electron emission is of secondary importance, whether from multiphoton photoemission, ablation or plasma formation. Key features of the process are its prompt response, measured to be in the sub-picosecond range, and its simple setup which promotes virtually any metallic edge to become a potential source of electrons. Indeed such technique has been proven using a plethora of different target materials and geometries, from needles~\cite{li_ultrafast_2010} to standard copper TEM grids ~\cite{scoby_effect_2013}, which makes it appealing as versatile method for searching time-zero. Laser fluence values used vary from $0.1$ to $10 J/cm^2$, larger than typical values for UED on solid-state sample, and require to increase the laser pulse energy and/or decrease the spot size. 

More recently another technique for electron-laser cross-correlation has been proposed and implemented, drawing from the examples of successful timing tools at FEL facilities (see for example~\citet{bionta_spectral_2011,harmand_achieving_2013}). In FELs the X-ray pulse is used to induce a transient change in the optical properties of a specimen. X-ray absorption instantaneously increases the free carrier density in the material, modifying the complex index of refraction, both in phase and amplitude. This process causes a sudden variation in the optical reflectivity of the material which can be probed by an optical pulse, providing accurate timing information. High energy electron beams traversing the same material can induce similar changes on its optical characteristics. 
Two main features of this technique make it very attractive for use in UED setups: first, when the method is applied in transmission geometry, electrons travel tens-to-hundreds of micrometers through the material, depositing large amount of energy and generating large absolute value of free carriers. The transmission of the subsequent probing optical laser will be sensitive to the total number of free carriers along the optical path. In comparison with X-rays, a lower number of electrons will be needed to induce similar changes in the optical transient reflectivity of the material. Second, the temporal delay information is encoded in the energy variation of the probing laser pulse, which can be easily measured with photo-detectors at very high speeds. Such high bandwidth measurement may allow characterization of fast temporal electron jitters, even at high repetition rates, opening the door to fast beam-based temporal feedback systems. 
The choice of the sensing material, its thickness and the geometry of the interaction determine the response time of the technique, with an ultimate limitation given by the time it takes for the energy absorbed to be transformed into electron-hole pairs and, therefore free carrier density modulation. In \citet{Cesar_2015}, a 1-mm thick Germanium slab was used, demonstrating measurable signal down to electron beam charges of 1 pC. Improved detection designs, such as the one demonstrated in~\cite{droste_high-sensitivity_2020}, hold the promise of improving the sensitivity of such technique well into the fC-range. 

\subsubsection{Laser-to-RF synchronization} 
\label{section:clocking}

When using time varying fields for acceleration and/or compression, phase locking between the different oscillators (RF and laser) in required.
The most use figure of merit to characterize the system phase stability is the cumulative rms time jitter ~\cite{scott_high-dynamic-range_2001} around the n-th harmonic of the laser repetition rate~\cite{du_precise_2011}. 
This can be promptly measured from characterization of the system in the frequency domain~\cite{tsuchida_wideband_1998}.

Once characterized, different signals can be phase locked to a reference with the use of a phase-locked-loop (PLL). A typical locking scheme includes a custom very-low-noise microwave oscillator as a common reference for all the subsystems. In order to perform laser phase locking, the oscillator cavity length can be adjusted controlling the position of the cavity end-mirror with voltage-regulated piezoelectric actuator, with typical bandwidth in the (tens of) KHz range limited by mechanical resonances the system.

\begin{figure}[ht]
\includegraphics[width=1\columnwidth]{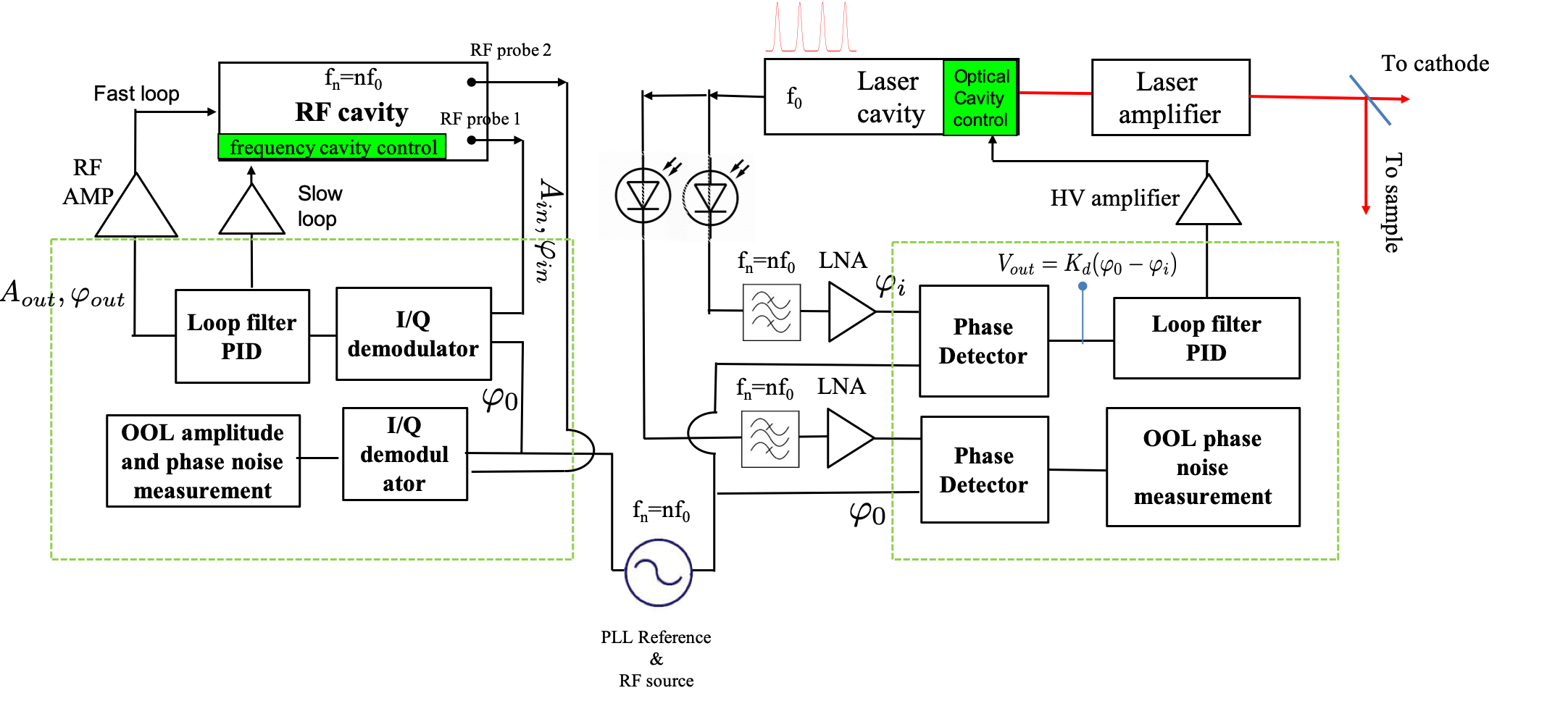}
\caption{An example schematic of a synchronization system for a UED setup including RF signals.}
\label{Fig:synch}
\end{figure}

A schematic of a typical synchronization setup is shown in Fig.~\ref{Fig:synch}. 
The right side of the schematic shows the laser-to-RF synchronization diagram.
After the phase detection a proportional-integral-derivative (PID) filter is applied to produce an output voltage control for the oscillator cavity. By changing the PID parameters of the filter, the spectral response of the PLL loop can be optimized. A second phase detection chain is used to perform out-of-loop (OOL) measurements on the system, and verify the performance. OOL measurements are essential part of a feedback system performance characterization, providing an independent measurement of the field and the total effect of the feedback loop, including unwanted spurious components.

The figure also presents a general diagram for RF cavity field control. Feedback loops in this case act on the field amplitude and phase, therefore the RF electronics in the loop will have to decouple AM from PM (I/Q demodulator). A vectorial PID loop will provide the output signal to the RF amplifier to stabilize the cavity. 

In implementing a PLL loop, both analog or digital electronic solutions can be used. In particular, Field-Programmable-Gated-Array (FPGA) technology is becoming very common in the field of particle accelerator controls. FPGAs-based boards are today equipped with ADCs, DACs, clocks and clocks distribution channels, and can perform all the functions highlighted in the green dashed boxes of Fig.~\ref{Fig:synch}. 

Depending on the particular application and on the specific environmental conditions, different phase-locking techniques have been applied to achieve sub-10 fs synchronization, maintained for extended periods of time~\cite{kim_drift-free_2008,yang_10-fs-level_2017}.
As an alternative solution for compact UED setups, the signal driving the RF cavity can be derived from the laser, using the optical oscillator as a direct reference for the the PLL loop~\cite{walbran_5-femtosecond_2015,otto_solving_2017}. This simple solution provides natural lock between the cavity driving signal and the laser system, while the phase of the field inside the cavity is stabilized by the feedback loop.
The drawback of this configuration is in not been able to pick an independent oscillator reference with optimized noise figure outside the feedback loop.

\subsubsection{Truly single shot measurements}

Temporal streaking can be used in UED experiments to obtain continuous temporal information over the duration of the incoming electron beam. In this setup the deflecting element is placed after the UED sample, obtaining a streaked image of the diffraction pattern. The technique was already proposed during early UED experiments~\cite{Mourou1982}  and has been successfully demonstrated more recently~\cite{musumeci_capturing_2010}, ultimately reaching $<$50 fs temporal resolutions with MeV-class electron beams~\cite{Scoby:2013doba}. In this operation mode the duration of the electron beam constitutes the temporal field of view of the experiment and is chosen to be much longer than the pump laser, in the (tens of) picosecond-range. A laser pulse initiates the process simultaneously to the passage of the electron pulse, and the temporal response of the sample is encoded in the electron beam temporal distribution. 
Temporal streaking of the electron beam downstream the sample provides coupling between the streaking plane and time and enables direct measurement of its temporal evolution at a subsequent screen.

\begin{figure}[ht]
\includegraphics[width=0.4\columnwidth]{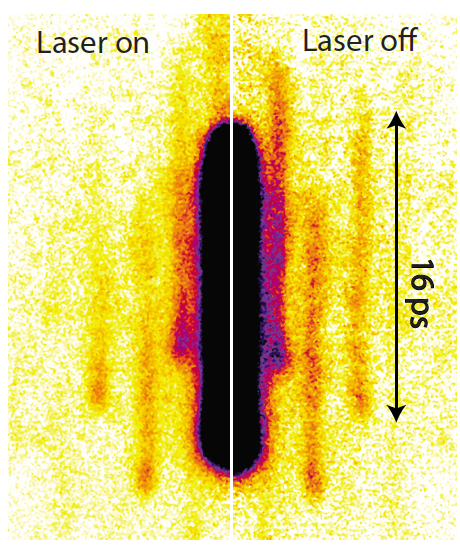}
\caption{Example of UED experiment with temporally streaked electron beam. From \cite{musumeci_capturing_2010}.}
\label{Fig:streakedUED}
\end{figure}

The advantages of this technique are demonstrated by the experiment results from \cite{musumeci_capturing_2010}  reported in Fig.~\ref{Fig:streakedUED}.
The image reports an example of streaked electron beam after passage through a single-crystal gold sample. The time axis (in vertical) shows peak intensity decrease due to Debye-Waller effect following laser excitation. The entire temporal information is compressed in one single image.

Due to the induced coupling between longitudinal and transverse planes, the main limitation to the temporal resolution of the method is the transverse emittance. Indeed the total beam size at the detector plane $\sigma_x$ is the convolution between the geometric beam size without streaking $\sigma_{x_0}$, and the streaking contribution, $\sigma^2_x= \sigma_{x_0}^2 + (K_{TCAV}\sigma_t)^2$ where $K_{TCAV}$ is the deflecting cavity calibration factor, measured in $m/s$.
At the same time the method requires a larger number of electrons in the beam. For a given temporal resolution, the electron number requirements in a matching time slice should follow the requirements for single-shot UED defined earlier, i.e. roughly 1e6 electrons per units of temporal resolution, setting a beam current requirement. For example, to obtain 100 fs resolution, an electron beam with a current of 1.6 A should be used. 
Also, spatial information along the streaking plane is lost, and overlapping between different streaked Bragg peaks should be avoided~\cite{Floettmann:2014jfba}. 
A complementary method to obtain truly single shot information without the use of an RF deflecting cavity, exploits large time-correlated energy spreads generated either by the longitudinal space charge effects or by the source itself as in the case of laser-wakefield accelerators~\cite{he_capturing_2016}.
The chirped beam is sent through a dispersive magnetic element after passing through the sample, obtaining energy streaked images at the detector. In the assumption of linear chirp, a direct correlation between energy and time axis is established.

\subsubsection{Control of lateral coherence and beam size }
\label{sec:beamsize}

The use of an optical system is critical to maximize the resolution in an optical setup. The scattering of the electrons in the sample changes their angle, and optics is used to convert this angular deviation into a transverse offset which can be detected using a beam profile monitor screen. An ideal optical system for this task is one for which the transverse position on the detector screen does not depend on the position of the electron at the sample so that a simple map exists between diffraction angles into position offsets. In beam optics formalism, this corresponds to setting the first element of the 6x6 transport matrix $R_{1,1}$ to be equal to zero. This could be accomplished by a series of round lenses as typically done when operating a transmission electron microscope in diffraction mode. Alternatively one could simply use a very long drift and settle for an equivalent condition where the transverse offset on the detector screen is dominated by the angular deviation at the sample plane (i.e. $R_{1,1}\sigma_x \ll R_{1,2}\theta_b$). For a drift of length $L$, $R_{1,1} = 1$ and $R_{1,2}= L$ so that this condition will be satisfied for a sufficiently long distance between the sample of the detector. 
$R_{1,2}$ is the so-called length of the diffraction camera and enters in the calibration of screen offset to angle which is essential to get quantitative information from the diffraction pattern.  If a combination of lenses is used, the diffraction patterns need to be calibrated and a known Bragg peak or a calibration target can be used for this scope.  

Before we go in detail on the subject of transverse beam control, it is important to clarify the definition we adopt to characterize the spread of a distribution which, following accelerator and beam physics, is the root-mean-square (rms). 
Such a definition can be used independently from the actual details of the distribution and transported along the beam line using linear equations. The relation of the rms size with other definitions, such as FWHM or FW50 (or full width containing 50 $\%$ of the beam) more common in other literatures, will depend on the particular shape of the distribution.

There are a variety of motivations to control the size and shape of the transverse distribution of the electrons illuminating the sample in UED experiments. For example, by increasing the transverse spot size at the sample (which can be done provided sufficiently large sample and pump area), one can reduce the uncorrelated beam divergence and therefore increase the coherence length $L_c$. Conversely, a very small spot size is needed to understand the role of local heterogeneities in structural dynamics and whenever large samples can not be used. 
In fact, in typical custom kev and MeV UED setups, the transverse probe size has been around 100 $\mu$m rms, and smaller local details are averaged out in the Bragg peaks. 

An exciting research and development opportunity is to combine the strengths of UED and TEM, i.e. femtosecond pulse duration/temporal resolution with $\mu$m and smaller probe size, to enable studies of ultrafast structural dynamics with very high spatial resolution. Using the formulas in Sec.~\ref{section_coherence_length}, we can estimate the beam quality requirements to achieve simultaneously desired probe size and momentum transfer resolution in micro and nano UED. If we target an rms probe size at the sample $\sigma_x=1~\mu$m rms and an uncorrelated beam divergence $\sigma_{x'}=100~\mu$rad rms yielding reciprocal space resolution $\Delta s=2\pi\sigma_{x'}/\lambda=0.26$\AA$^{-1}$ for $\gamma=10$ electrons, the corresponding normalized emittance requirement is $\epsilon_n=\gamma\sigma_x\sigma_{x'} < 1$~nm-rad, at the lowest end of what achievable with state-of-the-art electron sources. 

In these demanding cases, simply measuring how small the spot size is at the sample becomes a technological feat. Typically, a spot size measurement is obtained from the quantitative analysis of beam images from fluorescent screens or other 2D detectors (see below for general discussion). These work well for low charge beams with spot sizes down to 10 $\mu$m. At higher beam charges, effects like saturation or space-charge blooming \cite{murokh:blooming} can impede the measurement of smaller spots. Multi-shot techniques, such as moving a knife-edge (\cite{ji:knifeedge} or thin wires in the beam \cite{borrelli:submicron, orlandi:wirescanners}, are better suited for $\mu$m scale spot size measurements. 

\paragraph{Electron optics}

In the following we will discuss electron focusing, starting from the lens geometry, configuration and limits and then addressing the most common magnet technologies employed.
It is worth noting that that space charge effects enter in this discussion only at second order, mostly being responsible for emittance growth. Somewhat counter-intuitively, in tight focusing conditions the beam waist is ballistic and fully dominated by the emittance term and not by space charge forces (see envelope equation in Sec.~ \ref{section:spacecharge})~\cite{SandR}.

Both electrostatic and magnetic lenses can be used for focusing~\cite{TEM:book}, but in practice there is a clear advantage in focusing strength for magnetic lenses as soon as the electron velocity reaches a sizable fraction (~0.1) of the speed of light (Einzel or immersion lenses are used in some cases inside the accelerating gap \cite{ACHIP:source}). 

Solenoids are the most common electron optical element in UED beamlines. The focal length of a solenoid of effective thickness $L$ is $f = \frac{(4 B\rho)^2}{B^2 L}$ where $B\rho = m_0 c \beta \gamma / e_0 $ is the relativistic beam magnetic rigidity. Spherical and chromatic aberrations \cite{hawkes:aberrations} limit the smallest spot sizes that can be achieved. The coefficients are on the same order of the focal length \cite{reimer2013} and cause an effective emittance growth in the beam line. Spot sizes of few microns have been achieved using solenoid lenses~\cite{shen_micro_UED18}. The velocity spread inside these lenses has an interesting effect on temporal resolution discussed in \Citet{Baum:temporallenses}. For ultrashort electron bunches, off axis particles acquire large transverse velocities at the expenses of their longitudinal velocity, resulting in temporal distortion of the pulses at the exit of the lens. By carefully designing the optics to take into account the nonlinear terms in the transport, including the introduction of RF cavities serving as temporally varying lenses, it is possible to avoid or minimize these effects.

The quadrupole lens is another focusing element which focuses in one direction and defocuses in the other one. The focal length of a single quadrupole of effective thickness $L_q$ can be written as $f = \frac{B\rho}{g L_q}$ and has a much more favorable scaling with energy than the solenoid. $g$ is the quadrupole gradient and strongly depends on the gap size. For small gaps (mm-scale) quadrupole gradients approaching $g \simeq$ 1000 Tesla/mm are achievable \cite{ghaith:pmq}. In order to get focusing in both directions, the most common configuration is the quadrupole triplet where three quadrupoles with alternating orientations are used. Both the more traditional (2f -f 2f) \cite{ji_nanoued_2019} and (2f -f f) \cite{lim:pmq} configurations have been employed with the latter a preferred choice for large and collimated input beams. More exotic configurations have been proposed to improve the optical characteristics of the lens system. For example, the Russian quadruplet~\cite{zhou:quadruplet} is a highly symmetric optical configuration which satisfies the imaging condition with equal magnification in the x and y plane. This configuration uses 4 quadrupolar lenses with strength inverted about the symmetry plane (i.e. $f_1 f_2 -f_2 -f_1$). More recently a quadrupole quintuplet \cite{wan18uem} configuration has been discussed in order to minimize the effect of aberrations in high energy electron beamlines, although still not demonstrated in diffraction experiments. Note that in systems with a large number of independent optics,  keeping the axes of the lenses aligned to the tolerances required to minimize the aberrations and get the expected spot size is still an open challenge, and skewness and misalignment-induced aberrations are common.

Conventional electromagnets use current carrying coils and an iron yoke to bend the field lines and complete the magnetic circuit. The magnetic field depends linearly on the current density until saturation in the high permeability yoke takes place. For current densities below 1.5 A/mm$^2$ the magnet can be simply air-cooled \cite{tanabe:book}. For larger current densities, typically water-cooled hollow core conductors are employed. Rapid advances in superconducting technology (SCT) have enabled the development of superconducting magnets, especially useful for relativistic UED beamline which have higher field requirements \cite{fernandez:SCTEM}. Type II superconductors like Nb3Sn are capable of reaching higher fields and therefore focusing strengths, thanks to their larger critical magnetic fields~\cite{RossiBottura}.

\begin{figure}[ht]
\includegraphics[width=1\columnwidth]{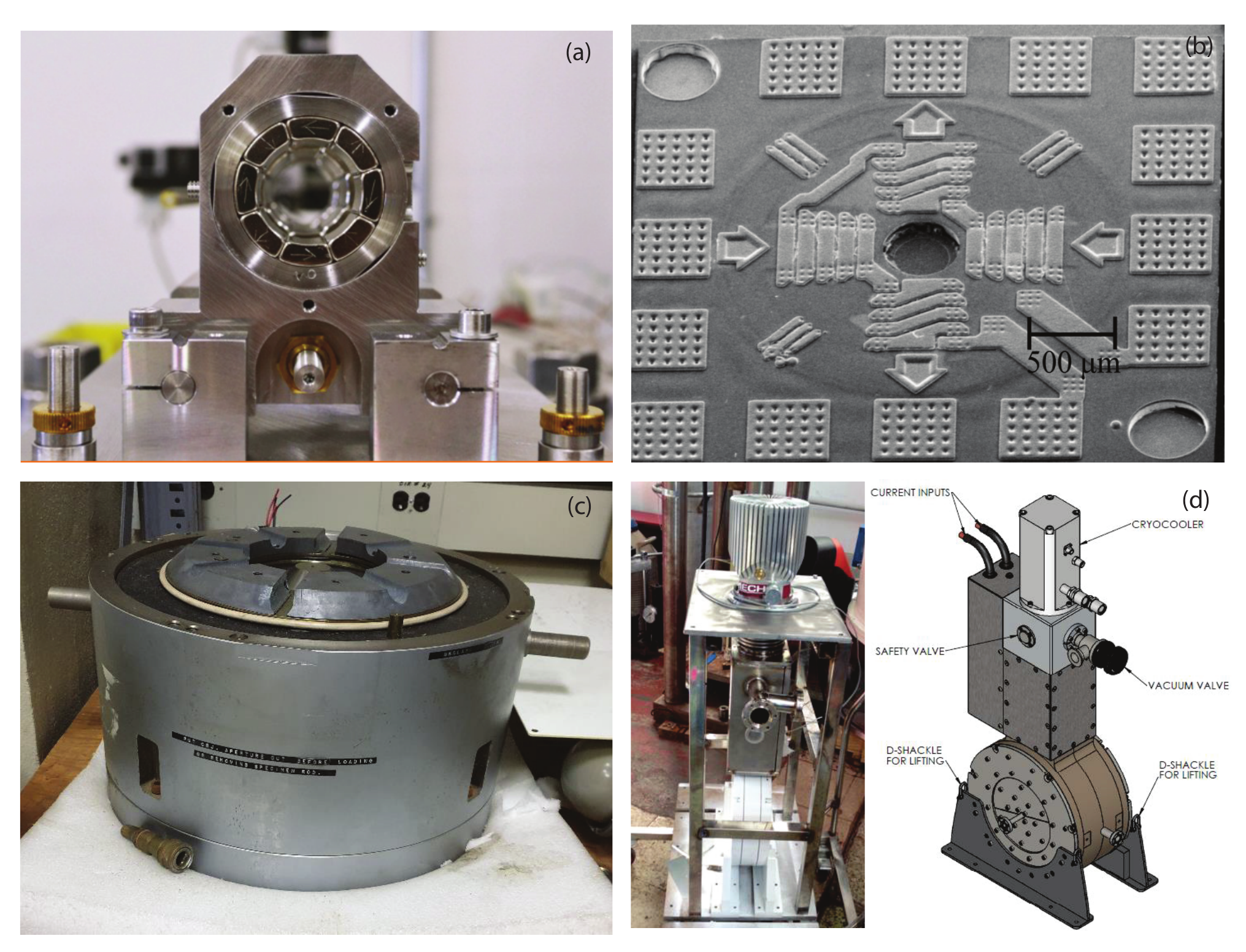}
\caption{(a) Pure permanent magnet quadrupole, (b) MEMS-based quadrupole (adapted from \cite{harrison:memsquads}), (c) normal conducting solenoid, and (d) superconducting solenoid lens (adapted from \cite{IHEP:HTS}).}
\label{Fig:electronlenses}
\end{figure}

Permanent magnet technology (either pure or hybrid) is a competitive candidate as it eliminates the need for the power supply and has no cooling requirement \cite{halbach:permanentmagnets}. Typically it represents a compact, vibration-free, vacuum compatible solution with the potential for larger focusing gradients. Long term demagnetization effects and the lack of tunability are the main challenges. Translating the lens along the beam axis is usually the only way to control the beam transport \cite{cesar16uem}. Another interesting opportunity driven by the rapid progress of MEMS technology is the possibility of growing on thin wafer an entire coil/yoke assembly (see \ref{Fig:electronlenses}(b)). The flat geometry significantly eases the cooling requirements. These magnets have been tested experimentally and hold the promise for very large field gradients~\cite{harrison:memsquads}.

\begin{figure}[ht]
\includegraphics[width=0.99\columnwidth]{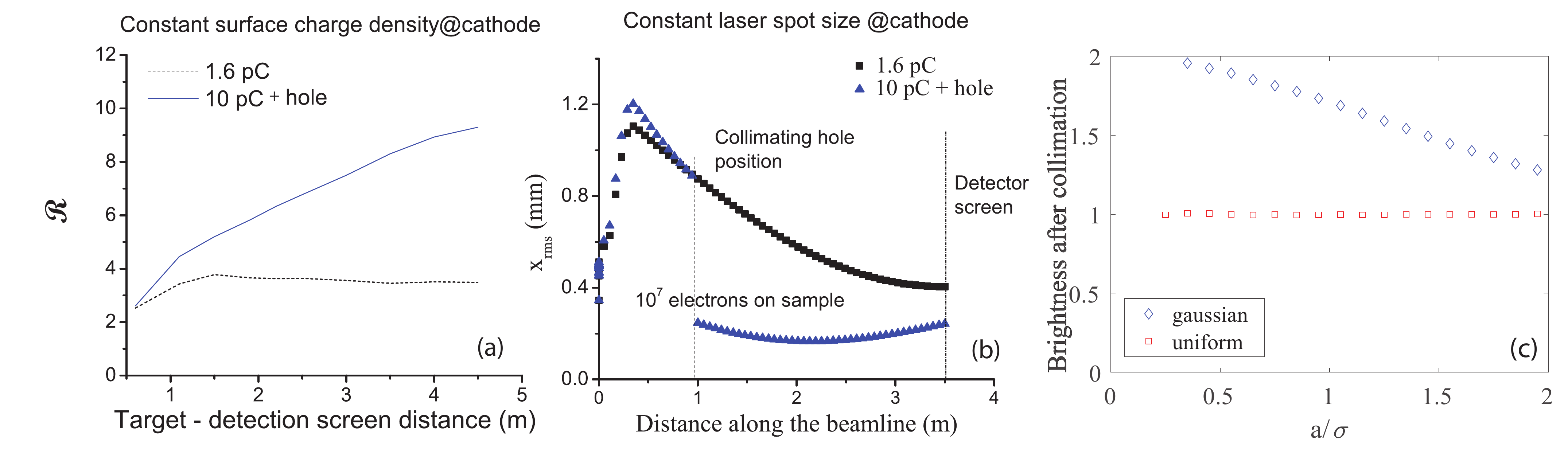}
\caption{a) UED resolving power as a function of the target-screen distance. The Bragg angle is assumed to be 3 mrad. b) Evolution of the transverse spot size along the beamline for the case with (blue) and without (black) the collimating hole. (adapted from \cite{musumeci2010high}) c) Average beam brightness improvements obtained by aperturing the beam for gaussian and uniform beam distributions as a function of hole size (normalized to rms spot size).}
\label{Fig:collimator}
\end{figure}

\paragraph{Collimation}
We conclude this section with a discussion on transverse collimation. Beam apertures have been employed in electron microscopes for a long time, both before and after the sample plane in the instrument, and can provide benefit to UED beamlines as well.  
Without the collimator, the dimensions of the probe beam depend on the beam dynamics and are sensitive to many operating parameters. A fixed aperture can decouple the probe area from the machine setup. Furthermore, depending on the spatial distribution of the beam, use of transverse collimation has been suggested to improve the beam quality mainly due to the fact that the beam brightness in the beam core is typically larger than the average beam brightness \cite{BazarovPRL}. Order of unity advantages can be obtained in this way as exemplified in Fig.~\ref{Fig:collimator}(c), where the ratio of the beam brightness before and after the collimation is shown as a function of aperture size (normalized to rms beam size). While for a uniform beam distribution, the amount of charge collimated balances the reduction in phase space volume keeping the total brightness constant, for a gaussian distribution an increase in brightness by a factor of two can be obtained. Such an effect becomes more evident in space charge dominated beams, where the fields and the forces at the center of the beam are quasi-linear. Collimation of the outer part of the beam, the so called buffer-charge, will eliminate most of space charge-induced emittance growth \cite{musumeci2010high}.

We can get a better understanding at how the collimator works to improve the quality of the patterns, by looking at the simulations in Fig.~\ref{Fig:collimator}. The cases reported start with different charge at the cathode, 1.6 pC and 10 pC, but have equal charge (1.6~pC) at the sample plane located 1 m from the cathode right after the collimator. 
In Fig.~\ref{Fig:collimator}(a) the simulation is performed by keeping the surface charge density at the cathode constant (i.e. the 10 pC beam has a larger spot size at the cathode). The diffraction camera resolving power $\mathrm{R}$ is generally improved using the aperture. The improvement is larger if we increase the sample-detector distance simply due to the fact that the apertured beam reaches a smaller spot size at the waist located at the detector screen. In another example Fig.~\ref{Fig:collimator}(b) shows the evolution of the spot sizes along the beamline comparing two cases where the cathode initial spot is kept constant at 500 $\mu$m (see Fig. \ref{Fig:collimator}(b). In this case the gain is approximately a factor of two in reciprocal space resolution at the detector screen. In both of these examples, this is due to the hole effectively removing the high-emittance particles from the beam, thereby cleaning up the transverse phase space. 

\subsection{Electron detection schemes}
\label{sectionII.e}

Electron detectors are a key element in a UED setup, as much as they are in electron microscopy. While most of the UED research efforts has been focused on beam generation and manipulation techniques, improvements of detection schemes, both in space and in sensitivity would have a tremendous impact on the technique, decreasing the integration times by decreasing the number of electrons needed for the experiments, and contributing to the elimination of the background and to an optimal SNR. 

To our advantage, electron detection has been studied for decades and produced a large amount of literature, mostly driven by electron microscopy. In the following we summarize the status of the field in UED. 

\subsubsection{Indirect electron detection schemes and efficiency}
\label{sectionII.d.1}

In conventional non-relativistic UED micro-channel plates (MCP) are used for direct amplification of diffracted keV electrons. The intensified electron flux is then converted by a scintillator to visible photons which are subsequently fiber-optically coupled to a high efficiency charge-coupled device (CCD) camera. It is relatively straightforward to achieve single-electron detection capability due to the large gain of the MCP and the high light collection efficiency of the fiber-optics coupling.

MCPs have also been tested for MeV electrons, obtaining high quality single-shot diffraction patterns \cite{Musumeci:2011ivba}. Blurring of the pattern was observed as a result of the large penetration depth of MeV electrons and the resulting excitation of secondary electrons in many surrounding micro-channels. It was also found that due to the active amplification process, the signal from the MCP has larger fluctuations which can be a concern in single-shot measurements where very small changes in the pattern are to be detected. Performance degradation of the MCP and fiber-optics after long term exposure to MeV electrons was not observed.

An effective alternative for the detection of MeV electrons is the use of optimized passive scintillator screens which are low cost and provide high electron-to-photon conversion efficiency and improved spatial resolution. A phosphor screen yields as many as a few thousand photons for each MeV electron due to the large penetration depth of MeV electrons. As an example, two recent papers reported calibration measurements showing greater than 10$^3$ photons per MeV electron from a Lanex Fine (a commercial version of phosphor P43) screen \cite{Glinec_2006, Buck_2010}. In fact, considering an energy loss rate of 1.2\textendash1.5 MeV cm$^2$/g for 1\textendash4 MeV electrons and a screen surface density of 34 mg/cm$^2$ corresponding to $\simeq$ 0.5 mm thickness, the total energy deposition by each electron is approximately E$_{loss}$=50 keV. For an optimal choice of phosphor material and screen composition, the efficiency in conversion of this energy into output visible photons is on the order of $\eta$=15$\%$\textendash25$\%$. Approximately half of these photons will exit from the screen side facing the CCD camera while roughly an equal amount exits from the back side. Since the photon spectrum is narrowly peaked at $h\nu=2.27$ eV (545 nm), we have $n_{scr}=(1/2) E_{loss} \eta/h\nu =1.7 $\textendash$ 2.8 \times10^3$ as an estimate of the number of photons emitted from each side of the screen per incident MeV electron. 

It becomes then important to maximize the collection efficiency of the optical system which images the detection screen onto the charge-coupled device. The collected solid angle of a lens with numerical aperture $N = f /D$ where $D$
is the diameter of the lens and $f$ its focal length is proportional to $1/N^2 (M+1)^2$ where $M$ is the 
magnification factor. At the same time in order to maximize the reciprocal space resolution, one would want to increase the magnification so that more pixels can be used to cover the same momentum transfer interval. For a given detector, the best situation is obtained when the size of the diffraction pattern at the screen is matched to the dimensions of the CCD array so that $M$ is close to 1 - and the collection angle maximized. For example, a 
scattering angle of 3 mrad from a 4 MeV beam energy corresponds to a momentum transfer $s$ up to 4  \AA$^{-1}$. If the CCD chip used has a vertical dimensions of 7~mm, then the diffraction pattern reaches its optimum width size 2.4~m downstream the sample.

With a properly designed lens coupling system whose collection efficiency is higher than 1$\%$ and a state-of-the-art CCD camera capable of single-photon detection, single-electron imaging is possible. This was demonstrated in the work from the UCLA group where diffraction spots from planes up to  $(800)$ were detected from a single crystal 20 nm gold sample in a single shot \cite{Li:2011ivba}. In order to further increase the photon yield per electron (and therefore use less sensitive cameras), fluorescent screens with larger phosphor density or thickness (higher electron-to-photon conversion efficiency) and still reasonably small point-spread-function (PSF) values could be used, for example the DRZ standard screen.

\begin{figure}[ht]
\includegraphics[width=0.9\columnwidth]{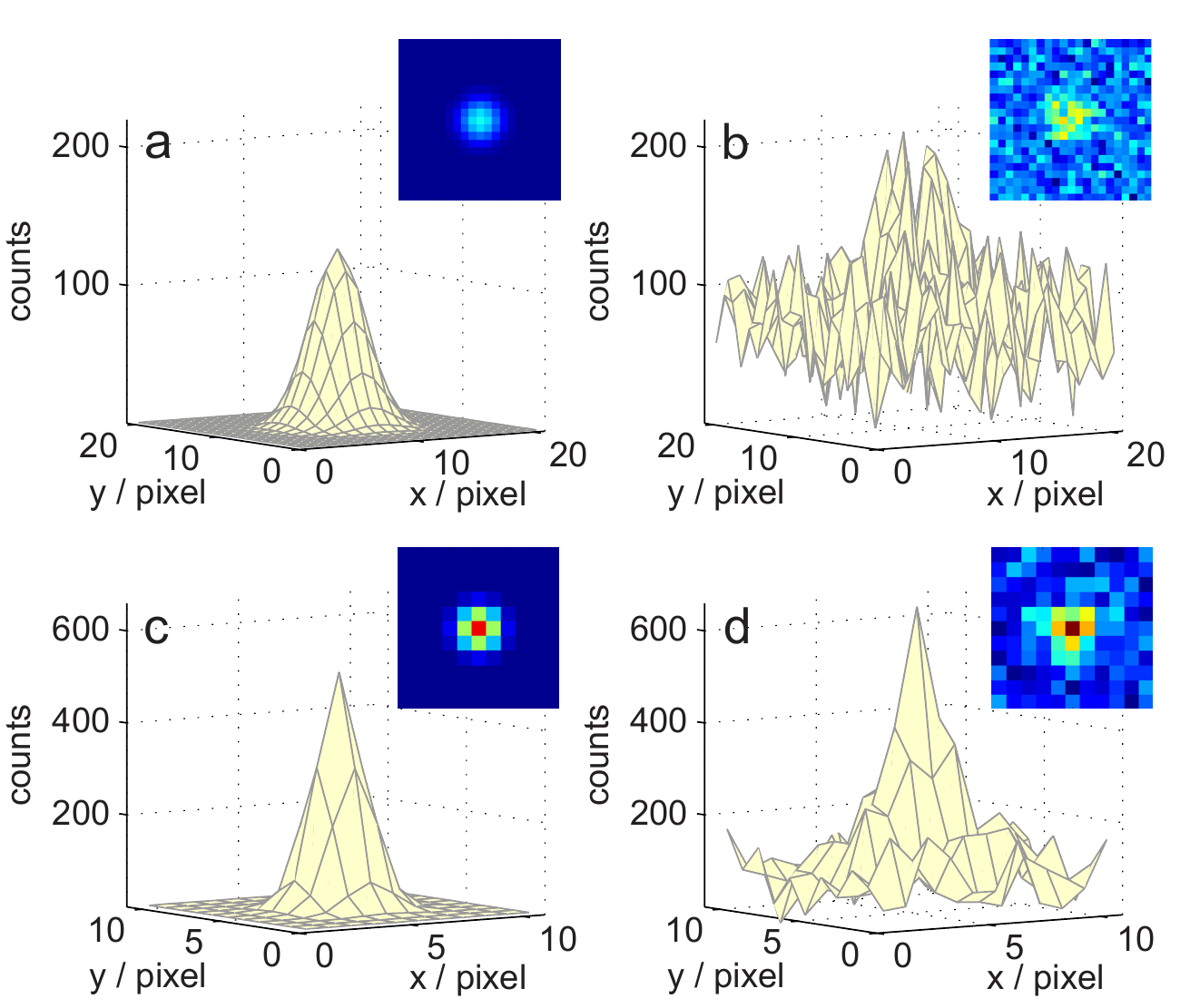}
\caption{The ideal PSF (left column) and its convolution with the camera readout noise (right column). (a) and (b) are for the no binning case, and (c) and (d) are for the 2 by 2 binning case. (adapted from \cite{Musumeci:2011ivba})}
\label{binning}
\end{figure}

Scintillator-based detection schemes offer high sensitivity, but also several shortcomings. First, the they suffers from image burn in. For example in P43, intense fluorescence can persist at a low level for minutes afterwards even though the fluorescence 1/e lifetime is 0.7 ms. This is disadvantageous when analyzing subtle differences in diffraction patterns. Faster scintillators are available, but generally exhibit low quantum efficiency. Second, and more important, a typical spatial resolution of a phosphor screens is on the order of 50-100 $\mu$m, limiting the reciprocal space (q-space) resolution of the system. For the detector employed  in the experiment from \citet{Li:2011ivba}, the PSF was around 64 $\mu$m, resulting from a combination of the phosphor grain size and the film thickness. 
High spatial resolution can be achieved at the expenses of detection efficiency, by utilizing very thin scintillating screens and high numerical aperture optics to collect the light. For example using a 20 $\mu$m YAG:Ce crystal with an in-vacuum infinity corrected microscope objective coupled to an in-air CCD recently the possibility of spatially resolving features in the beam down to 3 $\mu$m has been demonstrated \cite{maxson2017}. Trading off spatial resolution with sensitivity can be obtained by binning the image (see Fig. \ref{binning}).

The dynamic range of the imaging system is another important requirement, given the large intensity variation between different features in the diffraction pattern ( for example bragg peaks versus diffused scattering signal). An effective solution is to use a radially symmetric, variable neutral-density apodizing optical filter on the output side the phosphor screen, extending the system dynamic range by over 7 orders of magnitude. Similar large dynamic-range detection scheme has also been pursued and implemented for beam halos characterization in high electron accelerators~\cite{freeman_experimental_2019}. 

\subsubsection{Direct electron detection }
\label{sectionII.d.2}

\begin{figure}[ht]
\includegraphics[width=1\columnwidth]{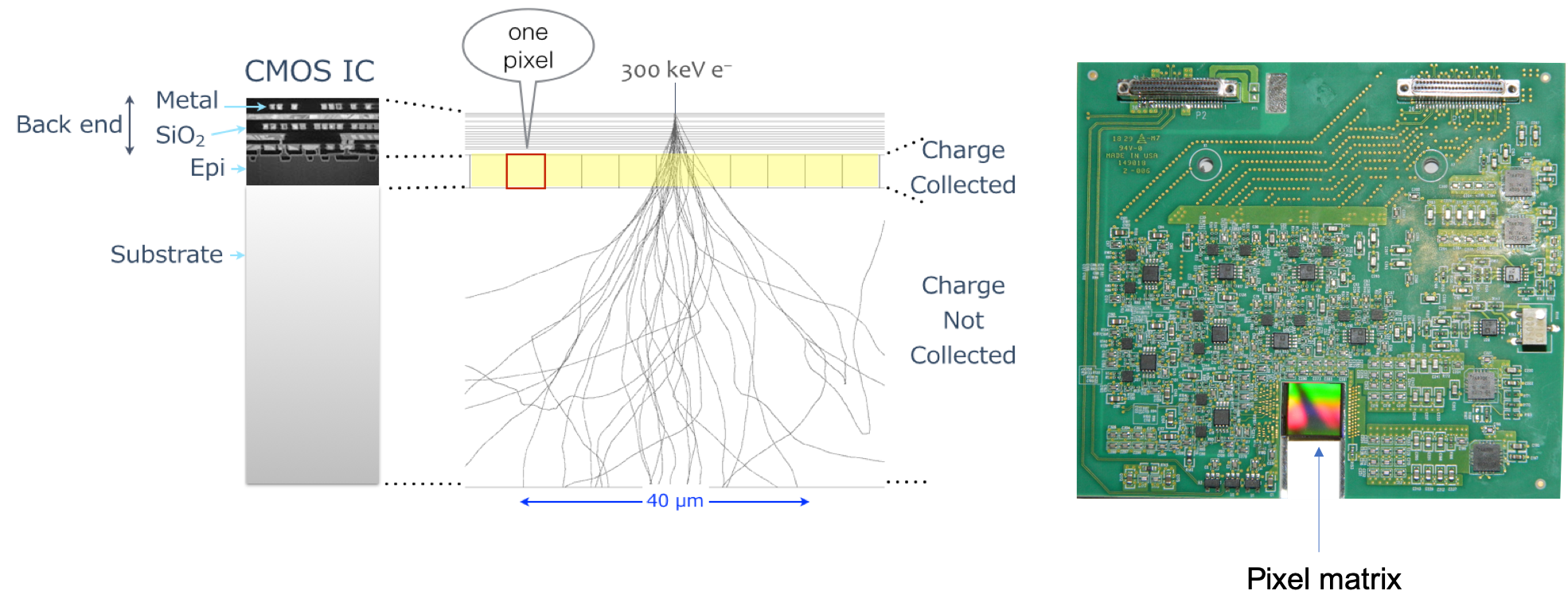}
\caption{Left: Principle schematic of direct electron detection. Electron-hole pairs formed in the p-doped epitaxial layer (Epi) by the beam passage form the image signal.
Right: A picture of the TEAM 1K direct detector assembly, including the detector and the in-vacuum electronics. Courtesy of Peter Denes.}
\label{DDD}
\end{figure}

Recently, active pixel sensor technology (APS) initially proposed for detectors in particle physics~\cite{turchetta_monolithic_2001} has been demonstrated and further developed for electron microscopy and diffraction~\cite{milazzo_active_2005}.
Here the electron beam impinges directly on the sensor (from top to bottom in Fig.~\ref{DDD}(a)), creating electron-hole pairs as it moves across. The charge created in the lightly p-doped epitaxial layer (Epi) diffuses towards a collection site (n-well diode). The signal level is proportional to the energy lost by the electron in the active p-doped epitaxial section (Fig.~\ref{DDD}).
In complementary metal-oxide-semiconductor (CMOS) APS (Fig.~\ref{DDD}(a)), transistors are implanted on the top of the Epi surface, and then connected through layers of metal and insulator (at the top of the structure) for pixel readout and zeroing. The entire structure is supported by a bottom (low-resistivity) thick substrate.
The thickness of the epitaxial layer defines the detector efficiency but also the transverse pixel size. The thicker the active region the larger the energy lost by the particle and the signal ($\sim$1000 $e-h$ pairs for 1 MeV beam through per 1 $\mu$m silicon). The same thickness also defines the spread of the electron lateral scattering, causing consequent broadening of the spatial response of single electron to clusters of pixels. The optimal thickness value depends on the electron beam energy. In MeV-class beams, with longer mean free path,  the epitaxial region is made as thick as 14~$\mu$m, in order to increase the detector efficiency~\cite{vecchione_direct_2017}, while for low energy electrons a few micrometers is enough. 

Direct electron detection provides unprecedented performance in terms of efficiency and resolution, which make it a very attractive technology for experiments with low illumination, such as electron microscopy and speficic UED modes, including gas-phase or nano-diffraction experiments. Thanks to the large number of e-h pairs for each electrons and the very low leakage current, the detector quantum efficiency (DQE) of such systems approaches 1~\cite{battaglia_characterisation_2010}. 
Furthermore, CMOS-based sensors have demonstrated spatial resolutions well below 10 $\mu$m, thanks to the development of back-thinned technology ~\cite{battaglia_characterisation_2010}. 
In UED-mode~\cite{vecchione_direct_2017}, the multi-channel electronics installed in very close vicinity with the sensor (Fig.~\ref{DDD}(b)) allows acquiring single-shot diffraction patterns at high speed and to correct for spatial (and temporal) jitters without compromising in acquisition times. Low dose images are accumulated and the undiffracted beam can be used for intensity calibration and shot-to-shot spatial alignment, optimizing resolution the same way the blurring from sample vibration is removed in TEMs.

A further advantage of high speed and single-electron sensitivity, is the possibility to perform cluster imaging  ~\cite{battaglia2009cluster}. In this mode, individual electron hits are counted. This modality assumes 
single-electron events per pixels and, therefore, require low dose per frame. Under this assumption, the image contrast and the LSF of the imaging systems can be considerably improved.

Another interesting development is the hybrid pixel array detector (EMPAD - electron microscope pixel array detector) developed at Cornell for scanning transmission electron microscopy \cite{Tate_EMPAD}. The 128$\times$128 pixel detector consists of a 500 $\mu$m thick silicon diode array bump-bonded pixel-by-pixel to an application-specific integrated circuit. The in-pixel circuitry provides a 1,000,000:1 dynamic range within a single frame, allowing the direct electron beam to be imaged while still maintaining single electron sensitivity.

\section{Measuring dynamics of matter in a solid state with bright electrons} \label{sectionIII}

    \subsection{Introduction}

The focus of this section is to provide quantitative tools and showcase the impact/potential of UED techniques in solid-state and material physics applications. For the sake of clarity, we will differentiate between ultrafast electron diffraction (UED) and ultrafast electron diffuse scattering (UEDS) signals, the former being associated with Bragg-peak electron scattering, and the latter, much weaker, resulting from electron scattering with the system phonon modes. 
The signals in these time-resolved crystallographic techniques 
have directly benefited from developments in enhanced beam brightness and shorter pulse duration that the last decade has brought. There is now a long list of extremely exciting examples of UED studies in solid state materials that probe a wide range of phenomena in most classes of materials, phases and microstructures (single crystal, polycrystal, monolayers and amorphpous/liquids). As we will show, increasingly complex and subtle phenomena have been visualized in recent years, and many important questions at the very center of condensed matter physics can now be addressed directly by UED.
 
Section~\ref{sec:scattering_theory} provides a summary of the theory of electron scattering in materials to provide a quantitative basis on which to understand both UED and UEDS signals.  In Sec.~\ref{sec:ued_experimental_req} we describe the main experimental requirements and constraints, including the determinants of signal-to-noise and issues of specimen preparation. 
Finally, in Sec.~\ref{sec:ued_examples} we present selected experimental results that exemplify some of the unique capabilities of the UED technique. 

\subsection{Summary of theory results for time-resolved electron scattering from crystalline solids}\label{sec:scattering_theory}

\subsubsection{Scattering from crystals including phonon excitations}

Under equilibrium conditions, atoms in crystalline materials fluctuate about their lattice positions in a manner that depends on the temperature and phonon band structure of the material. Following laser excitation, these atomic positions can change as a function of time in a number of distinct ways that have characteristic effects on the electron scattering intensity, $I(\mathbf{s})$.  Thus,  measurement of time-dependent electron scattering provides rich and detailed information on lattice transformations and phonon excitations as we summarize below. 

Following the most common perturbative treatment as given in ~\cite{Xu2005,warren1990}, the electron scattering intensity can be expanded in a Taylor series, $I(\mathbf{s})\approx I_0(\mathbf{s})+I_1(\mathbf{s})+\dots$, in the small atomic displacements associated with phonons.  The results of this expansion provide the framework that is most commonly used to analyze ultrafast electron scattering experimental data. 

\emph{Zeroth-order scattering: $I_0(\mathbf{s})$}

    The zeroth order term in the series expansion for $I(\mathbf{s})$ yields Bragg scattering modified by the lattice/phonon excitations:
    
    \begin{equation}\label{eqn:I0}
    I_{0}(\mathbf{s})\propto \delta\left(\mathbf{s-G}\right)\left|\sum_{\alpha}f_{\alpha}(\mathbf{s})\exp(-M_{\alpha}(\mathbf{s}))\exp(-i\mathbf{s}\cdot\mathbf{r}_{\alpha})\right|^2.
    \end{equation}
    where, as described in Sec.\ref{SSEDtheory}, $\mathbf{\alpha}$ is the index of each basis atom in the unit cell.  The \emph{anisotropic} Debye-Waller factor (DWF), $\exp(-M_{\alpha}(\mathbf{s}))$, depend on the  $M_{\alpha}(\mathbf{s})$ for each basis atom, which are given exactly by:
    \begin{equation}\label{eqn:DWF_anisotropic}
    M_{\alpha}(\mathbf{s}) = \frac{1}{4m_{\alpha}}\int\frac{\textup{d}\mathbf{k}}{(2\pi)^3}\sum_{j}|a_{j,\mathbf{k}}|^2|\mathbf{s}\cdot\hat{\mathbf{e}}_{j,\alpha,\mathbf{k}}|^2. 
    \end{equation}
    The phonon Eigenvectors $\mathbf{\hat{e}}_{j,\alpha,\mathbf{k}}$ describe the direction (or polarization) of the atomic displacements associated with the phonon mode of frequency, $\omega_{j,\mathbf{k}}$. The index $j$ specifies the phonon branch which labels the symmetry properties of the phonon mode (e.g. longitudinal or transverse and optical or acoustic modes). The mode amplitude $a_{j,\mathbf{k}}$ is related to the quantum number $n_{j,\mathbf{k}}$, the number of phonons with that index in the phonon field; $|a_{j,\mathbf{k}}|^2=\frac{\hbar}{m_{\alpha}\omega_{j,\mathbf{k}}}\left(n_{j,\mathbf{k}}+\frac{1}{2}\right)$.   The DWF depends on the amplitude of atomic motion associated with all phonon modes and suppresses the structure factor (and therefore scattering intensity).  This can be understood as a weakening of microscopic structural correlations due to vibrational atomic motion away from their average lattice coordinates in the material. The effect of the atomic displacements associated with phonon excitations on the intensity of Bragg scattering is to exponentially suppress diffraction peak intensities.  $M_{\alpha}(\mathbf{s})$ is a complicated expression in this general form, but its magnitude clearly scales as $\mathbf{s}^2$.  Thus, phonon excitation suppresses the intensity of peaks in a very characteristic way as a function of scattering vector.  In fact, Equation~\eqref{eqn:DWF_anisotropic} can be shown to reduce to $M_{\alpha}(\mathbf{s})$ = $2\pi^2\langle u_{\alpha}^2\rangle s^2$ in limit of isotropic atomic displacements. In this limit it is clear that the suppression of Bragg peak intensities depend on both the mean-square atomic displacements and the magnitude of the scattering vector squared. 
    
\emph{First-order scattering: $I_1(\mathbf{s})$}

    The first-order term in the expansion described above is called the \textit{thermal diffuse scattering} intensity and is given by the following expression:
    \begin{equation}\label{eqn:first_order_TDS}
    I_1(\mathbf{s})\propto \sum_{j}\frac{n_{j,\mathbf{s-G}}+\frac{1}{2}}{\omega_{j,\mathbf{s-G}}}\left| F_{1j}(\mathbf{s}) \right|^2.
    \end{equation}
     where $F_{1j}(\mathbf{s})$ is called the \textit{one-phonon structure factor} and is given by:
    \begin{equation}\label{eqn:1phonon}
    F_{1j}(\mathbf{s})=\sum_{\alpha}\frac{f_{\alpha}(\mathbf{s})}{\sqrt{m_{\alpha}}}\exp\left(-M_{\alpha}(\mathbf{s})\right)\left( \mathbf{s}\cdot\mathbf{\hat{e}}_{j,\alpha,\mathbf{k}} \right)\exp\left( -i\mathbf{s}\cdot\mathbf{r}_{\alpha} \right).
    \end{equation} 
   This term in the expansion has a very different character than $I_{0}$.  $I_{1}$ is non-zero at all scattering vectors, not just at scattering vectors that satisfy the Laue condition.  Equation~\eqref{eqn:first_order_TDS} shows that $I_{1}$ scattering at $\mathbf{s}$ is exclusively due to phonon excitations with wavevector $\mathbf{k=s-G}$, where $\mathbf{G}$ is the reciprocal lattice vector associated with the closest Bragg peak.  Thus $I_{1}$ provides momentum-resolved information on phonon excitations in the crystal.   
   
\begin{figure}[ht]
\includegraphics[width=1\columnwidth]{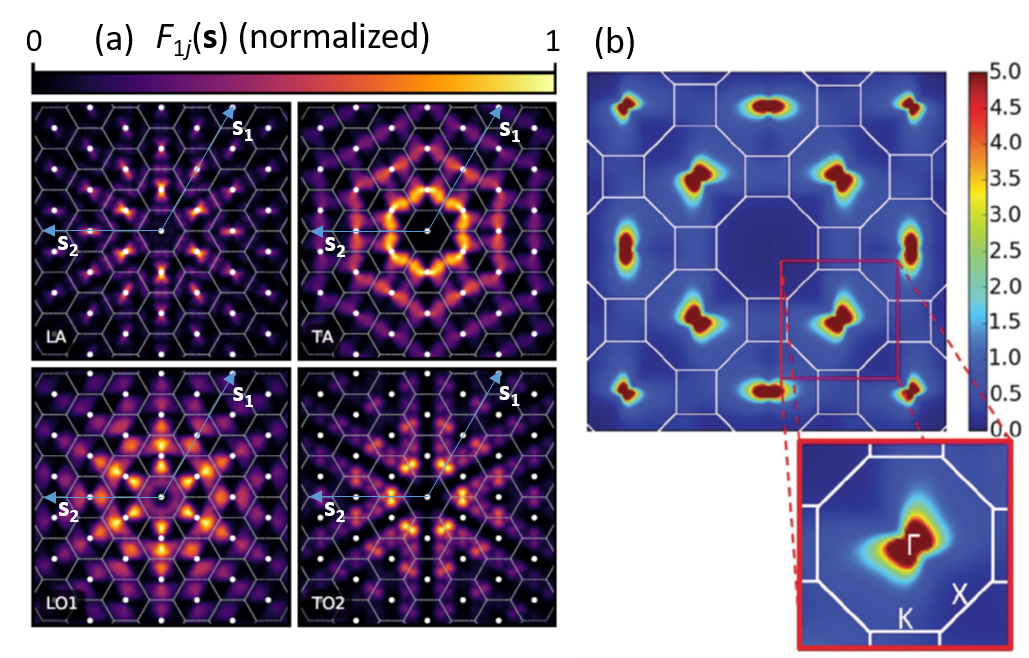}
\caption{Phonon diffuse scattering in materials a) The relative strength of the single phonon structure factor, $F_{1j}(\mathbf{s})$, as a function of scattering vector for 4 different in-plane phonon branches of graphite; longitudinal acoustic (LA), transverse acoustic (TA), longitudinal optical (LO1) and transverse optical (TO2). The hexagonal in-plane Brillouin zone surrounding each Bragg peak is indicated.  Two (aribitrary) scattering vectors, s1 and s2, are shown to indicate the tendency of diffuse scattering features to extend along the scattering vector direction for longitudinal phonons and extend orthogonal to the scattering vector direction for transverse phonons due to dot product in Equation~\eqref{eqn:1phonon}. Adapted from \citet{ReneDeCotret2019} b) Computed diffuse scattering from all phonon modes in crystalline Au at a temperature of 300 K.  Brillouin zone boundaries are indicated with white lines. Adapted from \citet{Chase2016}}.
\label{fig:1phonon_calc}
\end{figure}

   Thermal diffuse scattering $I_1(\mathbf{s})$ gives detailed, wavevector-resolved information about the lattice structural fluctuations in terms of the phonon mode amplitudes $\frac{n_{j,\mathbf{s-G}}}{\omega_{j}(\mathbf{s-G})}$. This term is weighted by $F_{1j}(\mathbf{s})$. The form of $F_{1j}(\mathbf{s})$ is very similar to Eqn.~\eqref{eqn:scattering_intensity} except for an additional factor of $\mathbf{s}\cdot\mathbf{\hat{e}}_{j,\alpha,\mathbf{k}}$.  This factor gives distinct structure to $F_{1j}(\mathbf{s})$ (and therefore also $I_1(\mathbf{s})$) through the set of phonon Eigenvectors, $\{\mathbf{\hat{e}}_{j,\alpha,\mathbf{k}}\}$,  leading to regions of reciprocal space where $F_{1j}(\mathbf{s})$ vanishes if $\mathbf{s}\perp \mathbf{\hat{e}}_{j,\alpha,\mathbf{k}}$. The $\mathbf{s}$ dependence is contained through its relation to $\mathbf{k}$ and $\mathbf{G}$ which is $\mathbf{s}=\mathbf{G}+\mathbf{k}$. The single-phonon structure factor is a $\mathbf{s}$--dependent weight for each phonon contribution to the total diffuse intensity $I_1(\mathbf{s})$. Generally, the polarization vectors $\hat{\mathbf{e}}$ are best computed using density-functional methods for real material systems. Some examples of calculations of $F_{1j}(\mathbf{s})$ and thermal diffuse scattering are shown in Fig.~\ref{fig:1phonon_calc}. 
   
   The perturbative single phonon scattering theory presented above is an attractive starting point for understanding electron scattering from materials, but is an approximation.  The limits of this approximation and the more general multi-phonon theory has recently been fully described by Zacharias and coworkers ~\cite{zacharias2021efficient, zacharias2021multiphonon}.

\subsubsection{Time-dependent factors in Bragg scattering: $I_0(\mathbf{s},t)$}
\label{SStheorytime}
The results presented above provide a quantitative basis on which to understand ultrafast electron scattering signals from single crystal materials.  Specifically, how UED provides a window on nonequilibrium structural dynamics within a well defined phase and can also provide details on the dynamics associated with the transformation between phases. Here we identify how various materials physics processes lead to qualitatively distinct changes in electron scattering intensity.

    \emph{Order and periodicity:} Phase transitions that yield a change in lattice, charge or orbital order will tend to modify the set of reciprocal lattice vectors; $\mathbf{G\cdot R}_{n}=2\pi\times\textup{Integer}$. Transformations that change the space group/symmetry result in a different set of reciprocal lattice vectors and the appearance/disappearance of Bragg peaks from a diffraction pattern. Transformations that only modify the lattice constants (e.g. thermal expansion or strain), but not the space group/symmetry, re-scale the existing set of reciprocal lattice vectors and result in shifts of Bragg peak positions, \emph{not} new peaks. Strain can also be probed in electron diffraction patterns through peak broadening and asymmetry.

    \emph{Directed and coherent motion:} Optical excitation can result in the coherent, directed motion of atoms across many or all unit cells in a material without necessarily changing the space group/symmetry of the crystal.  This motion may be associated with a coherently excited vibration (oscillation) or the structural pathway along which the material evolves between two phases. Motion of this type changes the atomic coordinates, $\mathbf{r}_{\alpha}$, which modulate the interference condition in the structure factors, $|F_{0}(\mathbf{s})|^2$. Changes in structure factor due to atomic motion like these are directly observed as changes in the intensity of Bragg peaks across the entire detector in a manner that is characteristic of the motion.  The impacts are not confined to a single Bragg peak;  relevant information is distributed throughout the pattern. Thus, a full characterization of the motion will --in general-- require the time-dependence of a sufficiently complete set of diffraction peaks, not just a single one. For example, a coherent optical phonon will modulate the $\exp(-i\mathbf{s\cdot r}_{\alpha})$ phase term of the structure factor $F_{0}(\mathbf{s})$. This effect will yield a characteristic intensity modulation at the frequency of the phonon, but only in diffraction peaks associated with reciprocal lattice vectors with a non-zero projection onto the atomic motion $\mathbf{u}_{\alpha}$; i.e. those $\mathbf{G}$ for which $\mathbf{G\cdot u}_{\alpha}$ is nonzero.  

    \emph{Bonding, valency, orbital order and atomic form factors:} In the solid-state, atomic scattering factors are not necessarily isotropic due to chemical bonding and orbital ordering that is present. Atomic form factors for electron scattering, $f_{\alpha}(\mathbf{s})$, are sensitive to details of the valence charge distributions, in particular at small scattering vectors where these changes tend to be largest \cite{zheng2009aspherical}. The charge state (valency) of an atomic species also impacts strongly on the form factor.  Thus, photo-induced changes to bonding, orbital occupation and valency can yield distinct and measurable changes in scattering intensity through changes to the atomic scattering factors themselves. Such effects are distinct from a rearrangement of the atomic coordinates within the unit cell and can, in principle, be distinguished by the very different characteristic dependence on $\mathbf{s}$ that is manifested through the structure factors (Eq. 29) \citet{Otto2019}.   
    
    \emph{Debye-Waller factor:} Thermal fluctuations in atomic position have a characteristic impact on Bragg peak intensities through the Debye-Waller factor.  These effects are given rigorously by Eq. 29 and 30, but are difficult to physically interpret in this form. In UED one often considers the term $I_0(\mathbf{s},t-t_0)/I_0(\mathbf{s},t_0)$ which can be given by $\sim\exp(-2 M_{\alpha}(\mathbf{s}))$. The average change in $\langle u^2 \rangle$ can then be determined in the simple isotropic case using the following
    \begin{equation}\label{eqn:DW_intensity}
    -\ln\left(\frac{I_0(\mathbf{s},t-t_0)}{I_0(\mathbf{s},t_0)}\right) =
    2\pi^2\left(\langle \Delta u(t-t_0)^2 \rangle \right)s^2.
    \end{equation}
    Equation~\eqref{eqn:DW_intensity} provides a detailed view of the average transient lattice heating of the material following laser excitation. The measurement of Bragg peak intensities can be converted to an average change in the statistical distribution of $u^2$, the time-scale of which is commonly of interest in addition to the magnitude.

\subsubsection{Time dependent factors in the diffuse intensity: $I_1(\mathbf{s},t)$}
\label{SSdiffusetheorytime}
    \emph{Phonon mode amplitudes in $I_1(\mathbf{s})$:} Unlike the DWF, diffuse intensity provides a momentum-resolved picture of phonon mode amplitudes $\frac{n_{j,\mathbf{k}}}{\omega_{j,\mathbf{k}}}(t-t_0)$ if the single-phonon structure factors $F_{1j}(\mathbf{s})$ are known to a reasonable degree.  The diffiuse intensity at scattering vector $\mathbf{s}$ reports exclusively on phonons with wavevector $\mathbf{k = G - s}$.   Changes in diffuse intensity report on the changes in phonon mode amplitude that can result from either changes in occupancy $\Delta n_{j,\mathbf{k}}(t-t_0)$ (usually phonon \emph{emission}) and/or changes in mode frequency $\omega_{j,\mathbf{k}}(t-t_0)$ everywhere in the Brillouin zone. For the typical case where mode frequencies are relatively unchanged by photo-excitation the transient diffuse intensity at the detector $\Delta I(\mathbf{s},t-t_0)$ is given by
    \begin{equation}
    \Delta I_1(\mathbf{s},t-t_0)\propto \sum_{j}\frac{\Delta n_{j,\mathbf{k}}(t-t_0)}{\omega_{j,\mathbf{k}}(t_0)}|F_{1j}(\mathbf{s},t_0)|^2.
    \end{equation}
    In the time-domain, the measured rate of phonon emission $\Delta n_{j,\mathbf{k}}(t-t_0)$ initiated by photo-excited electrons contains information about the electron-phonon coupling vertex at that wavevector. Diffuse intensity measurements, when appropriately related to the phonon system, has the potential to yield dynamics of phonon modes and band structures analogous to way the angle-resolved photo-electron spectroscopy yields dynamics of electronic states and bands.  
    
    \subsubsection{Electron beam requirements and considerations}
    
    In UED and UEDS experiments there are three primary practical considerations related to electron beam parameters. First, the electron beam spot-size at the sample determines the spatial resolution of the probe and may limit the maximum momentum resolution (see Fig.\ref{fig:B4dVsSS}).  As a minimum requirement, this resolution must be finer than the laser pump spot-size by at least a factor of two to maintain relatively homogeneous excitation conditions throughout the probed volume (specific experimental considerations can make this requirement more stringent). However, the in-plane grain/crystal size may effectively set the required spatial resolution in single crystal experiments. Crystal, grain or domain sizes can be as small as a few nanometers, 
    Second, the electron beam spot size at the detector (placed at a post-specimen diffraction plane) effectively determines the momentum-resolution in single crystal experiments. In a UED experiment momentum resolution must be sufficient to resolve/differentiate Bragg peaks; i.e. the momentum resolution at the detector, $\Delta s$,  must be a fraction of the separation of adjacent reciprocal lattice vectors. $\Delta G$. In UEDS experiments, the Bragg peaks need to be well resolved, occupying a minimum of the Brillouin Zone that surrounds each peak.  Phonon-diffuse intensity, $I_1(\mathbf{s})$ (Eq. ~\ref{eqn:first_order_TDS}), is much weaker than Bragg peak intensity and is difficult to seperate from $I_0(\mathbf{s})$ where they overlap strongly.  That is, phonons with wavevector $k<\Delta s$, are typically not measurable in a UEDS experiment. Third, bunch charge and accumulation conditions place limits on signal detection.  We treat this third consideration at some length in Sec. ~\ref{sec:ueds_snr} below.
    All three primary electron beam considerations are interdependent and determined by the source brightness, as described earlier in Sec.~\ref{SS},~\ref{section_coherence_length},
 \ref{sec:brightness} and \ref{sourcesize}.

\subsection{Experimental requirements}\label{sec:ued_experimental_req}
In this section we will introduce the important considerations regarding UED experiments on solid-state specimens. These are sample preparations methods (Sec.~\ref{sec:ueds_sample_prep}), laser excitation conditions (Sec.~\ref{sec:ueds_laser_excitation}), signal detection and noise considerations (Sec.~\ref{sec:ueds_snr}), sample reversibility considerations in multi-shot experiments on the same sample (Sec.~\ref{sec:ueds_reversibility}) and details pertaining to the handling and processing of UED measurement data (Sec.~\ref{sec:ueds_data_handling}).  

\subsubsection{Sample preparation methods}\label{sec:ueds_sample_prep}
\begin{figure}[ht]
\includegraphics[width=1\columnwidth]{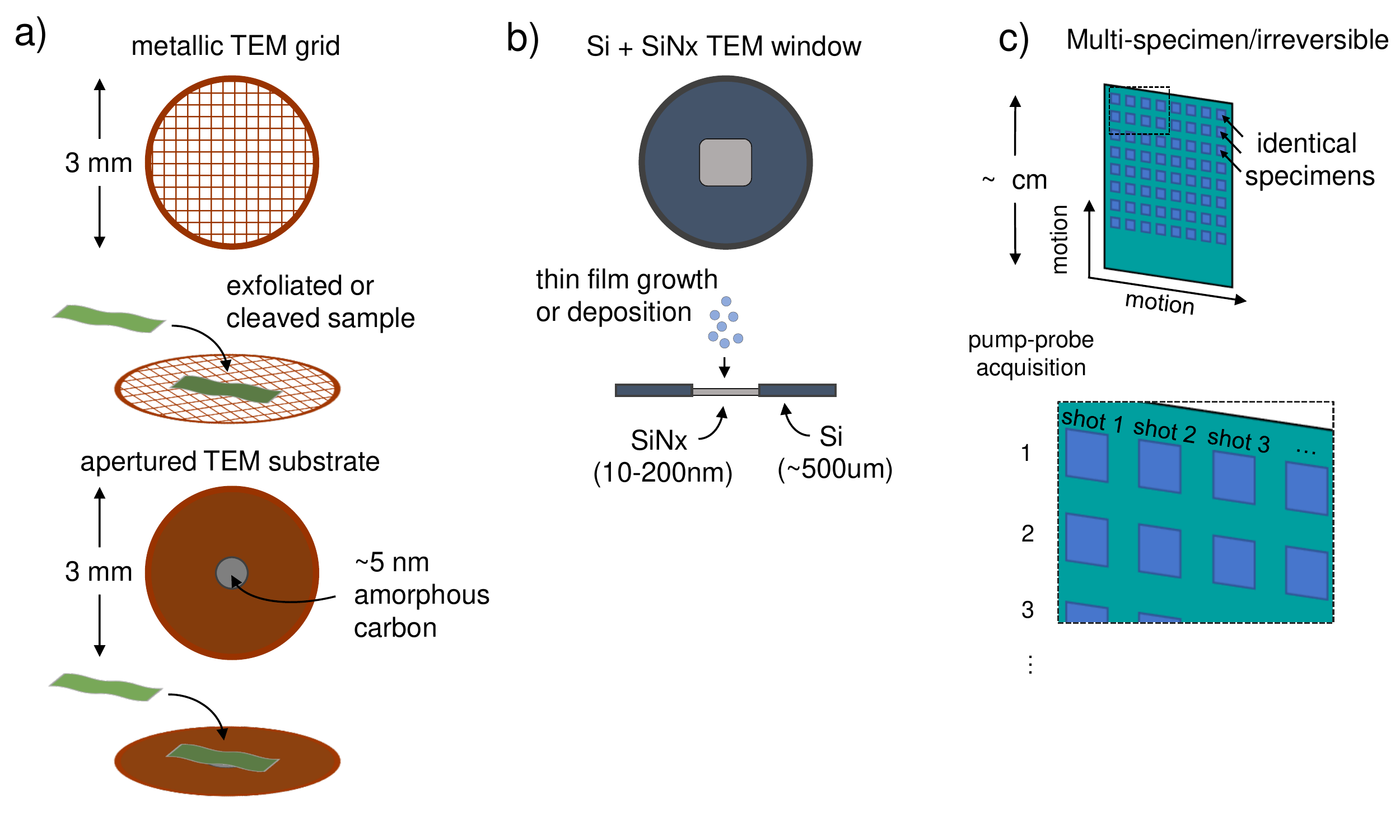}
\caption{Examples of common sample types in solid state ultrafast electron scattering. a) Metallic wire grids (usually Cu) provide a mesh substrate onto which thin single crystal flakes can be placed. b) Etched Silicon window with Silicon Nitride forming a thin transparent region.  Powder samples can be grown using various deposition techniques. c) Large scale sample concept for a single-shot or irreversible experiment, where each individual and nominally identical sample region can only be pumped and probed for one shot.}
\label{fig:ueds_samples}
\end{figure}

UED experiments build on many decades of developments in conventional electron microscopy and have similar sample requirements.  The kinematical approximation for Bragg peak intensities presented above is in quantitative agreement with those measured in electron diffraction patterns of single crystalline specimens only for nm-scale thicknesses.  Thicker specimens require dynamical (multiple scattering) diffraction calculations if a truly quantitative determination of the changes to structure factors is desired.  Thus, to obtain easily interpreted results there is a strong incentive to perform UED experiments on very thin specimens. Such specimens typically make use of standard substrates that have been developed and employed to support samples in transmission electron microscopes. Some typical examples are shown in Fig.~\ref{fig:ueds_samples}. Generally the substrates must be transparent to electron beams at the relevant energies. Examples include metallic wire grids to support films and crystalline flakes, silicon nitride membrane windows and amorphous carbon apertures. Depending on the exact substrate details the overall electron beam transmission can be in the range of 20-90\%. The main requirements for the substrate is that they are sufficiently large in area to accommodate the relatively large beams employed in UED and thus maintain sufficient scattering intensity signal and adequate thermal conductivity to transport heat out of the excited area sufficiently fast (discussed further below). Recent developments in ``nanoprobe" UED~\cite{ji_nanoued_2019} have produced nanometer scale beams which are expected to be a significant step forward in effectively probing small area samples while maintaining beam brightness. Irreversible or single-shot experiments often require larger-format sample configurations with the in-situ ability to translate the sample between shots so that a new area of sample is pumped and probed (Fig.~\ref{fig:ueds_samples}c).  More delicate samples such as organic crystals, air-sensitive materials and those for which the management of thermal dissipation is critical, may require completely customized solutions for sample preparation and mounting. 

Thin film deposition techniques are well suited to grow material specimens for UED interrogation.  Electron beam deposition, pulsed laser deposition, plasma-enhanced chemical vapour and atomic layer deposition (amongst other techniques) have been used to grow materials ranging from elemental metals to complex oxides.  However, these approaches tend to yield fine-grained polycrystalline films that give Debye-Scherrer type powder electron diffraction patterns. Single crystal specimens, by contrast, are usually prepared by mechanical exfoliation or ultra-microtomy \citet{eichberger2013sample, liu2020disassembling}, which can yield large area samples down to single monolayer thicknesses. Layered materials are particularly well suited to these methods. Certain materials (commonly semiconductors) where extensive nanofabrication progress has been made can be precision etched over a sufficiently large area down to sub-100 nm thicknesses (Si, Ge, GaAs). Some of these are in fact commercially available but are very expensive and fragile. A current technical limitation on the epitaxial growth of single crystal samples for UED is the lack of electron beam transparent single crystalline substrates that are compatible with these techniques (molecular-beam or other). Further work in this area holds the promise of both producing more single crystals to be studied, but also consistent sample-substrate interfaces for heat dissipation. 

\subsubsection{Laser excitation conditions}\label{sec:ueds_laser_excitation}
    One of the primary advantages of ultrafast electron scattering (compared to x-ray scattering) in transmission experiments on solid-state materials is the excellent match between typical optical absorption depths and the sample thicknesses for which kinematical (or quasi-kinematical) scattering applies. At near-IR and visible wavelengths skin depths are on the order of 10 nm in metallic films, with absorption lengths increasing to 100s of nm for above bandgap excitation in semiconductors and insulators.  Thus, it tends to be rather straightforward to design transmission geometry experiments in which the electron beam probes a nearly homogeneously excited volume of material.  Large signals from homogeneously excited volumes significantly simplifies data analysis and interpretation. 

\subsubsection{Determinants of signal detection; shot-noise limits}\label{sec:ueds_snr}
    Beam brightness has been a primary motivator behind the development of new pulsed electron beam sources for UED. This is because of signal-to-noise (SNR) in a UED experiment is fundamentally limited by beam brightness. We briefly discuss SNR considerations at a general level as they apply to the measurement of pump-induced changes in ultrafast electron scattering intensity from solid state samples. These considerations will serve to provide further motivation for continued improvements in electron beam brightness.
    
    In time-resolved scattering and diffraction, the differential intensity $\Delta I/I$ is almost always considered and the SNR of a measurement places a limit on the magnitude of the optically induced change in scattered intensity, $\Delta I$, that can be reliable determined~\cite{Kealhofer2015}. The average number of electrons detected at a given scattering vector, $\langle N_e \rangle$, is given by $\langle N_e \rangle = \eta p_{\mathbf{s}} Q N$, where $\eta$ is the quantum efficiency of the detector, $p_{\mathbf{s}}$ is the probability of scattering at vector $\mathbf{s}\propto |f(\theta)|^2$, $Q$ is the number of electrons per pulse (bunch charge) and $N$ is the number of accumulated pulses. $N$ is the product of the experimental repetition rate $f_{\textup{rep}}$ and the total signal integration time $T$. $\langle N_{e} \rangle$ describes the available ``signal" mapped at $\mathbf{s}$ on the detector and is primarily determined by the source brightness and the scattering cross-section $p_{\mathbf{s}}$. The signal is subject to a number of relevant noise terms, which are discussed next.

\underline{Shot noise:} This is determined directly from the counting statistics $\sigma_{\textup{shot}}(Q,f_{\textup{rep}},T) = \sqrt{\langle N_e \rangle}$. The relationship between detector counts and ``single electron detection instances" varies depending on the detector type, but Poisson statistics on a per pixel or per region-of-interest basis usually still apply. 

\underline{Source noise:} This term depends on the noise properties of the electron source used for the experiments, characterized by a noise spectral density $\alpha_{\textup{source}}$ and is given by $\sigma_{\textup{source}}(Q,f_{\textup{rep}},T)=\alpha_{\textup{source}}\langle N_e\rangle/\sqrt{T}$.

\begin{figure}[ht]
\includegraphics[width=1\columnwidth]{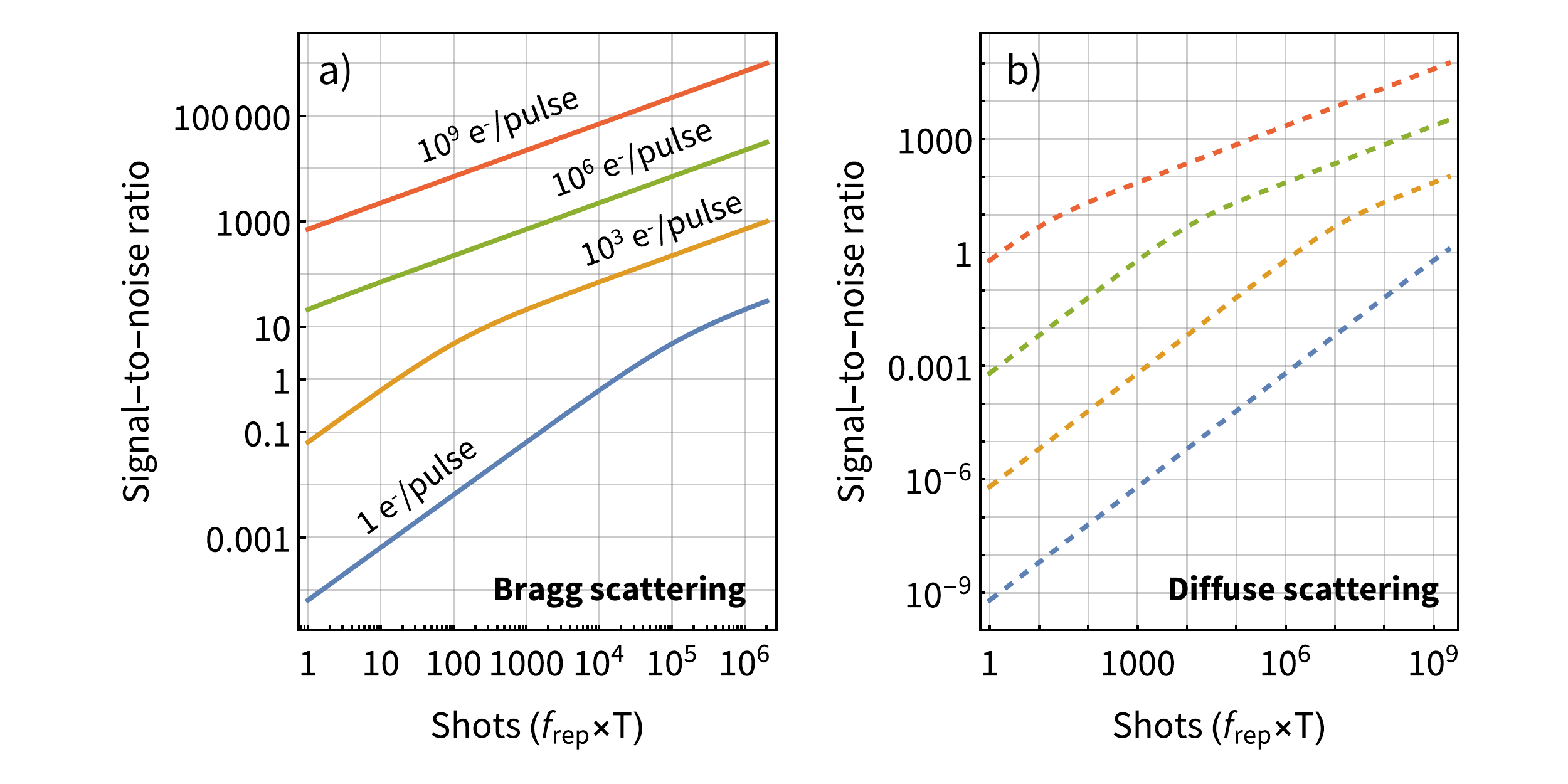}
\caption{Signal to noise considerations in typical solid state scattering experiments. a) SNR of Bragg scattering as a function of total collected shots ($f_{\textup{rep}}\times T_{\textup{experiment}}$) for various bunch densities using typical scattering and SNR paramters.  b) SNR for diffuse scattering using identical SNR parameters as a) but with a scattering probability $p_{\mathbf{s}}$ 10$^{-5}$ smaller than Bragg scattering.}
\label{fig:ueds_snr}
\end{figure}

\underline{Detector noise:} This term includes, gain noise $\sigma_{\textup{gain}}$, pixel integration/binning noise $\sigma_{\textup{int}}$ and readout noise $\sigma_{\textup{readout}}$. All of the relevant noise terms add in quadrature. The total signal to noise is expressed as
\begin{equation}
\textup{SNR} = \eta p_{\mathbf{s}}Q N\Big/\sqrt{\sigma_{\textup{shot}}^2 + \sigma^2_{\textup{source}} + \sigma_{\textup{gain}}^2+\sigma^2_{\textup{int}}+\sigma^2_{\textup{readout}}}     
\end{equation}
In typical solid-state samples with thickness in the range of 10-100 nm, the Bragg scattering probability is $p_{\mathbf{s=G}}=I_{\mathbf{s=G}}/I_{\textup{tot}}\sim 10^{-3}$ (for a single Bragg peak). Figure~\ref{fig:ueds_snr} a) shows the SNR as a function of accumulated electron bunch shots ($f_{\textup{rep}}\times T$) for Bragg scattering for various bunch charges $Q$. Single-shot Bragg scattering becomes possible with $10^5$ electrons. For the typical diffuse scattering, shown in Figure~\ref{fig:ueds_snr}~b), scattering probabilities are $p_{\mathbf{s=G+k}}=I_{\mathbf{s=G+k}}/I_{\textup{tot}}\sim 10^{-7}-10^{-8}$, many orders of magnitude lower than Bragg scattering. For these intensities, many shots must be collected to achieve necessary SNR. Interestingly single-shot UED requires bunch charges on the order of $10^9$ electrons per pulse. As practical consideration, in typical UED setups the shot noise contribution dominates over source stability high-performance detectors noise.

\subsubsection{Heat dissipation and limitations in multi-shot experiments}\label{sec:ueds_reversibility}

Transmission ultrafast electron diffraction experiments are performed on thin-film specimens that are susceptible to heat accumulation effects.  In pump-probe spectroscopy, thin film specimens are often deposited onto thick optically transparent substrates to efficiently remove heat from the laser excited film. In UED experiments the same approach cannot be employed, since the total film thickness must typically be kept below $\sim$100 nm;  heat must be removed via transport in the plane of the film rather than normal to the film. Convective cooling via air is also not effective in a vacuum environment. In a pump-probe experiment, energy is deposited at a rate equal to $F\times f_{\textup{rep}}$, where $F$ is the absorbed pump-fluence (mJ/cm$^2$) and $f_{\textup{rep}}$ is the pulse repetition rate. For a given $F$, the rate of thermal transport of pump-laser deposited energy out of the excited region will, in practice, set some limit on the laser-excitation repetition rate that can be used in an experiment. As a result, SNR in solid-state UED cannot be increased arbitrarily through the use of higher repetition rate sources.  Given the maximum repetition rate determined by heat dissipation consideration, SNR improves directly with Q and T as described in the previous section.  This provides a strong argument for continued improvements in electron beam brightness and stability as primary enablers of future advances in UED.      

There are, however, a number of effective and proven strategies to enhancing the rate of in-plane thermal transport. For truly ``free-standing" thin samples in the quasi-2D limit, a useful model to understand the tradeoffs is provided by the equation~\cite{Jager2018ata}:
\begin{equation}\label{eqn:t_recovery_2D}
    t_{r} = \frac{w^2}{\kappa}\left(\frac{T_0}{T_f}-1\right).
\end{equation}
In Eqn.~\ref{eqn:t_recovery_2D}, $t_r$ is the relaxation or recovery time, $w$ is the width of the pump-beam (excited region), $\kappa$ is the thermal conductivity, $T_0$ is the initial excited effective temperature and $T_f$ is the final temperature. The cooling time $t_r$ scales with square of the width of the excitation region, $w^2$. Thus, nanoprobe setups promise a step forward in this regard because the laser-deposited energy can diffuse out of the probes region on potentially a nano-second time-scale, allowing for repetition rates into the several MHz and potentially into the GHz range. In addition, more complex specimen geometries can be used to dramatically increase thermal transport out of the laser excited region and reduce cooling times between laser shots. It is only necessary that the probed region be electron beam transparent.  The region surrounding this 'window' can be as thick as desired and thermally engineered. TEM sample supports based on Si:SiN nano-membranes provide an excellent solution in this respect.  Window sizes and membrane thicknesses can be chosen to optimize SNR and thermal transport conditions leading to cooling rates somewhere between a truly 2D film and the conditions typically employed in spectroscopy.  

To ensure that appropriate steady-state conditions are present in solid-state samples during pump-probe UED experiments one can follow the evolution of the UED patterns at negative pump-probe time delays (i.e. probe arriving before the pump) over the course of an experiment.  Changes in these patterns as a function of lab time can clearly indicate that the sample is deteriorating due to repeated laser shots.  In addition, negative time delay patterns can indicate if an inappropriate or unexpected steady state condition is achieved at the pulse repetition rates being used in the experiments.  If so, modifications to the accumulation conditions can be made accordingly.

\subsubsection{Data processing for solid-state scattering}\label{sec:ueds_data_handling}

    Efficient handling of large experimental datasets is essential for UED experiments.  The raw data typically comprises a sequence of pump-probe delay time stamped diffraction images that can easily exceed hundreds of Gigabytes.  
    Basic data reduction steps include the removal of artifacts specific to the camera and experiment geometry that are not associated with the desired signals (e.g. detected laser light or dead pixels) and the determination of suitably averaged, differential (pump-on minus pump-off) images at each pump-probe time delay.  Typically, this can be accomplished by subtracting appropriate reference images on a per scan or per time-point basis and stacking repeated measurements. Shot to shot or scan to scan normalization of the signals can be used to diagnose and correct for some systematic changes during the experiment (e.g. source noise, beam intensity and position drifts).
    In some cases it can be desirable to remove background signals that result from the sample substrate or heating effects that are not removed by straight forward image subtraction.  Methods to accomplish this vary and have been developed by researchers on a case by case basis, although various approaches to background subtraction have been previously published (\citet{rene2017general, Siwick2004}).
    
    It is unlikely that the processing of UED and UEDS data and subsequent extraction of dynamical structural information will ever obtain the level of automation that is common in conventional/static xray or electron crystallography.  However, the further development of software tools that facilitate both the processing and exploration of time-resolved data, and the reliable, standardized, and quantitative extraction of meaningful structural information from it is urgently needed by the community.  Some recent progress on developing an open-source software ecosystem for UED and UEDS has been made (\citet{RenedeCotret2018}) and methods of time-resolved structural refinement have been published (\citet{liu2020chemistry}), but these efforts are in their infancy.  The development of codes that are capable of time-resolved structural refinement from data sets in which multiple scattering is not negligible is also highly desirable, but not yet available.   
    
\subsection{Examples from literature}\label{sec:ued_examples}

In this section is to present a selection of experimental results showcasing the unique capabilities of ultrafast electron diffraction tools. Owing to their short wavelength and large elastic cross-section, and thanks to technological development in acceleration, compression and control of dense high-brightness beams, electron probes can today efficiently capture the temporal evolution of irreversible processes, sample micrometer-sized areas, and deliver high reciprocal space resolution and signal-to-noise ratio for detection of weak signals such as thermal diffuse scattering, while at the same time maintaining a temporal resolution of 100 fs or below~\cite{cheng_light-induced_2022}\cite{}. As a consequence, an increasingly broad range of phenomena in the solid-state can be directly observed in single crystal, polycrystalline, monolayer and heterostructured specimens.
For a survey of the landmark works in the field, we refer the readers to previous reviews (\citet{zewail20064d, Sciaini2011, sciaini2019recent}).  

    \subsubsection{Following ultrafast evolution of irreversible processes with high brightness beams}
    
\begin{figure}
\includegraphics[width=1\columnwidth]{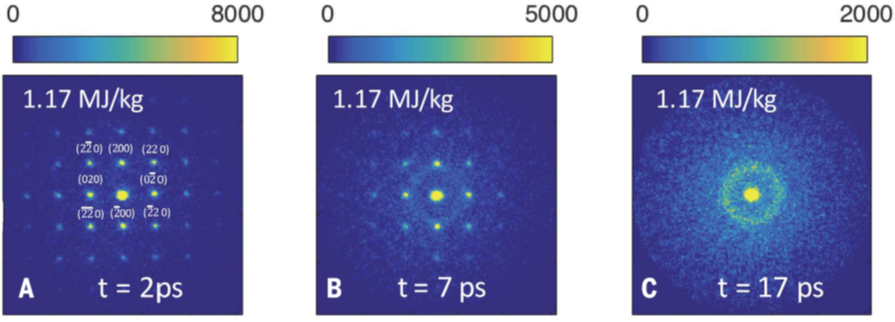}
\caption{Ultrafast photoinduced melting of Au as observed with single-shot UED. A) - C) MeV UED patterns of a 35 nm freestanding single-crystal gold film at three different time delays (indicated) relative to the arrival of a femtosecond laser pulse (400 nm) that deposits 1.17~MJ/kg of electronic excitation energy into the material. The initial period of lattice heating, driven by electron-phonon coupling, is evident in the suppression of the Bragg peak intensities at early times (panels A to B). The loss of crystalline order, or melting, is evident at later times as the Bragg spots are replaced by the diffuse ring pattern expected for the liquid phase (panels B to C). Adapted from ~\cite{Mo2018}.
}.
\label{fig:ued_example_order_disorder}
\end{figure}

    Some of the earliest work that applied UED to solid-state systems were performed to interrogate the irreversible processes involved in laser-induced melting and ablation of solids~\cite{Mourou1982, siwick_atomic-level_2003,Sciaini2009}. These processes have enormous practical relevance for laser machining and materials modification, and studies of matter under extreme conditions (e.g. warm dense matter), but also to questions of fundamental importance like the stability limits of crystalline solids (Lindeman vs Born), entropy catastrophe, heterogeneous versus homogeneous nucleation mechanisms\cite{siwick_atomic-level_2003, Lin2006, Mo2018} and non-thermal (or electronically-induced) melting~\cite{zier2015signatures}. Precise measurements of the material transformation requires at the same sub-picosecond temporal resolution and large diffraction signals generated from individual electron probes, i.e. high charge.
    UED signals are able to distinguish between lattice heating, which preserves long-range order (crystallinity), and the phase transition dynamics (order-disorder transition).  Lattice heating increases in the mean-square amplitude of atomic vibration about their lattice sites is associated with a characteristic reduction in the intensity of Bragg peak intensities in the UED patterns. As described in Sec.~\ref{SStheorytime}, this Debye-Waller effect is associated with a suppression of peak intensity that depends linearly on increases in $\langle u\rangle^2$ but quadratically on scattering vector.  Bragg peaks are not broadened by simple lattice heating, but are by a breakdown in the long-range order described by the reciprocal lattice vectors.  As crystalline order is lost through the course of a melting transition, Bragg peaks at large scattering angle are lost completely and those at small scattering angle are replaced by the diffuse rings of scattering intensity that are expected of the liquid/amorphous/disordered phase where only short range pair-correlations are present (similar to those of gas-phase samples described in Sec.~\ref{sect:GUED}). This is illustrated for laser-excited gold in Fig.~\ref{fig:ued_example_order_disorder}~\cite{Mo2018}, The diffraction patterns were each taken with a single 20 fC electron pulse, required due to the irreversible nature of the process.  High brightness, ultrafast electron beams are the primary enabler of such studies since SNR improves directly with bunch charge (see Eq.~\ref{B4Dmin} and Fig.~\ref{fig:ueds_snr}).
    
    Gold has both weak electron-phonon coupling and exhibits bond hardening following photo-excitation (~\cite{Ernstorfer2009}) so the melting transition takes $>$10~ps.  Aluminium has much stronger electron-phonon coupling and the same process was observed to occur in $\sim$3~ps via a homogeneous nucleation mechanism at sufficient pump fluences~\cite{Siwick:2003gb}. Strong photo-excitation of semiconductors was predicted to lead to non-thermal melting transition that is driven by purely electronic excitation from bonding-type valence band states to anti-bonding type conduction band states, not lattice heating.  This was observed directly in silicon by \citet{Harb2008}. Spin-lattice coupling has also recently been interrogated from the lattice perspective using UED ~\cite{windsor2021exchange, tauchert2022polarized}.
    
    \subsubsection{Exploring the dynamics of low-dimensional quantum materials}
    
    Reduced dimensionality can induce the emergence of quantum behaviour in materials through electron confinement. Quantum materials provide a rich playground for light-induced control of material properties, but direct access to the lattice dynamics is complicated by the faint signal associated with the small numbers of atomic layers (1-to-few).  The changes in lattice and charge order that is associated with the transformation can now be followed in remarkable detail with UED, as is illustrated by the example below. Thanks to the strong interaction of electrons with the lattice 
    even monolayer ~\cite{Mannebach_2015, HeXing2020} and few-layer heterostructures ~\cite{Luo2021} are accessible.  
 
\begin{figure}
\includegraphics[width=1\columnwidth]{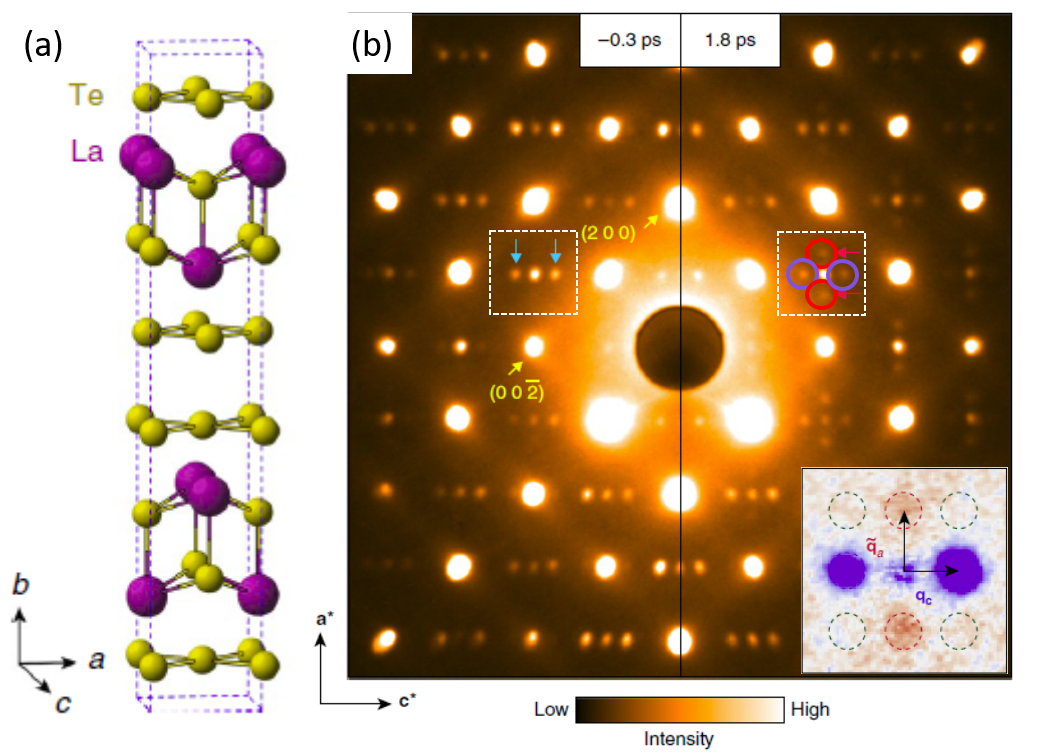}
\caption{Light-induced charge density wave order in LaTe$_3$: a) Structure of LaTe$_3$ showing two unit cells b) Diffraction patterns of LaTe$_3$ before (LHS, -0.3 ps) and after (RHS, 1.8 ps) photoexcitation showing various Bragg and superlattice peaks.  The superlattice peaks before photoexcitation (cyan arrows) results from a periodic lattice distortion along the c-axis that is associated with the equilibrium CDW phase (LHS, -0.3 ps). Following photo excitation new superlattice peaks appear (red circles), indicating the formation of a new CDW order along the a-xis at the expense of a weakened CDW order along the c-axis (purple circles).  Inset, the changes in superlattice peak intensities indicate that there is a competition between CDW order along these two axes at equilibrium and that this balance can be tipped by photoexcitation. Adapted from ~\cite{Kogar2020}}.
\label{fig:ueds_LaTe3}
\end{figure}  
    
    As one example, UED setups can be used to reveal symmetry breaking transitions, a concept that is central to condensed matter physics.  Whether such symmetry breaking can be controlled by optical excitation is a question of fundamental importance for the "properties on demand" type approaches described in Sec.~\ref{sec:ss_material_phenomena}. As an example, LaTe$_3$ is a layered compound in which a small lattice anisotropy in the $a-c$ plane results in a uni-directional charge density wave (CDW) along the $c$ axis (Fig. ~\ref{fig:ueds_LaTe3} a)). The periodic CDW lattice distortion yields superlattice peaks in the diffraction pattern that are distinct from the Bragg peaks of the undistorted structure (Fig. ~\ref{fig:ueds_LaTe3} b) (-0.3 ps));  i.e. new reciprocal lattice vectors. Using ultrafast electron diffraction ~\citet{Kogar2020} found that, after photoexcitation, the CDW along the $c$ axis is weakened and a different competing CDW along the $a$ axis subsequently emerges (Fig. ~\ref{fig:ueds_LaTe3} b) (1.8 ps)). The timescales characterizing the relaxation of this new CDW order and the re-establishment of the original uniaxial CDW are nearly identical, which points towards a strong competition between the two orders. The new density wave represents a transient non-equilibrium phase of matter with no equilibrium counterpart.

UED enables studies aimed at revealing how light can be used to control the structure of quantum materials by probing lattice and charge order directly.

    \subsubsection{Ultrafast electron diffuse scattering with high momentum resolution and SNR}

Ultrafast electron probes provide a unique tool for measuring the coupling between electron and phonons, and the evolution of phonon population in non-equilibrium scenarios. Such signal appears through patterns in the diffuse scattering background (UEDS). Accurate measurement of UEDS intensity across the momentum space requires high resolution in reiprocal space, to separate the Bragg and phonon diffuse scattering, and at the same time large momentum space field of view. Furthermore, SNR requirements are orders of magnitude more higher than for the case of Bragg peak detection, since the phonon diffuse intensity is (in general) several order of magnitude weaker and, therefore, competing with the measurement background floor.

\begin{figure}
\includegraphics[width=1\columnwidth]{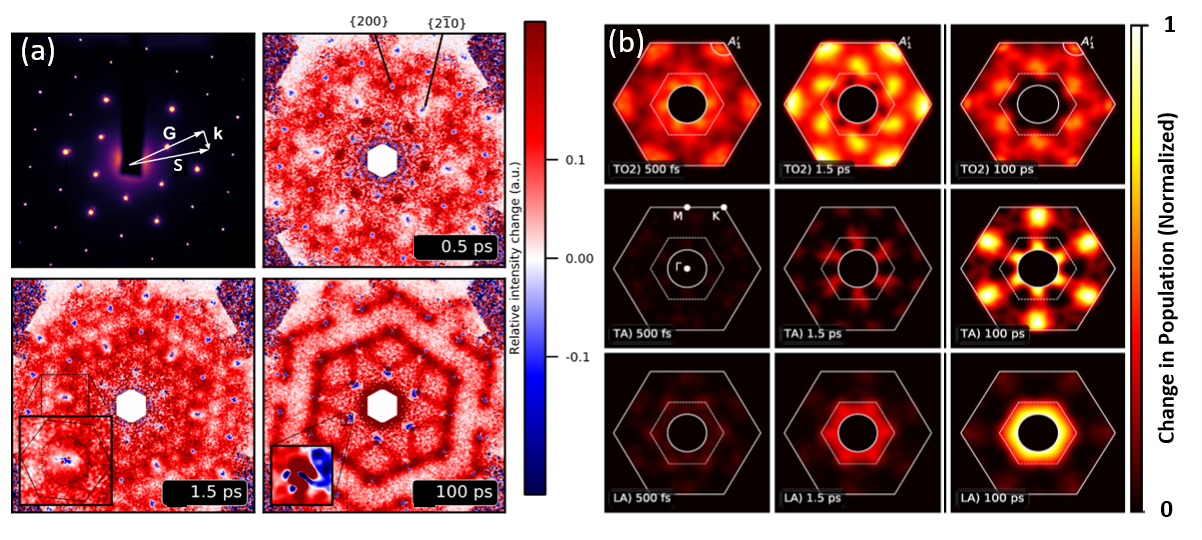}
\caption{Ultrafast electron diffuse scattering of electron-phonon coupling and nonequilibrium phonon relaxation in Graphite a) Following excitation at 800 nm, diffuse scattering provides direct information on the time-dependent changes in phonon occupancy.  Top left panel, raw electron scattering pattern of the graphite flake indicating the relevant vectors; $\mathbf{s}$, $\mathbf{G}$ and $\mathbf{k}$. Other panels show the change in electron scattering intensity $\Delta I(\mathbf{s},t) = I(\mathbf{s},t)-I(\mathbf{s},0)$ following photoexcitation for a few representative time delays (indicated). The data is remarkably rich. b) Time, wavevector and band-dependent changes in phonon population can be extracted from the data shown in a).  Those changes, everywhere in the hexagonal BZ of graphite, are shown for three phonon bands (TO, TA and LA).  Adapted from  \cite{ReneDeCotret2019}}.
\label{fig:ueds_example_graphite_phonons}
\end{figure}

Figure~\ref{fig:ueds_example_graphite_phonons}(a) shows an example of differential UEDS patterns in grafite, covering delay times between 0.5 – 100 ps following laser excitation. The impinging laser pulse drives vertical electronic transitions on the Dirac cones that provide an approximate Derun scription of the electronic bandstructure.  This excitation impulsively `photodopes' the material with a non-equilibrium electron-hole plasma of carrier density controllable by excitation fluence. UEDS has been used to show, from the perspective of the lattice, how these hot carriers come back into equilibrium with the phonon system and how the phonon system subsequently thermalizes through phonon-phonon relaxation and anharmonic decay. 
The evolution of the diffuse scattering  following photoexcitation is dramatic and striking.  An attractive feature of this technique is that a discrete, strongly coupled mode yields a peak in the differential scattering pattern at the BZ momentum position associated with that mode at short delay times due to the preferential (rapid) heating (see Eq. 31). This can be seen in the 0.5 ps  pattern at K-points around the (2$\bar{1}$0) peak and is also the explanation for the ‘star-like’ pattern of diffuse intensity that can be seen around the (200) peak. The data shown effectively provides a wavevector resolved map of the electron-phonon coupling strength in graphite ($g_{\mathbf{s}}$), which can be quantitatively extracted using the nonthermal lattice model described in the previous section (\cite{ReneDeCotret2019}). The diffuse scattering pattern at 1.5 ps reveals the decay channels for this population of strongly coupled optical phonons as they relax through anharmonic coupling into primarily mid-BZ acoustic phonons (a mix of LA and TA modes). On longer timescales the processes involved in the thermalization of this profoundly nonequilibrium, hot acoustic phonon system through momentum conserving phonon-phonon scattering processes are observed.  By 100 ps the acoustic phonon system appears to be thermalized, but a more detailed investigation revealed otherwise as described below.  
    
By complementing the UEDS with first principle density-functional theory calculations of the phonon polarization vectors, $\mathbf{e}_{j,\alpha,k}$ it is possible to transform the measured data into a map of the phonon populations for each mode as shown in Fig.~\ref{fig:ueds_example_graphite_phonons} b). The ability to obtain such information across the entire reduced BZ on ultrafast timescales is an important new capability for materials physics.  At 500 fs it is clear that optical phonons are primarily differentially excited.  At intermediate timescales, the anharmonic decay pathways of these strongly coupled optical phonon into acoustic phonons are seen.  At 100 ps it appears that the LA phonon branch is in a quasi-thermal state, with phonon occupancies following the expected 1/$s^2$ dependence.  However, the TA phonon branch is still in a profoundly non-thermal state even at 100 ps. There is a quasi-thermalized population of TA phonons around the zone center, but there is also a large population of high wavevector TA phonons near the M-points of the BZ that result from the momentum conserving relaxation pathways for phonons in the acoustic branches.  An unexpected observation.

UEDS provides rich time, momentum and branch resolved information on the state of the phonon system and has yielded insights into inelastic electron and phonon scattering ~\cite{Chase2016, Waldecker2017,maldonado2020tracking, Helene2021}, soft phonon physics ~\cite{Otto2021}, charge density wave ~\cite{cheng2022light} and polaron formation ~\cite{RenedeCotret2022} in materials.  Further improvements in time-resolution should enable an electron-based analog of fourier-transform inelastic xray scattering ~\cite{trigo2013, teitelbaum2021measurements}.

\section{Techniques and challenges in gas-phase time-resolved electron diffraction}

\label{sect:GUED}
    \subsection{Introduction}
        \subsubsection{Laser driven dynamical processes}
        Molecules can be thought of as atomic-scale machines that convert light into chemical energy and heat through the motion of atoms and the destruction and creation of chemical bonds. This intricate dance takes place on the picometer scale, with the speed of the moving atoms determined by internal forces. The fast motion, combined with the small distances over which they take place, results in structural changes taking place over tens to hundreds of femtoseconds. The accurate observation of these structural dynamics is essential for elucidating the reaction mechanisms, which has motivated the development of instruments capable of probing reactions with sub-Angstrom spatial resolution and femtosecond temporal resolution. First observations of these dynamics were enabled by the development of femtosecond lasers, which could be used to precisely trigger reactions and probe changes in their energy landscape, giving rise to the field of Femtochemistry \cite{Zewail2000}. These first experiments, however, lacked the spatial resolution that can be provided by scattering and imaging probes with sub-Angstrom de Broglie wavelengths. This section will focus on a method capable of spatially resolving nuclear dynamics in photo-excited molecules with femtosecond temporal resolution: Gas Phase Ultrafast Electron Diffraction (GUED).

        \subsubsection{Milestones in GUED}
\begin{figure}[ht]
    \centering
    \includegraphics[width=0.8\columnwidth]{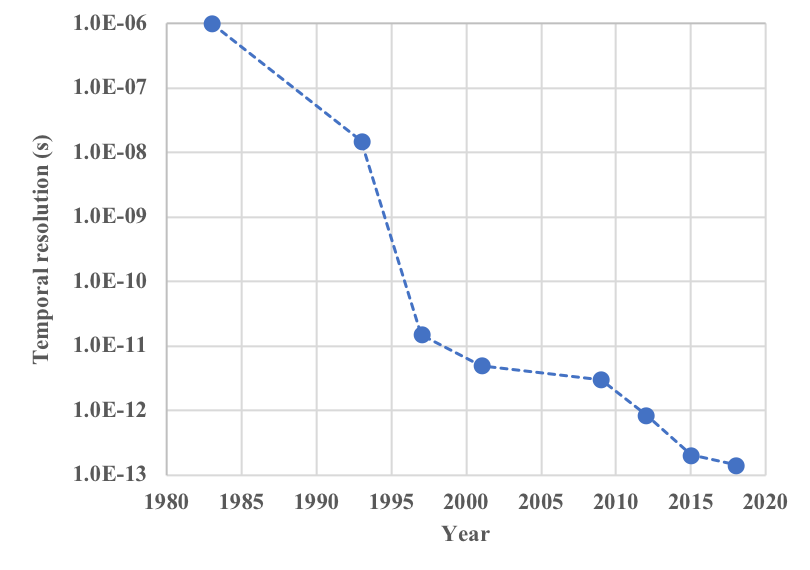}
    \caption{Temporal resolution in GUED over time, shown by a few representative experiments. The data taken from representative experiments corresponding to references \cite{Ewbank1993,Ihee2001,Ischenko1983,Reckenthaeler2009,Williamson1997,Yang2018_CF3I,Yang2016_N2,Hensley2012}.}
    \label{fig:GUED_tempRes_vs_year}
\end{figure}

        In a GUED pump-probe experiment, molecules in the sample volume are excited by a short laser pulse (the pump) and then probed by a short electron pulse which arrives at a predetermined time delay with respect to the pump. The resulting scattering pattern of electrons is recorded in a two-dimensional imaging detector, typically after accumulation of multiple electron pulses. Multiple snapshots of the changing molecular structure can be recorded by adjusting the relative time delay between the laser and electron pulses. Time resolved gas electron diffraction experiments where a sample was excited by a laser and probed by an electron pulse can be traced back to early experiments with microsecond resolution \cite{Ischenko1983}. From there the temporal resolution improved very rapidly, as shown in Fig. ~\ref{fig:GUED_tempRes_vs_year}. It was improved to 15 ns by incorporating photocathodes that were triggered by the same laser that excited the sample \cite{Ewbank1993}. Soon after, GUED experiments reached a resolution of a few picoseconds through the use of femtosecond lasers and improvements in detector technology, which were applied to capturing the structure of short-lived reaction intermediates \cite{Williamson1997}. These picosecond experiments relied on a DC acceleration of photoelectrons to energies between 30 keV to 60 keV and were extremely challenging, as the charge of the electron pulses was kept purposely low, on the order of few thousands or tens of thousands of electrons per pulse, in order to minimize the Coulomb broadening of the pulse duration. In addition, at the level of a few picoseconds, the velocity mismatch between laser and electron pulses starts to play a role in the temporal resolution, as the time delay between laser and electrons pulses changes as they traverse the sample. By further reducing the electron pulse charge and minimizing the distance to the sample and the size of the interaction region, subsequent GUED experiments were able to reach 850 fs resolution, which enabled the retrieval of the 3D structure of laser-aligned isolated molecules \cite{Hensley2012}. In these experiments, the reduction in electron flux and sample volume was compensated by operating at higher repetition rates. At this stage, the temporal resolution was limited as much by the duration of the electron pulses as by the temporal blurring that results from the velocity mismatch between laser and electron pulses. The next breakthrough came with the implementation of relativistic MeV RF photoelectron guns in GUED experiments, which improved the resolution to 230 fs \cite{Yang2016_N2} and more recently to 150 fs \cite{Yang2018_CF3I}. The use of relativistic electrons significantly lowered the space charge induced temporal broadening of the electron pulses, and reduced blurring due to velocity mismatch to the level of a few femtoseconds. This was a very significant technological advance because it enabled the direct observation of coherent nuclear motion in molecular reactions which takes place on time scales of a few hundred femtoseconds, resulting in the observation of vibrational and dissociative nuclear wavepackets \cite{Wilkin2019,Yang:2016I2}, spatially resolving the passage of a nuclear wavepacket through a conical intersection \cite{Yang2018_CF3I}, the observation of a ring-opening reaction\cite{Wolf2019} and of coherent dynamics in the reaction products \cite{Wilkin2019,Wolf2019}.

    \subsection{Pump probe requirements}
    \label{section:pumpprobeGas}
    GUED pump-probe experiments are directly sensitive to the relative positions of the nuclei in a molecule, which allows for probing both reaction kinetics and dynamics given sufficient temporal resolution. Reaction kinetics are concerned with the rate with which a product is formed, and the time scales can vary from femtosecond up to milliseconds and beyond. Reaction dynamics are concerned with the actual path that the nuclei take during the reaction, $i.e.$ the motion of each atom during a structural rearrangement. In most cases, this motion takes place on timescales ranging from tens to hundreds of femtoseconds. If the reaction dynamics are coherent, $i.e.$ all excited molecules undergo the transformation simultaneously, then the full motion of the nuclei can, in principle, be mapped using GUED. If the reaction is thermally driven, each molecule will still go through the reaction in a short time, but different molecules will undergo the transformation at different times, thus the GUED measurement can only capture the reaction kinetics, along with the structure of intermediate and final products. Recent improvement in the temporal resolution of GUED, which is currently in the order of 100 fs, have been transformative to the field, as they enable GUED to capture reaction dynamics.
    In addition to the temporal resolution, the pump laser and the sample delivery are crucial aspects of a successful experiment. Ideally the pump pulse will be designed to produce a specific excitation condition. This requires control over the duration, wavelength and fluence of the laser pulse. 

        \subsubsection{Temporal resolution}
    
    The temporal resolution of a GUED experiment has many contributions, as expressed in Eq.~\ref{eq:temporalresolution}. With the advent of commercial laser systems capable of delivering sub-30 fs pulses , the laser pulse duration $\tau_{pump}$ is seldom the limiting term in the overall temporal resolution of a GUED experiment (see Sec.~\ref{lasers}). 

\begin{figure}[ht]
        \centering
        \includegraphics[width=1\columnwidth]{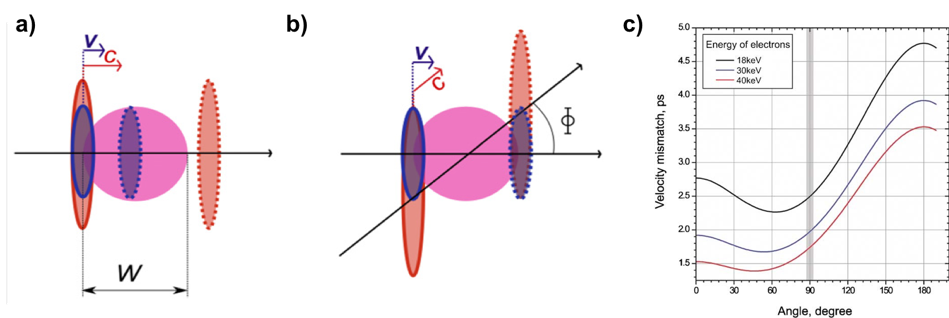}
        \caption{Panels a) and b) show a diagrammatic representation of the effect of velocity mismatch at two different interaction region geometries, with the optical (red) and electron (blue) pulses traversing the target sample volume (pink). W represents the width of the sample volume, and v and c the velocities of the electron and optical pulses. The temporal broadening induced by the geometry in panel a) can be calculated as: $t_{\mathrm{VM}} = (W/v)-(W/c)$. In panel b) the pulse front tilt angle and relative angle between the optical and electron beam, $\Theta = \arccos(v/c)$, compensates the effect of velocity mismatch and preserves the temporal resolution of the experiment. Panel c) shows the angular dependence of the temporal resolution. Panels a) and b) are adapted from Ref.~\cite{Sciaini2011} and panel c) is adapted from Ref.~\cite{Srinivasan2003}.}
         \label{fig:GUED_GVM}
        \end{figure}

The electron bunch length $\tau_{probe}$ is dependent on the pulse length of the drive laser, the initial energy spread in the electron bunch and the space-charge induced pulse broadening during propagation. Control and manipulation of $\tau_{probe}$ to obtain short electron pulses has been reviewed in Sec.~\ref{sectionII.c}.
In GUED experiments using non-relativistic electrons, the temporal resolution is typically dominated by the velocity mismatch between the pump laser and the probe electron, $\tau_{VM}$\cite{Williamson:1993gcba}. For electron beams with energies around 100 keV traversing a target volume a few hundred micrometers in diameter, the $\tau_{VM}$ term can be as large as 500 fs. Laser pulse-front tilting and non-collinear interaction geometries can be used to mitigate this contribution\cite{Shen2019,Zhang2014,Xiong2020}. A diagram illustrating the loss of temporal resolution due to velocity mismatch and how to overcome it are shown in Fig.~\ref{fig:GUED_GVM}. 

In GUED experiments with relativistic electrons the $\tau_{VM}$ term is typically less than 10 fs, and the overall resolution is rather affected by $\tau_{\Delta_{pp}}$ term, a consequence of fluctuations in timing and energy of the electron bunches, typically attributed to instabilities in the launching field and/or the timing of the drive laser system (see Sec.~\ref{section:longitudinalDynamics}).These effects are more pronounced in setups with RF cavity electron sources-based setup or those fitted with RF bunch compressors, as the use of time-dependent fields requires extremely precise synchronization between the drive laser and the cavity fields (see Sec.~\ref{section:clocking}). In principle, shot-by-shot data acquisition and time-stamping could enable the temporal sorting of the signal in post-processing, thus mitigating contributions from the $\tau_{\Delta_{pp}}$ term (see Sec.~\ref{section:timestamping}).

        \subsubsection{Laser pump pulses}
        \label{section_laserpump}
        The pump laser pulse parameters are selected to excite the molecule to a specific state or states. In many molecules energies above 4 eV are needed to reach the first excited state.
        The most commonly used laser source for UED experiments is a Ti:Saphire laser, which has a central wavelength of around 800 nm, which corresponds to 1.55 eV in photon energy. Higher photon energies can be reached using nonlinear optical processes to generate the second, third and fourth harmonics at 3.1 eV, 4.65 eV and 6.3 eV, repectively. An optical parametric amplifier (OPA) can be used to produce tunable wavelengths in the visible and UV down to 200 nm, which gives more flexibility to select the excitation wavelength but generally produces less pulse energy than the harmonic conversion.  
        The laser pulse duration must be short enough as not to impact the temporal resolution of the experiment. On the other hand, if the laser pulse is very short, this would result in a broad spectrum and simultaneous excitation of multiple energy levels which in some cases might not be desirable. 
        
        The fluence of the laser pulse is often a critical parameter in the experiment due to two competing requirements: excite a sufficiently high fraction of the sample volume and avoid multiphoton excitation. The signal to noise ratio (SNR) of a GUED experiment is directly proportional to the fraction of excited molecules, since the measured signal increases proportionally to the number of excited molecules while the noise remains unchanged. For comparison, the SNR increase associated with using higher bunch charges is, at best, proportionally to the square root of the electron beam current, since a higher number of electrons increases both the signal and the noise. For a 1-photon transition, the excitation fraction is proportional to the product of the absorption cross section of the molecule and the laser fluence. Most experiments require excitation of at least a few percent to achieve an adequate signal level. As the laser intensity increases, multiphoton channels need to also be considered. Having both single and multiphoton excitation in a single experiment is often undesirable as it makes the data interpretation much more complex. In some cases, although, it is possible to separate the dynamics arising from single and multiphoton channels \cite{Yang2018_CF3I}. Thus, the laser fluence must be optimized to yield the highest possible excitation percentage at the desired one-photon channel, while keeping multiphoton excitation to a minimum.
        In most GUED experiments, the nature of the excitation (single vs multiphoton) can be determined using a power scan, where the power of the laser is varied while monitoring a strong feature in the diffraction signal. If the changes in the feature are linear with laser intensity, the excitation is most likely single-photon, while a quadratic or higher dependence often indicate multiphoton excitation. This method, however, is far from ideal, as it needs to be carried out before the experiment and with little knowledge of the structural changes underlying different features in the diffraction signal. Moreover, a power scan takes up valuable beam time that could otherwise be used to acquire pump-probe data. Ideally, a separate experiment would take place before the GUED measurement to determine the laser intensity at which multiphoton effects become significant. 
        
        Finally, the pump laser must overlap the electron at the interaction region and allow for a uniform excitation of the sample volume. This can be achieved by either making the spot size of a laser beam with a gaussian profile slightly larger than the electron beam, or by shaping the laser beam into a flat top spatial profile. The required laser fluence is determined by the spot size of the electron beam and the absorption cross section of the target molecule at excitation wavelength. With typical electron beam sizes ranging between 100 $\mu$m and 300 $\mu$m, most experiments are performed with laser pulse energies between 10 $\mu$J and 100 $\mu$J in the UV, which requires a few-mJ laser pulse at 800 nm to drive the OPA. 

    \subsection{Sample delivery requirements}
    Careful design of a sample delivery system is necessary to ensure that an adequate number of intact molecules is delivered to the interaction volume. The upper bound to the sample density is set such that multiple scattering is avoided, while the lower limit is set such that there is a sufficient current of scattered electrons to overcome the noise. In most UED experiments the fraction of scattered electrons is limited by the achievable sample density, and is only a few percent, far from the regime where multiple scattering becomes an issue. In UED experiments, this minimum viable number of scattering events must reflect the fact that only photoexcited molecules undergoing structural changes contribute to the difference-diffraction signal. In UED, the percentage of molecules excited is kept deliberately low, around 10 $\%$, in order to minimize the likelihood of multi-photo absorption and the inadvertent capture of multiphoton dynamics, which are often challenging to assign and interpret. Time-resolved gas-phase UED experiments typically require 10$^7$ scattering events from photoexcited molecules per data point (time delay) in order to achieve a publishable signal level over a 10 \AA$^{-1}$ momentum transfer range. A smaller number of scattering events, $\sim6\times10^6$, is, however, required to resolve the static structure of non-photoexcited ground state species. To achieve a minimum viable 10$^8$ scattering events per data point, from a volume where 10 $\%$ of the molecules is photoexcited requires, precise control over the sample pressure at the interaction region, as well as the dimensions and geometry of the interaction volume. For example, for MeV UED experiments on small organic molecules, such as cyclohexadiene, with scattering cross-sections on the order of 10$^{-18}$ cm$^2$, achieving the minimum viable 10$^8$ scattering events for a single data point requires a sample density of $3\times10^{16}$ molecules cm$^{-3}$. This can be achieved using a 120 Hz electron source delivering 2 fC per pulse with an acquisition time of approximately 1 hour per data point, which equals $\sim5\times10^9$ incident electrons in total\cite{Wolf2019}. This equates to $\sim$250 scattering events per electron pulse, of which $\sim$25 arise from photoexcited species. For target molecules with larger scattering cross sections, on the order of 10$^{-17}$ cm$^2$, such as those containing heavy atoms, $e.g.$ 1,2-diiodotetrafluoroethane, the minimum viable sample density is commensurately lower, around $3\times10^{15}$ molecules cm$^{-3}$. An alternative is to increase the interaction length of the sample gas and electron beam, but this runs into practical limitations due to the focusing conditions and spatial overlap of the excitation laser, in addition to velocity mismatch and sample consumption issues.  Limitations in sample availability and vapor pressure and chamber pumping speed often limit sample density to 10$^{17}$ molecules cm$^{-3}$. The percentage of excited molecules in the sample volume is determined by the optical pump fluence, which may itself be limited by the available pump power and geometry constraints around the interaction region, in other words, the distance between interaction region and incoupling and focusing optics. Experiments at lower signal levels due to lower excitation fraction or lower sample density will result in a data set with a more limited range of momentum transfer, essentially reducing the spatial resolution. 
    
    Different nozzle designs have been developed and successfully employed in the study of samples with a wide range of scattering cross-sections, vapor pressures and thermal decomposition properties. When studying samples with low vapor pressure, sample density is often limited by the temperature of the sample. In these cases, precise control of the temperature across the sample delivery system is important, along with the ability to flow carrier gases through the sample. Sample delivery strategies commonly used in gas-phase UED experiments can be categorized under two main classes: pulsed and continuous nozzles. In this section we compare the main features of these nozzle types and discuss the considerations of nozzle selection. All dimensions, temperatures and pressures described hereafter refer to, or have been obtained at, the gas-phase MeV UED instrument at SLAC \cite{ASTA_RSI_2015,Shen2019}.

        \subsubsection{Continuous nozzles}
        Continuous nozzles have been used both as effusive and supersonic nozzles in GUED experiments. Effusive nozzles consist of a circular orifice a few tens of micrometers in diameter at the end of a gas transport line. Typically built out of stainless steel, without moving parts or sealing surfaces, these nozzles are extremely reliable and can be easily heated to facilitate the delivery of low-vapor pressure samples. When delivering samples with vapor pressure of less than 5 Torr at room temperature, it is recommended to keep the sample reservoir in vacuum and as close as possible to the heated nozzle. This minimizes the potential for cold spots along the transport line and reduces the risk of clogging the nozzle orifice with condensates. To achieve the highest possible sample density from the expanding gas plume, nozzles are typically placed one electron beam diameter away from the center of the interaction volume. This close proximity between the nozzle and intersection region can, in some cases, result in the ablation of the nozzle by the pump laser, leading to orifice damage and/or clogging. 
        
        An alternative approach to compensate the rapid decrease in gas density away from orifice is to elongate the interaction region, for example, by using flow-cells or nozzles with elongated orifices. Flow cells, consist of a transport tube with two circular orifices aligned with each other and oriented perpendicularly to the tube and parallel to the propagation axis of the electron. These cells enable higher sample densities and longer interaction region at the cost of a more angularly constrained interaction region and increased background pressure upstream of the chamber. For a flow cell with a 4 mm path length and 500 $\mu$m orifices, sample densities between of $3\times10^{16}$ and $1.6\times10^{17}$ molecules cm$^{-3}$ can be achieved for sample pressures of between 1 and 5 Torr. Most samples which are liquids at room temperature can achieve these vapor pressures with gentle heating ($<$ 60 degree Celsius). Nozzles with elongated orifices, also known as slit nozzles, often require high sample pressures to achieve identical density over the interaction volume, due to the rapid expansion of the gas plume. A 60 by 1000 $\mu$m slit nozzle requires a sample pressure of around 20 Torr to achieve a density of 10$^{16}$ molecules cm$^{-3}$.
        
        Supersonic nozzles are useful for producing a beam of rotationally cold molecules, and also result in a more collimated gas beam. These nozzles have a deLaval profile, with a small internal orifice followed by a conical opening. The sample is mixed with a noble gas at high pressure, to collisionally cool the target molecules as they go through the nozzle. In GUED experiments this gas is typically Helium as it offers the smallest scattering cross section and minimizes background scattering, even though heavier noble gases cool more efficiently. Using an internal hole of 30 $\mu$m and a backing pressure of 1 to 3 atmospheres, rotational temperatures in the range of 20 to 50 K can be achieved a short distance from the nozzle exit.
        
        Continuous effusive and supersonic nozzles typically require large amounts of sample, typically 1 mL per hour, and thus are better suited to higher repetition rate UED instruments.

        \subsubsection{Pulsed nozzles}
        Electromagnetic pulsed nozzles have been routinely used in GUED experiments at repetition rates up to 360 Hz. Although piezo actuated pulsed nozzles are theoretically capable of repetition rates above 1 kHz, their use in GUED experiments has not yet been demonstrated. GUED experiments often require nozzles to operate over a large temperature range to accommodate different samples, which has been a challenge for the piezo valves. In brief terms, electromagnetic pulsed nozzles use a solenoid and a set of springs to move a plunger inside the nozzle body in an oscillatory fashion. At the end of the plunger, a gasket material seals against the nozzle orifice. When the plunger moves away from the orifice, gases in the nozzle body move through the orifice into the chamber. The reciprocating motion of the plunger results in a pulsed flow of gas into the chamber. Orifice diameters in electromagnetic pulsed nozzles range between 50 and 200 $\mu$m. In order for the electron beam to transverse the highest possible sample density, the trigger delay and opening time of the pulsed nozzle must be adjusted to match the arrival of the electron beam at the interaction region. This is typically done by maximizing the scattering intensity of a known sample. Opening times are in the order of 175 $\mu$s. By only delivering sample when the electron beam is present, pulsed nozzles reduce the background pressure in the chamber and the sample consumption. This not only positively impacts the signal to noise ratio of the scattering signal, but also decreases the downtime associated with the emptying or replacing of the sample trap and the reloading of sample. For example, in a GUED experiment running a 360 Hz, the use of a pulsed nozzle equates to 16-fold decrease in sample usage compared to a continuous nozzle. Typical sample usage rates for a continuous nozzle are in the order of 2 mL per hour. To aid in the delivery of sample with low vapor pressure, pulsed nozzles can be backed with a few bars of helium. In these cases, the nozzle is heated to prevent sample condensation. However, heating is limited by the thermal decomposition properties of the sealing materials and solenoid wire coating. Known failure modes of pulsed nozzles include wearing of the sealing surfaces and/or plugger, solenoid damage from poor heat dissipation and orifice clogging with sample condensates and/or materials from nozzle wear. Positioning the nozzle horizontally can reduce the likelihood of clogging. In experiments where the rotational temperature of the sample molecules is not critical, the pulse nozzle is positioned as close as possible to the interaction volume without clipping the electron or optical pump beams. The distance between the pulse nozzle tip and the center of the interaction volume often lies in the 150 and 250 $\mu$m range. For a pulsed valve with a 100 $\mu$m orifice, sample pressures exceeding 40 Torr are required to achieve the minimum sample density of 10$^{16}$ molecules over the interaction volume sampled by a 200 $\mu$m FWHM electron beam.

        \subsubsection{Chamber design considerations}
\begin{figure}[ht]
        \centering
        \includegraphics[width=1\columnwidth]{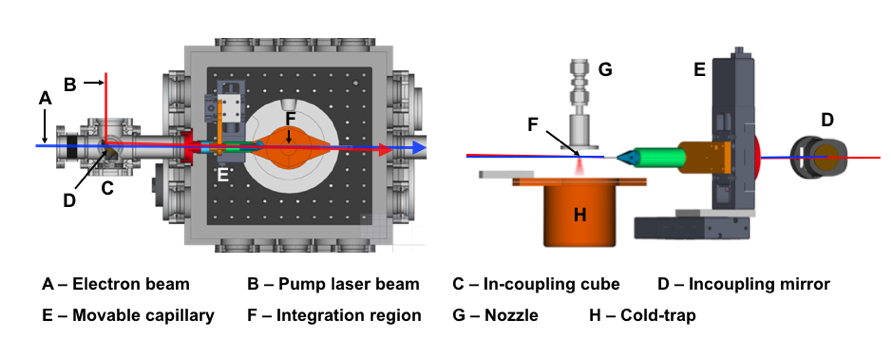}
        \caption{Schematic of the MeV GUED instrument at SLAC. Adapted from Ref. (Shen et al., 2019)}
         \label{fig:GUED_MeV_GUED}
        \end{figure}

        The design of a target chamber for GUED must address five major considerations: maintain adequate vacuum isolation between the target chamber and the electron source, establish the interaction region geometry, allow diagnostic tools to be moved into the interaction region, trap exhausted sample away from the interaction region and allow quick access to the nozzle and sample trap. These design considerations are addressed in the paragraph below. Figure~\ref{fig:GUED_MeV_GUED} shows an example of a GUED target chamber.

        \underline{Vacuum system}
        
        The gas load in GUED experiments is managed by a combination of turbo molecular or diffusion pumps and cryogenically cooled high surface area structures (cold-traps). Pumping speeds in excess of 1000 Ls$^{-1}$ are typically required to maintain chamber pressures in the order of 1$^{-5}$ Torr. Vacuum isolation between the electron gun, and sample chamber is maintained by a series of differentially pumped chambers. MeV RF guns require an operating pressure of 10$^{-10}$ Torr , while DC keV guns and RF compression cavities operate at pressures around 10$^{-7}$ Torr. Nevertheless, the use of gate-valves placed either side of the samples chamber is recommended. Gate valves allow the chamber to be vented independently of the rest of a setup, a welcome feature in GUED experiments where the nozzle needs to be frequently serviced and the cold-trap emptied. Moreover, these gate-valves, when interlocked to pressure gauges, protect the electron gun from contamination due to sudden pressure spikes in the target chamber.

        \underline{Interaction region geometry}
        
        The interaction region marks the overlap of the pump and probe beams with the target sample. The nozzle system is typically placed within a few hundred micrometers of the interaction region using a 3-dimensional translation stage. In setups using a collinear pump probe geometry, an incoupling 90-degree holey-mirror is placed inside a differentially pumped chamber positioned immediately upstream of the sample chamber. This chamber is kept at two orders-of-magnitude lower pressure than the chamber and thus preventing the mirror surfaces from being contaminated by sample molecules. A copper shower-stopper placed behind the holey-mirror protects its substrate from stray high energy electrons. In colinear incidence setups, both electron probe and laser pump beams are delivered to the interaction region using a long capillary a few millimeters in diameter, as shown in Fig.~\ref{fig:GUED_MeV_GUED}. The position of the capillary must be adjustable to allow the overlap between pump and probe beams while maintaining adequate clearances between the beams and the inner walls of the capillary.
        
        In setups using keV electrons, the laser and electron propagation directions are typically set at an angle between 60 and 90 degrees. This configuration is simpler in that the focusing optics can be kept outside the chamber, with the laser coupled in and out through viewports. In addition, the laser pulse front can be tilted to compensate for the velocity mismatch (Zhang et al., 2014), which for the case of 100 keV electrons requires and angle of 60 degrees between the beams.

        \underline{Diagnostics}
        
        The ability to verify the dimensions of the probe and pump beams, as well as, their spatial overlap is key to the success of GUED experiments. This can be achieved by imaging the beams onto a YAG screen and/or performing knife edge measurements using blades placed at the interaction region plane. These devices can be introduced to the interaction region by either independent translation stages or by being mounted to the nozzle system. By adding a crystalline sample to the diagnostic devices, one can also assess the temporal overlap of the pump and, based on the Debye-Waller response of crystalline sample following photoexcitation, produce a rough-estimate of the time-zero position of the instrument. Alternatively, the plasma lensing effect \cite{Dantus1994} can be used to determine the spatial and temporal overlap of the laser and electron pulses. Here the laser pulse energy is increased to ionize a sample gas, and the plasma produces a distortion in the electron beam.

        \underline{Sample trapping}
        
        Immediately adjacent to the interaction region, a cryogenically cooled high surface area trap is used to condense the exhausted sample. The use of a sample trap not only significantly improves the background pressure of the chamber and by extension the SNR of the diffraction data, but also increases the longevity of the pumping system. However, these improvements come at the cost of increased downtime from repeatedly venting the sample chamber in order to empty or replace the trap. When using a helium compressor based cryo-pump to cool the trap, steps must be taken to damp the propagation of vibration unto the chamber. The use of bellows to mount the cryo-pump and the use of flexible thermal straps to connect the cold-trap to the cryo-pump interface is recommended.

        \underline{Accessibility}
        
        Unhindered and ease of accessibility to the interaction region is key to an efficient GUED experiment. Therefore, the use of large access flanges or preferably doors, is recommended. During the experiment, samples have to be replenished and the cold trap cleaned. Additionally, windows and optics might need to be cleaned and the nozzle serviced.

        \subsection{Signal analysis}
\begin{figure}[ht]
        \centering
        \includegraphics[width=1\columnwidth]{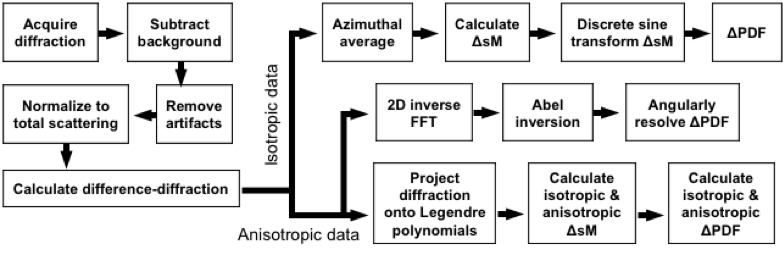}
        \caption{Block diagram of the data analysis methodologies used in GUED.}
        \label{fig:GUED_block_diagram}
        \end{figure}

        The analysis of GUED data follows the principles established by the diffraction difference method developed by Ihee et al.\cite{Ihee1997}. In this method, a reference signal acquired prior laser excitation, is subtracted from the overall time-dependent signal. The resulting difference-diffraction signal accentuates any features associated with photo-induced structure changes by removing the contribution of the atomic scattering and other background counts that are not time dependent. The resulting difference signal can then be further processed using one of the methods summarized in the block diagram in Fig. ~\ref{fig:GUED_block_diagram}. These methods allow the retrieval to time-dependent structural information.
    
        \subsubsection{Signal processing}
\begin{figure}[ht!]
        \centering
        \includegraphics[width=0.5\columnwidth]{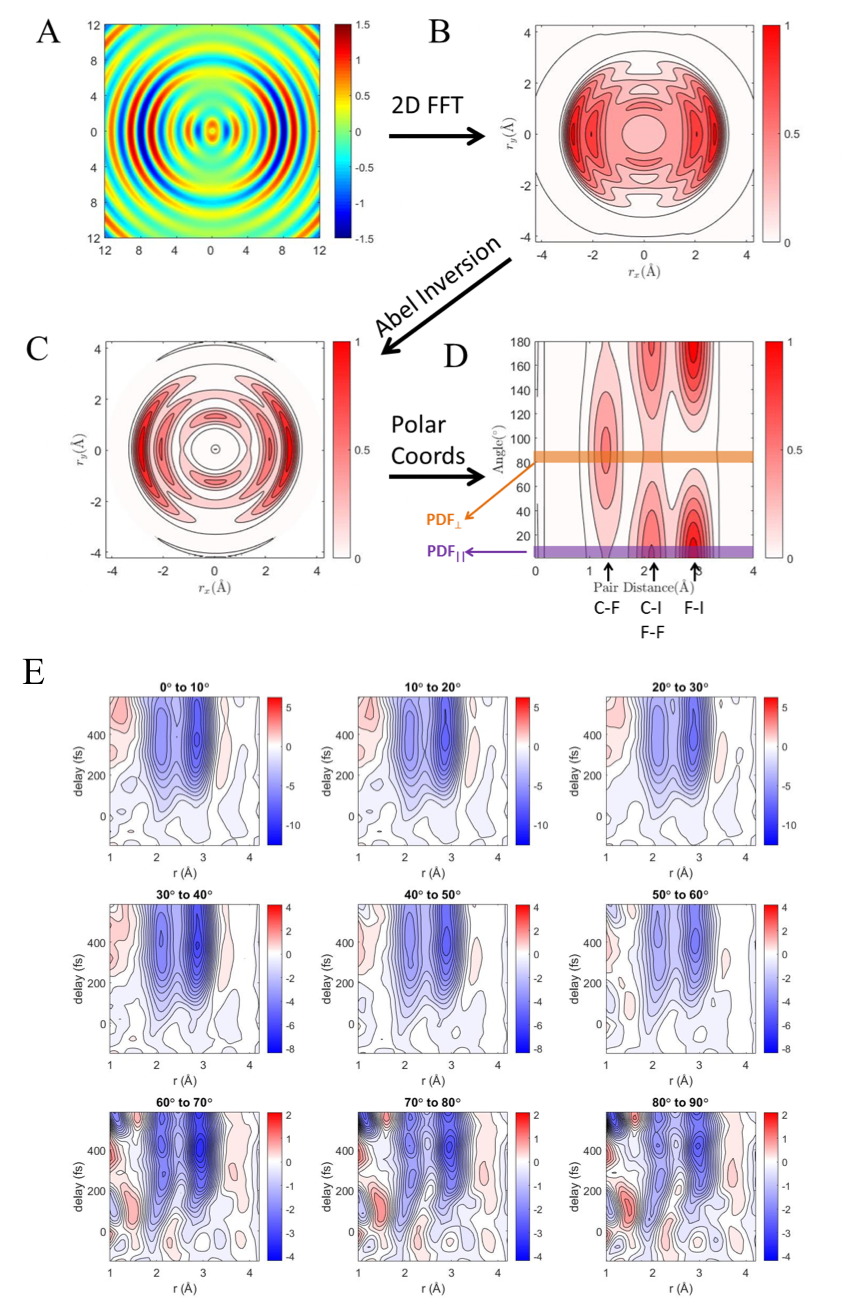}
        \caption{Example data analysis of angular dependent UED data (A) Simulated diffraction pattern for ground state CF3I with a cos$^2 \Theta$ distribution. (B) Projected PDF (C) Slice of the PDF obtained the Abel inversion of projected PDF. (D) PDF in polar coordinates with each peak corresponding to an atom pair, marked in the bottom. (E) Panels showing the $\Delta$PDFs obtained from a consecutive 10$^{\circ}$ cones of the difference diffraction pattern.}
        \label{fig:GUED_CF3I}
        \end{figure}

        The signal processing of a typical GUED experiment begins with the removal of detector and X-ray induced artifacts. Following the normalization of each diffraction pattern to the total scattering intensity, diffraction patterns recorded at the same delay stage position are averaged together and subtracted from a reference diffraction pattern recorded without the presence of a pump laser. The reference dataset is typically obtained by acquiring data a few picoseconds before the arrival of the pump pulse at the interaction region, $i.e.$ before time-zero. In isotropic datasets, difference diffraction patterns can be averaged azimuthally into a series of scattering curves, one per time delay. In datasets where photoexcitation results in an anisotropic distribution of excited species, difference diffraction patterns are further decomposed into an angle dependent scattering. These scattering curves are converted to modified scattering intensity curves, $\Delta$sM(s), using Eq.~\ref{Eq:sM}. The resulting time-dependent $\Delta$sM(s) can then be transformed into a time-dependent $\Delta$PDF, which provides a more intuitive representation of the structural dynamics at play. In anisotropic GUED datasets, a 2-D inverse Fourier transform followed by Abel inversion of diffraction-difference images results can be used to produce angularly resolved $\Delta$PDF. This method was successfully employed in the study of the photodissociation dynamics of CF$_3$I and CH$_2$I$_2$ \cite{Liu2020,Yang2018_CF3I}. Figure ~\ref{fig:GUED_CF3I} shows data analysis steps used to generate angular dependent $\Delta$PDFs for CF$_3$I. An alternative method to extract structural information from GUED data involves projecting scattering intensities onto Legendre polynomials in order to separate contributions from isotropic and anisotropic distribution of excited species. The 0$^{th}$ order Legendre polynomial encodes signals similar to that of an isotropic distribution of excited species, while higher order Legendre polynomials contain the information from the anisotropic part of the signal\cite{Baskin2005,Baskin2006}. These projected scattering intensities can then be converted into $\Delta$sM(s) and transformed into $\Delta$PDFs. This method was successfully employed in the UED study of C$_2$F$_4$I$_2$~\cite{Wilkin2019}.
        
        \subsubsection{Structural information retrieval methods}
        Several methods have been developed to extract structural information from experimental $\Delta$sM(s) and $\Delta$PDFs. For example, the structure of photo-products and reaction intermediates can be determined by using a least-squared fitting algorithm to find the set structural parameters which minimize the statistical $\chi^2$ between a calculated and experimental $\Delta$sM(s)\cite{Ihee2002}. This method is akin to the structure refinements used 
        in static GUED and is not suitable for the study of systems with multiple reaction pathways. In systems undergoing large changes structural information can be extracted from the $\Delta$PDFs directly. A photo-dissociation, for example, is expressed in the $\Delta$PDF as a localized bleach in the amplitude of a discrete set of distances, with an increase in the amplitude of distances commensurate with an increase in internuclear separation between photo fragments. By following the amplitude of $\Delta$PDF as a function of time, one can determine the timescale of structural changes, as well as the relative delay between their onset. This method was used to determine the timescale of carbon recoil and onset of the CF$_3$ fragment umbrella motion induced by the C-I bond fission in CF$_3$I\cite{Yang2018_CF3I}. Moreover, oscillations in $\Delta$PDFs 
        amplitude can also encode information on the structural dynamics of photo-products, as per illustrated in the study of C$_2$F$_4$I$_2$.\cite{Wilkin2019} With the aid of simulations these oscillations can be assigned to specific motions by comparing simulated and experimental 
        lineouts. This comparison can also be carried out in frequency space by Fourier transforming the $\Delta$PDF lineouts. This method was used to assign the rotation dynamics of CH$_2$ fragments produced during the photodissociation of CH$_2$I$_2$\cite{Liu2020}. Shifts and modulations of the $\Delta$PDF center-of-mass can also yield structural information once their origin is assigned 
        with the help of simulations. This is particularly relevant when exploring the dynamics of very broad and/or delocalized wavepackets and the relaxation dynamics of vibrationally excited photoproducts. This approach was successfully employed in the assignment of motions and 
        structural motifs to the major photoproducts of the photo-induced ring-opening of cyclohexadiene\cite{Wolf2019}.

\section{Opportunities and outlook for ultrafast electron diffraction} 
\label{Outlook}

Although UED has reached a high level of maturity as an experimental technique with several beamlines operating in a ‘user facility’ mode, continuous advancements in detection, acceleration and measurement techniques have the potential to enable further leaps in instrument performance. The development of new instrumentation in this area is far from complete. Different communities, from electron microscopy to particle accelerators, materials, condensed matter and atomic-molecular-optical sciences, have coalesced around this technique, using their skills and talent to advance its scientific breadth and impact into new areas, such as bio and catalysis related fields, for which novel methods of liquid sample delivery (nanofluidic cells or liquid micro-jets) have already been developed~\cite{Nunes2020,YangNunes_2020,Ledbetter2020}, showing an improved sensitivity to hydrogen bonds compared to ultrafast X-ray scattering measurements.

The cross-fertilization between ultrafast electron-based and x-ray-based science is expected to continue, introducing novel approaches to better harness the distinct characteristics of each approach. Different UED technologies will continue to advance in parallel, developing complementary advantages such as compactness, compatibility with sample environment, high temporal resolution and high average flux, disproving the concept of \textit{`one setup fits all'}, where a single superior technology clearly emerges. 

In the following we briefly discuss more in detail some of the possible future directions for UED instrumentation.

 \paragraph{Probe size}

The probe size is a key aspect of UED, one in which electrons have an important edge with respect to the x-rays as in principle an e-beam can be focused to a much smaller spot than what available at x-ray facilities. In practice, though, the limited average brightness of the electron sources has direct repercussions on the minimum probe spot-size that can be achieved at the specimen while maintaining sufficient resolution in reciprocal space. One solution is to modify electron microscope columns for photoemission to provide nanometer-scale beam sizes, but such setups suffer from limited temporal resolution and low electron flux. Custom setups based on particle accelerator technology can provide higher flux electron beams, but with typical probe sizes in the 50-100 $\mu$m range, orders of magnitude larger than those achievable in conventional electron microscopes.

Many relevant applications such as, for example, material engineering for energy harvesting and improved solar-to-electrical energy conversion efficiency, require a deep understanding of the energy flow in heterogeneous specimens (bulk, two-dimensional, nano-materials, organic and hybrid organic-inorganic compounds, etc...) as function of their local topographical and morphological properties, and demand probe sizes commensurate with grain sizes. Similarly, in Quantum Materials, spatially heterogeneous states and nanodomain formation appear to be common-place. More generally, nanometer-scale probes will enable access to key scientific problems related to local variations in dynamics response of materials due to variations of phonon spectra and density of states in vicinity of defects, impurities or boundaries, to discriminate the micro-texture in complex heterogeneous materials in space and time, and the observation of the energy transfer in the specimen in real-time.

Ongoing efforts to develop lower MTE photocathodes, higher accelerating fields, increased repetition rates and compact lenses with strong focusing gradients promise to reduce the spatial resolution gap existing between static and dynamics ED setups in the next decade, providing robust and stable (relativistic) femtosecond electrons packets of nanometer-size, with high average currents. 

\paragraph{Temporal resolution} 

The temporal resolution is currently limited by the electron pulse duration and the timing stability between the pump laser and the probe electron pulses. Higher accelerating fields, higher energies and RF frequencies and finer phase space manipulation techniques are likely to lead to shorter bunch lengths. In terms of arrival time stability, improvements are expected either through advancements in high speed electronics and controls, and due to the development of high precision time stamping tools combined with a new generation of fast detectors.

Compression of MeV electron pulses has recently been demonstrated to less than 30 fs, with significant improvements also on the timing stability \cite{Kim2020,maxson2017,Qi2020}. One can envision that at the current rate of progress it will not take long to reach below 10 femtosecond resolution in a UED experiment. Pushing the temporal resolution below 10 femtoseconds will provide access to electrically driven dynamics and high frequency optical phonon modes. For gas-phase, there are still faster dynamics that are out of reach. For example, capturing proton transfer or roaming reactions requires a temporal resolution on the order of 10 femtoseconds. Recent measurements have demonstrated that GUED is sensitive to electronic dynamics in addition to nuclear dynamics \cite{Yang2020_Pyridine} and a further, longer-term goal, is to reach attosecond resolution, a barrier which has been already achieved in X-rays \cite{xleap}. Note that for GUED, this would require revisiting the velocity mismatch even for MeV electrons, in addition to the pulse duration and timing jitter. 
        
\paragraph{Signal to noise ratio}
     
To elucidate the general rules that govern ultrafast dynamics, it is essential to carry out systematic studies where excitation conditions (laser wavelength, fluence,etc.) are varied, along with studies of comparison samples (either in solid state or gas form). Currently this is not possible due to the low probe beam average current, which results in low SNR and long acquisition times. The low SNR makes the data interpretation difficult, reduces the amount of information that can be extracted and so far has prevented the study of samples with low vapor pressure. A significant improvement could be made by introducing detectors with single electron sensitivity and fast readout such that an image could be read out after each electron pulse, as opposed to the current setup where many shots are accumulated at the detector. 

Increasing the electron beam current can be done either raising the charge per pulse or the repetition rate. While the first approach can be challenging, as it degrades the pulse duration and emittance, increasing the repetition rate can be done without degradation of the beam properties, but it is only effective if the sample is left enough time to relax in between shots. A hybrid DC-RF 90 keV UED setup has been demonstrated to operate at a 5 kHz repetition rate with a beam current that is one to two orders of magnitude higher (but temporal resolution much lower) than the current state of the art MeV-setup~\cite{Zandi2017,Xiong2020}. Using a MHz MeV electron gun can increase the electron beam current by an additional two orders of magnitude~\cite{Filippetto2016}. Any increase in the repetition rate must be accompanied by a proportional increase in the available average laser power to allow for efficient pumping of the sample volume. To a certain extent this could be achieved by reducing the effective volume of the interaction region.

An area for future growth is in the algorithms used to extract structural information from the UED patterns. As detailed in the previous sections, several methods have been developed and employed to analyze and interpret UED data in solid and gas phases, each optimized for a particular experiment and designed to extract a specific subset of the information content. The field could benefit significantly from the application of more advanced data science methods to maximize the amount of retrieved information, correct for multiple scattering and standardize analyses. This will be particularly important as new detectors are introduced with the possibility of recording diffraction patterns at much higher repetition rate, generating large datasets that will require at least some part of the analysis to be automated. It is also possible that in some cases the timing instabilities could be corrected as part of the data analysis itself~\cite{Fung2016}.

\paragraph{Beyond diffraction}
           
An area of great opportunity exists in carrying out multimodal measurements on a single photoinduced process to build a more complete picture of the dynamics. Each measurement can be thought of as a projection of some observables of the system, and thus usually requires modeling and theoretical input to be interpreted\cite{DeLaTorre:RMP}. Combining different measurements would provide more information and further constrain theoretical modeling for data analysis and interpretation, and thus provide a more rigorous comparison between experiment and theory. As an example, UED and time-resolved photoelectron spectroscopy have been recently combined to capture the nuclear motion together with the changes in the electronic state~\cite{Liu2020}. An enticing option would be  to pair up UED and ultrafast X-ray diffraction to disentangle the nuclear and electronic dynamics, since UED is sensitive to both electrons and nuclei, while X-rays scatter almost exclusively from electrons. This approach can be further expanded to combine UED with other laser-based experiments or with XFEL based spectroscopic or scattering measurements with nuclear and electronic sensitivity. 

Due to the similarity of the technology used, it is easy to envision RF-based UED beamlines close to XFEL experimental stations, with the electron pulses being naturally synchronized with the X-ray pulses. A variety of configurations can be imagined, where electrons and X-rays are used in the same chamber to study the same system, either to provide complementary information (elastic and inelastic scattering) or to access exotic  excitation modalities(X-ray pump electron probe or electron probe X-ray pump)~\cite{piazza:micron}.

Finally, while in this review article we focus on transmission electron diffraction, other electron-based scattering techniques will most likely benefit from the advances in electron sources and laser technology which have been discussed here, especially for what relates the transport and control of ultrashort electron pulses in the optical column \cite{limusumeci:prapplied, lu_uem_2018, denham:scaberr}.

The discussion of ultrafast electron imaging deserves a longer discussion that goes beyond the scope of this review. We only note here that the ability to obtain diffraction and microscopy information inside a single instrument could be a game-changer as recently showcased by  Ropers et al. \cite{danz2021ultrafast}, where the authors used a properly shaped mask to perform dark-field imaging of a crystalline specimen and follow the order parameter of a phase transition in an heterogeneous sample.

Similarly, being able to resolve in momentum space the electron energy loss spectrum would provide a wealth of information on the excited states of a specimen. Improved understanding of the longitudinal phase space manipulation techniques, such as minimization of the beam energy spread and the use of RF cavities as temporal lenses together with novel high resolution spectrometer diagnostics (time-of-flight or magnetic-based) are poised to make a significant impact here \cite{verhoeven2016time}.

\section{Acknowledgements}
\label{Ack}
The authors are grateful to J. Maxson, A. Kogar, Peter Denes, Dan Durham and Khalid Siddiqui for fruitful discussions. 
D.F. acknowledges support by the U.S. Department of Energy (DOE) Office of Science, Basic Energy Sciences, under the Contract No. DE-AC02-05CH11231.
P.M. acknowledges support from STROBE Science and Technology Center funded by National Science Foundation under Grant No. DMR-1548924.
R.K.L. acknowledges support by Tsinghua University Initiative Scientific Research Program No. 20197050028.
B.J.S. and M.R.O. acknowledge support from the National Science and Engineering Research Council of Canada, the Canadian Foundation for Innovation and les Fonds de recherches natures et technologies.
J.P.F.N. and M.C. were supported by the U.S. Department of Energy, Office of Science, Basic Energy Sciences under Award DE-SC0014170.

\newpage

\bibliographystyle{apsrmp}
\bibliography{RMP_UED}
\end{document}